\documentclass[pra,harvard,floatfix]{revtex4}

\usepackage{epsfig}
\usepackage{graphicx}
\newcommand{\beq}{\begin{equation}}
\newcommand{\eeq}{\end{equation}}

\begin{document}

\title{The Barkhausen effect}

\author{Gianfranco Durin$^1$
 and Stefano Zapperi$^{2}$
}

\affiliation{$^1$ Istituto Elettrotecnico Nazionale Galileo Ferraris,
 strada delle Cacce 91, I-10135 Torino, Italy\\
        $^2$ INFM SMC and UdR Roma 1, Dipartimento di Fisica,
    Universit\`a "La Sapienza", P.le A. Moro 2,
        00185 Roma, Italy}

\begin{abstract}
We review key experimental and theoretical results on the Barkhausen effect, focusing
on the statistical analysis of the noise. We discuss the experimental methods and the
material used and review recent measurements. The picture emerging from the
experimental data is that Barkhausen avalanche distributions and power spectra can be
described by scaling laws as in critical phenomena. In addition, there is growing
evidence that soft ferromagnetic bulk materials can be grouped in different classes
according to the exponent values. Soft thin films still remain to be fully explored
both experimentally and theoretically. Reviewing theories and models proposed in the
recent past to account for the scaling properties of the Barkhausen noise, we conclude
that the domain wall depinning scenario successfully explains most experimental data.
Finally, we report a translation from German of the original paper by H. Barkhausen.
\end{abstract}

\maketitle


\tableofcontents

\section{Introduction}
\label{sec:intro}

"Eisen gibt beim Unmagnetisionen ein Ger\"{a}usch; bei ganz gleichm\"{a}\ss igen
\"{A}nderung der magnetomorischen Kraft klappen die Molekularmagnete sprungweise in
ihre neue Lage und erzeungen dadurch in einer dar\"{u}ber geschobenen Spule
unregelm\"{a}\ss ige Induktionsst\"{o}\ss e, die sich im Telephon als Ger\"{a}usch
erkenntlich machen" \cite{BAR-19} [As the magnetomotive force is smoothly varied, the
molecular magnets flip in jumps to their new position. Because of this, they generate
irregular induction pulses in a coil wound around the sample, that can then be heard as
a noise in a telephone.](See the Appendix for a complete translation of the original
paper).

All the essential elements of the magnetization noise as discovered by H. Barkhausen
are contained in his words introducing the famous paper of 1919. A piece of iron, the
simplest magnetic material, produces a jerky noise when magnetized by a field smoothly
changing in time as, for instance, given by the slow movement of a magnetic yoke driven
by a hand. This noise is irregular, in contrast with the regularity of the applied
field, and can be easily heard with a microphone. This was the first indirect evidence
of the existence of magnetic domains, postulated a few years before by Weiss
\cite{WEI-07}: actually, Barkhausen believed that the noise was produced by the sudden
reversal of the entire magnetic domains (the "molecular magnets"). Only several years
later, in 1938, Elmore \cite{ELM-38} made the first observation of the motion of domain
boundaries in a cobalt crystal (described theoretically in 1932 by Bloch and thus
referred as Bloch walls) under the application of an external field. Curiously, Elmore
did not recognize this motion as the source of the noise heard by Barkhausen. This
instead was successively evidenced by Williams and Shockley measuring a SiFe single
crystal \cite{WIL-49}. It is worth noting that in this very year, Kittel in his famous
\textit{Physical theory of ferromagnetic domains}, warned "it may be well to correct
the widespread notion" that the Barkhausen effect is connected to the complete domain
reversal, as Williams and Shockley showed "quite clearly [...] that the Barkhausen
discontinuities correspond to irregular fluctuations in the motion of a domain boundary
(Bloch wall)" [from \cite{KIT-49}, pg. 9].

Once the origin of Barkhausen (BK) noise was understood, it was soon realized that it
could be used as an effective probe to investigate and understand the magnetization
dynamics in soft magnetic materials, and explain some of their hysteresis properties.
At the same time, it was also clear that the task was far from being easy. As
significantly noted "the Barkhausen effect research is both simple and hard"
[Ref.~\cite{MCC-76}, pg. 46], in the sense that it is relatively easy to perform a
measurement, but it is much more difficult to interpret it. This is due to the
stochastic character of domain wall (DW) motion, which proceeds in jumps, or
avalanches, when slowly magnetized. This dynamics is strongly affected by material
microstructure, and other effects such as the demagnetizing field, external stress,
etc. For this reason, earlier studies were focused on materials having the simplest
domain structure, like the SiFe single crystals (see for instance \cite{TEB-53}), where
it is possible to obtain a single DW in frame-type samples.

The intense activity of Barkhausen studies up to the 70s, both from the theoretical and
experimental side, are extensively described in the reviews of Rudyak \cite{RUD-71},
and of McClure and Sch{\"o}eder \cite{MCC-76}. Both papers report a large series of
experimental data on the statistical properties of the noise: distribution of duration
and size of the avalanches, power spectra, pulse propagation along the sample, etc. The
approach followed to interpret the data is fully phenomenological, without a precise
connection with the DW dynamics. For instance, power spectra \cite{MAZ-65,GRO-77} and
cross spectra using two separated coils \cite{GRO-78b} are described by a proper
superposition of \textit{elementary} independent jumps, whose spectral characteristic
are defined \textit{at priori}. These elementary jumps are not strictly related to well
specified DW displacements. Another approach tried to mimic the active interactions on
a DW considering the effect of a series of springs on a rigid wall, and calculating the
resulting power spectrum \cite{BAL-72,BAL-72b}.

Both reviews also present an exhaustive discussion on the possible effects of the
measuring setup on the detected signal, considering the effect of the dimensions of
both the solenoid and the pick-up coils, the different time constants related to the
electrical circuits, and, more importantly, the effect of the eddy currents. This
detailed analysis testifies the accuracy and capability in doing BK experiments in that
period. Most of the methodology is still valid and should be strictly followed, even if
this has not always happened in the recent literature.

The difficulties to correctly interpret the experimental data were partially explained
by the observation of Wiegman and ter Stege \cite{WIE-78} first, and later of Bertotti
\textit{et. al} \cite{BER-81}, who noticed as the statistical properties of the noise
typically vary along the hysteresis loop. Only considering a region where the DW motion
is the dominant or better the unique magnetization process, as around the coercive
field, it is possible to obtain a stationary signal and to correctly perform a
statistical analysis. This observation triggered a new series of detailed experiments,
which were quite well described by a new DW model, known as ABBM \cite{ALE-90,ALE-90a}.
Even if still phenomenological, this model has the merit to describe the statistical
properties of the noise on the basis of some simple assumptions verified
experimentally, and to be analytically tractable.

This new experimental determination of the noise properties put in evidence a series of
interesting characteristics. The BK noise is self-similar, and shows scaling invariance
and power laws. In other words, it has the typical features of a critical phenomenon,
whose detailed nature has been the subject of an intense debate. Following this line of
interpretation, in 1991, Cote and Meisel \cite{COT-91,MEI-92} claimed that the
Barkhausen effect is a particular example of a "self-organized critical" system
\cite{BAK-87}, in which there is no need to fine tune a control parameter, as in usual
second-order transitions. Following this suggestion, new experiments and models were
published quite regularly in the literature. After nearly 15 years, it is necessary to
review this large production of data and ideas and, taking into account the important
result previously appeared in the literature, to propose a single framework to
interpret the phenomenon. This is particularly important in order to deal with recent
observations of the Barkhausen noise lacking a coherent interpretation, such as for
measurements in thin films, hard materials, and low dimensional systems.

\section{Experiments and materials}
\label{sec:exp} In the vast literature about the Barkhausen noise, there is such a
large production and analysis of experimental data that there is a serious risk of
confusion and misunderstanding. Data have been collected on nearly all possible soft
magnetic materials, and considering many different properties, so that one can, in
principle, assume them as sufficiently exhaustive and reliable. The absence of a
coherent interpretation could then be ascribed to the lack of a proper theoretical
model.

Paradoxically, the situation is very different. Despite the large production of data,
the number of reliable results is quite small, and is restricted, with only a few
exceptions, to last decades of studies.  There are many reasons for this, as we will
discuss in detail in the following. One of the main points is that there is no general
agreement about a 'standard' experimental setup which would allow for a meaningful
comparison of the results. In fact, different experimental procedures can all produce
``nice'' Barkhausen data, but their interpretation could widely differ. The way the
measurement is performed influences significantly the details of the magnetization
process. Thus, there has been a strong effort in the last years to set experiments in
the simplest and most controlled way, avoiding unnecessary complications. This opened
the way to a convincing theoretical interpretation based on relatively simple models.
It is only through well defined experiments that parameters can be estimated and used
in the theory.

As discussed in the introduction, a crucial observation is that the noise properties,
such as the power spectrum, change along the hysteresis loop \cite{WIE-78,BER-81}. As a
consequence, it is important to measure the noise only around the coercive field in the
steepest part of the loop, where domain wall motion is the dominant (and often the
unique) magnetization mechanism, and the Barkhausen signal is stationary. This point
questions the reliability of many experimental results previously reported in the
literature.

It is convenient to critically discuss the experimental procedures proposed in the
recent literature, in order to identify a series of practical rules to establish a
'standard' setup. This will enable the direct comparison of different experimental
data, and the development of a common theoretical framework in which to analyze and
understand the magnetization process.

\subsection{Experimental setup}
\label{sec:exp_setup} As shown in the introduction, earlier experiments, starting from
the original one by Barkhausen, are all based on the induction of a flux change into
some concatenated coils in response to a slowly varying external field. The regularity
of the applied field contrasts with the irregular character of the induced pulses, as a
result of the random motion of the DW. More recent experiments make use of
magneto-optical methods based on the Kerr effect (MOKE) and are more suitable to
investigate the noise properties of thin films, as we will describe in detail (see
Sec.~\ref{sec:films}).

\subsubsection{Inductive measurements}

\begin{figure}
\centerline{\psfig{file=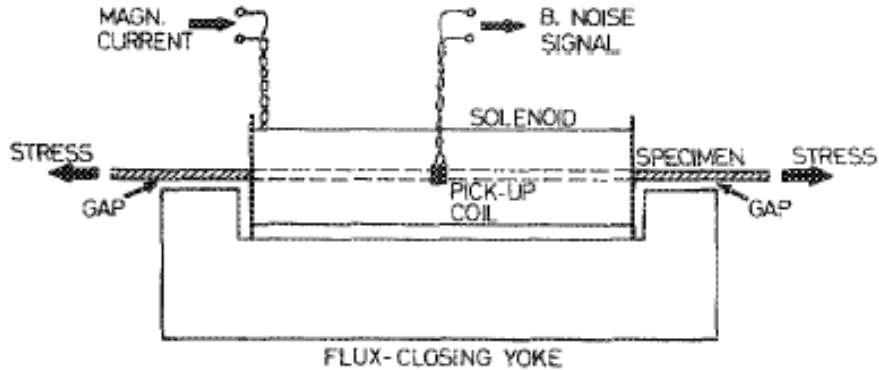,width=12cm,clip=!}} \caption{Schematic
representation of a setup for inductive measurements. In this case, the sample is
longer than the solenoid, so that the applied field is not uniform. The flux-closing
yoke is used to change the apparent permeability [From \cite{ALE-90a}, Fig. 1, p.
2909]} \label{fig:ale90af1}
\end{figure}

The essential elements of a inductive Barkhausen noise measurement are simply
identified: an external solenoid or a Helmholtz coil, able to produce a sufficiently
homogeneous field along the sample, and a certain number of pick-up (secondary) coils
wound around the sample to detect the induced flux. In general, it is a good practise
to use a solenoid much longer than the sample to ensure a uniform applied field. Also
in the case of application of an external tensile stress, it is preferable to use a
non-magnetic ribbon glued to the sample and attached to a weight (as done in
Ref.~\cite{DUR-00} using amorphous materials), so that the sample remains in the center
of the solenoid. To make measurements at different permeability there are two
alternatives with both advantages and disadvantages: as in \cite{ALE-88b}, a magnetic
yoke with a variable gap can be used, but, obviously, with a sample longer than the
solenoid (see Fig.~\ref{fig:ale90af1}); otherwise, the sample can be progressively cut,
as in Ref.~\cite{DUR-00}.

The induced signal $\dot{\Phi}$ both detects the contribution of the applied field $H$
across the pick-up coil (having a cross section $A_{pickup}$) and of the material
itself, so that $\dot{\Phi} = A_{pickup} \mu_o \dot{H} + A \dot{I}$, where $A$ is the
material cross section, and $I$ the magnetization. This expression is valid as long as
the eddy current shielding is negligible, as happens for the usual measurements
performed at very low frequency (usually up to no more than a few Hz). In high
permeability magnetic materials and for $A$ not too small in respect to $A_{pickup}$,
the contribution of the external field can be neglected for frequency up to several
kHz, thus the induced flux contains all the information about the material properties.
On the contrary, with low permeability materials, or when $A \ll A_{pickup}$, as for
instance, in thin films, it is necessary to compensate for the induced flux in air in
order to have $\dot{\Phi} \approx A \dot{I}$. This is usually performed using another
pickup coil with the same number of turns and the same cross section, but wound in the
opposite direction. A more refined solution requires a further coil put in air to get a
finer compensation \cite{WIE-78,WIE-79}. This solution is particularly important when
it is necessary to measure the hysteresis properties of thin films using inductive
methods \cite{SAN-03}.

In the simplest experimental conditions, as for a material having only two opposite
domains and thus a single DW, the induced flux is simply proportional to the DW
velocity $v$, as $\dot{\Phi} = N (2 I_s d v)$, where $I_s$ is the saturation
magnetization, $d$ the sample thickness, and $N$ is the number of turns. In the more
common case of a multiplicity of DWs, one can assume that $v$ measures the velocity of
the active DWs, even if this must be considered just as a rough approximation. In fact,
the induced signal indirectly collects, in a very complicated and unpredictable way,
the changes coming from all the rest of the sample because of long range spatial
interactions, and/or propagation of local deformations of DWs. This latter effect has
been quantitatively studied, showing that the flux variation decades roughly
exponentially with a typical scale of the order of a few cm \cite{GRO-78b}. This
suggests to keep the width of the pickup coil as small as possible (a few mm, in
practice). A measurement obtained with $N$ turns is thus a good approximation of a
measurement in a single cross-section of the material, with the advantage of the
amplification of the signal by the factor $N$. There is also an additional effect
suggesting to keep the width of the coils small. The magnetostatic field in an open
sample produces a counterfield known as demagnetizing field, which essentially depends
on the sample shape and on the domain structure. This field is spatially constant only
in a sample having the form of ellipse and \emph{at saturation}, otherwise it varies,
as for instance, when a domain structure is present. The demagnetizing effect has a
strong effect on the DW dynamics, limiting for instance the maximum size of a jump (see
Sec. \ref{sec:aval-distr}): it is thus convenient to consider a constant demagnetizing
field, and this is again obtained limiting the width of the secondary coils. A notable
example not following this rule is the setup used by Spasojevi\'{c} \textit{et al.}
\cite{SPA-96} where the pickup coil has a length of 5.5 cm, while the sample is only 4
cm long. This seriously questions the interpretation of the data and the obtained
scaling exponents given the strong spatial variation of the demagnetizing field.

Even following all these experimental prescriptions, we must clarify that the inductive
measurement is always a collective detection of many moving DWs, where propagation and
long range interactions still play a significant role. We emphasize it as the most
successful models proposed in the literature (see Sec. \ref{sec:theory}) are based on
the dynamics of a single domain wall. It is thus necessary to clarify what is the
reason of their success and, at the same time, understand their limitations. In
particular, we will describe in detail how the open problems, still under severe
theoretical and experimental investigation, could be possibly understood introducing
more complicated magnetization models which have not been considered so far.

A strict consequence of this collective detection of DWs motion is that the measured
signal cannot distinguish between single BK jumps and a superposition in space and/or
time of them. This limitation has strong consequences on the interpretation of the
statistical distributions of BK jumps, and in particular on the effect of the external
field rate (Sec.~\ref{sec:aval-distr}). Quite reasonably, it has been noted
\cite{WHI-03} that at zero driving rate (the so called adiabatic limit) the signal
should be considered a sequence of Barkhausen \emph{avalanches}, i.e. of elementary
jumps without any superposition. On the contrary, at finite driving rates, the signal
appears a sequence of \emph{pulses} containing a certain number of avalanches. We adopt
this distinction, as it is necessary to understand theoretically the effect of the
driving rate on the scaling exponents. Clearly, there is a practical difficulty to
establish the low frequency regime which approximates the adiabatic limit.

In order to further simplify the measurements, we finally note that the applied field
has usually a triangular shape instead of the common sinusoidal waveform used in
hysteresis loop measurements. As the noise is taken around the coercive field, as said,
this ensures to have constant magnetization rate, which is the fundamental parameter to
describe the variation of the BK statistical distributions (Sec.~\ref{sec:aval-distr}),
and of the power spectra (Sec.~\ref{sec:ps}).

\subsubsection{Magneto-optical Kerr effect measurements}

As described above, earlier measurements on thin films were done using inductive
setups, with fine compensation of air flux \cite{WIE-77,WIE-78,WIE-79}. These
measurements focused both on the noise power spectra \cite{WIE-77}, and on the
avalanche statistical distributions, detected along the hysteresis loop \cite{WIE-78}.
The authors were able to observe the noise in films down to a thickness of 40 nm. This
is a very surprising result, as we must considered that the size of avalanches roughly
diminish with the sample thickness, and thus can be considerably small at such
thickness. In ref.~\cite{WIE-78} the largest avalanche size detected is of the same
order of magnitude ($\sim 10^{-8}$ Wb) found in ribbons (cfr. Fig.~\ref{fig:dur00f1}),
while the largest duration is two order of magnitude smaller. This means that the 83-17
NiFe film used has DWs with an extremely high mobility. In other materials, the
situation cannot be so favorable. For instance, we have recently used a similar setup
to detect the noise in Co-base thin films. For thickness below 1 $\mu m$, the BK signal
becomes undistinguishable from the instrumental background noise, so that the
measurement is practically impossible. Thus the only valid alternative is to use
optical methods, based on the magneto-optical Kerr effect (MOKE), which is the change
of polarization of the light reflected by a magnetic material. This technique is
extensively used to investigate the domain structure of most magnetic materials
\cite{Hubert}. There are fundamentally two variations of the basic MOKE setup applied
to the investigation of the Barkhausen noise and in particular to the statistical
distributions. The first was introduced by Puppin \textit{et al.} \cite{PUP-00}, who
added an optical stage to tune the laser spot size onto the film, with values ranging
from 20 $\mu m$ up to 700 $\mu m$. The reduction of the background noise is obtained
using a photoelastic birefringence modulator working at 50 kHz, while the signal is
detected with a lock-in amplifier at 100 kHz to enhance the signal-to-noise ratio. This
system allowed to estimate the distribution of avalanche sizes in a Fe thin films
\cite{PUP-00a,PUP-00b} (see Sec.~\ref{sec:films}).

\begin{figure}
\centerline{\psfig{file=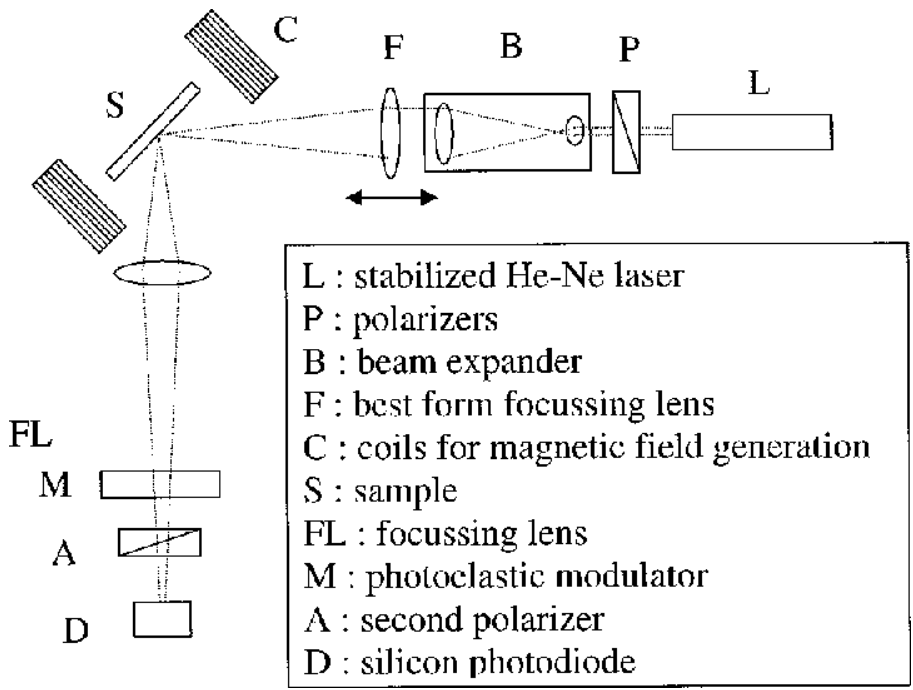,width=12cm,clip=!}} \caption{The focused
magneto-optical Kerr magnetometer with variable laser spot sizes used to investigate
the avalanche size distributions in thin films [From \cite{PUP-00}, Fig. 1, p. 1753]}
\label{fig:pup00}
\end{figure}

The second solution makes use of an advanced video processing technique to directly
observe the domains and their motion \cite{KIM-03,KIM-03a}. The main instruments used
are an objective of numerical aperture of 0.9 having a $\times$1000 magnification. The
domain images are obtained at the rate of 30 frames/s in real time and a spatial
resolution of 400 nm \cite{KIM-03a}. In this particular setup, the external field is
applied with an electromagnet up to nearly 99\% of the coercive field starting from the
saturation. Barkhausen jumps are thus triggered only by thermal fluctuations, so that
any sweep rate effect is completely eliminated. The experiment is performed repeatedly
imaging the same film area, using up to 1000 repetitions. The BK jump size is defined
as the area having two end points \emph{outside} the image; this choice raises some
problems in the interpretation of the experimental data, as we will discuss later
(Sec.~\ref{sec:films}).

\subsection{Magnetic materials}
\label{sec:materials} Nearly all soft magnetic materials have been studied using the
Barkhausen noise technique. But despite the large amount of published papers, a
systematic identification of material properties appears intrinsically hard to perform,
because of the large differences between the same type of material, the possible
annealing procedures, etc. Up to now, the only identification of different classes of
behavior (the "universality classes" deeply discussed in Sec. \ref{sec:depin}) has been
reported for SiFe polycrystalline ribbons with high Si content and amorphous FeCoBSi
alloys \cite{DUR-00}, which show two distinctive scaling exponents ruling the
Barkhausen distributions. This difference is understood in terms of a different DW
dynamics, as it will be discussed extensively.

Earlier fundamental studies have been performed in frame-type single crystals to
eliminate the effect of the demagnetizing field, where a single DW is present (see
Ref.~\onlinecite{COY-74,POR-79,VER-81,POR-81} for 3\% wt. SiFe, ref.
\onlinecite{WIL-49} for 3.8\% SiFe, Ref.~\onlinecite{WIL-50,TEB-53} for 4\% SiFe, and
Ref.~\onlinecite{GRO-77} for 4.5\% and 6.5 \% SiFe). The most studied polycrystalline
materials are the traditional non-oriented 3\% wt. SiFe strips
\cite{LIE-74,ALE-89,ALE-90a,GEO-94,BER-82,BER-81}, or the 1.8\% ones \cite{DUR-95}, as
well as the GO SiFe alloys \cite{STO-66,GRO-78b,KOM-85,KOM-86,RAU-86,JAN-82b,BER-81}.
Other interesting SiFe alloys are the ribbons (thickness about 50-60 $\mu$m ) with high
Si content (\% 6.5-7) produced by planar flow casting, which reveal interesting noise
characteristics \cite{CIZ-97,ZAP-98,DUR-00}. A large attention has received also the
NiFe alloys, such as the Permalloy (Ni$_{80}$Fe$_{20}$) and similar
\cite{LIE-72,BAL-72,LIE-77,WIE-78,GEO-94} or the Perminvar (30 \% Fe, 45 \% Ni, 25 \%
Co) \cite{URB-95,URB-95a}, particularly when a strong disorder is induced by
precipitation of some crystalline phases.

In more recent years, many studies have been focused on amorphous materials, especially
the iron-based family: some commercial ribbons, like the Metglass 2605
\cite{COT-91,MEI-92,OBR-94,PET-96}, the Metglass 2826 \cite{GRO-77b}, or the Vitrovac
6025 \cite{SPA-96}, and the vast production of various laboratories with different
composition and additives
\cite{JAN-82b,YAM-92,DUR-96,CIZ-97,ZAP-98,DUR-00,ZHE-02a,ZAN-03}. On the contrary,
Co-based amorphous alloys have received a very limited attention \cite{GEO-94}, due to
their high permeability values connected to the usual presence of a single large BK
jump.

Despite the large variety of compositions used, we should stress here that not all the
soft magnetic materials are appropriate for BK measurements. Especially, materials
having very nice 'soft' properties (high permeability, small losses, etc.) show reduced
or no BK noise, like the Co-base amorphous alloys or a well prepared Permalloy, with
limited structural defects. In fact, the essential ingredients for a 'nice' noise are
the presence of a consistent structural disorder, hindering the motion of DW, and a
non-negligible demagnetizing effect, limiting the size of avalanches. Also the
thickness appears to be an important factor to take into account, as it sets the limits
of eddy current shielding effect. For instance, SiFe ribbons having thickness of the
order of hundreds of $\mu m$, have to be magnetized at very low frequencies to detect
well defined BK jumps \cite{ALE-89,ALE-90a}.

\section{Experimental results}
\label{sec:exp_res}

\subsection{Barkhausen signal and avalanches distributions}
\label{sec:aval-distr}

\begin{figure}
   \begin{center}
   \begin{tabular}{ccc}
   \includegraphics[width=8cm]{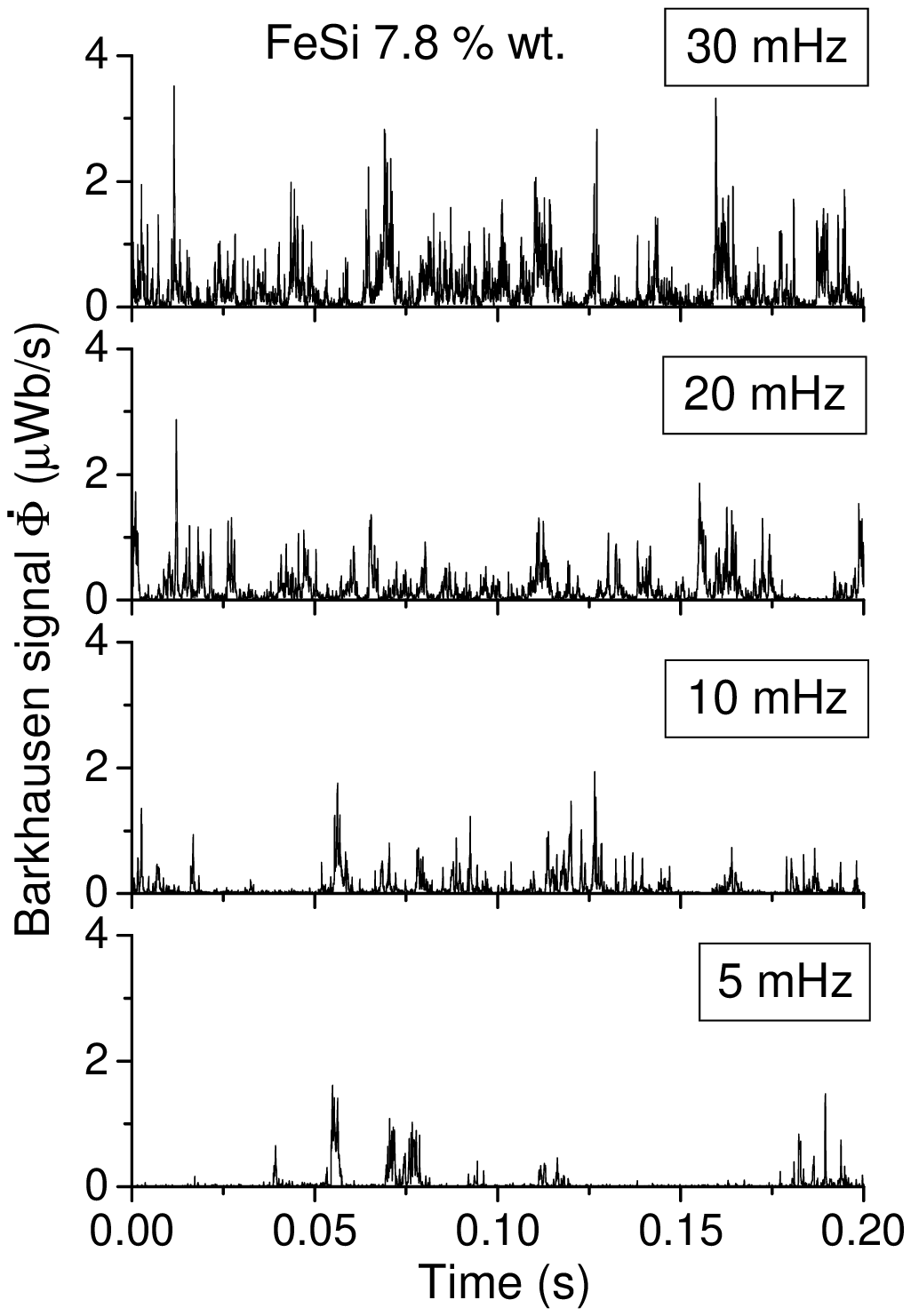}& &
   \includegraphics[width=8cm]{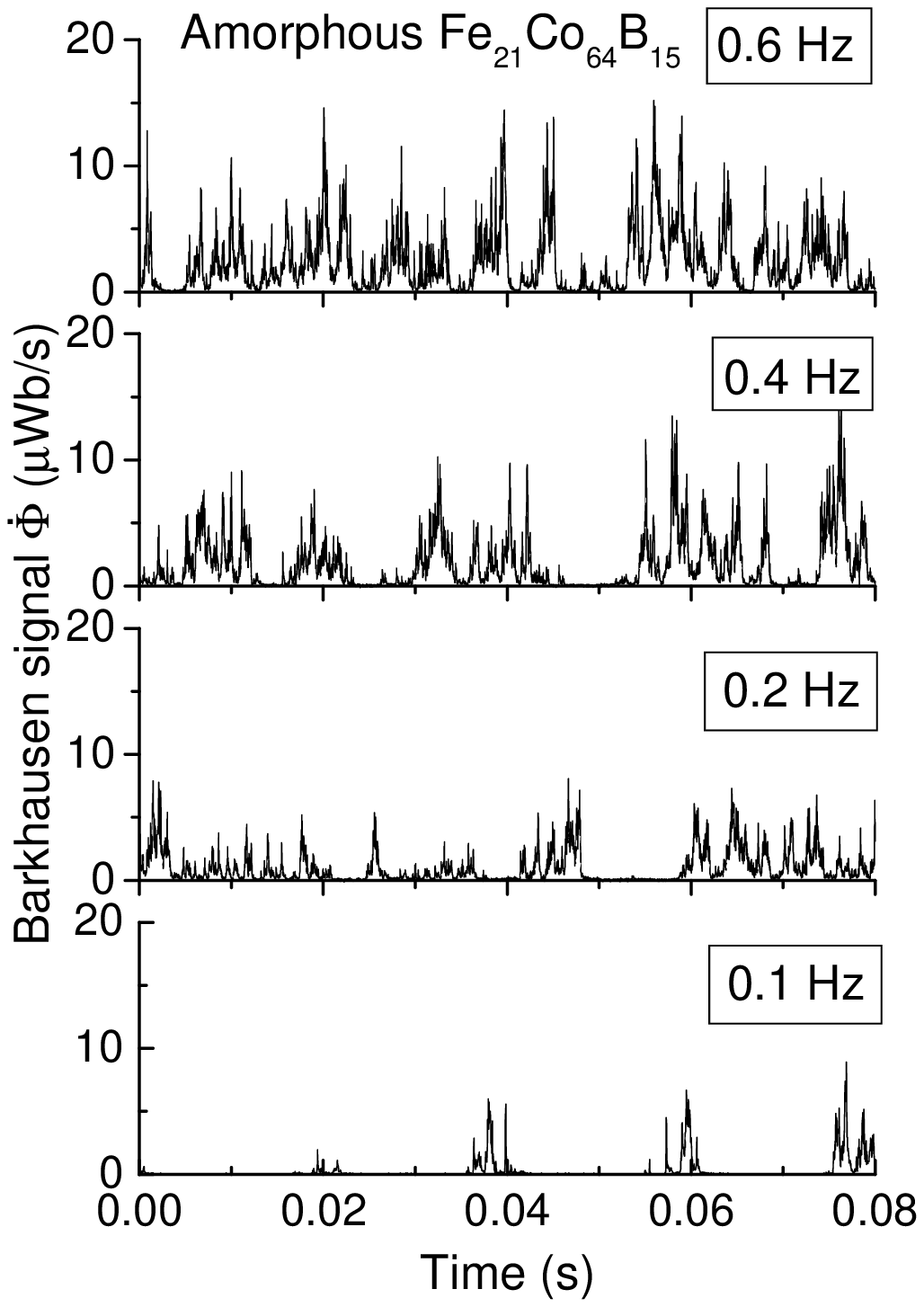}
   \end{tabular}
   \end{center}
    \caption {Time sequences of Barkhausen noise measured in a polycrystalline
    FeSi 7.8 \% wt. ribbon (left) and an amorphous
   Fe$_{21}$Co$_{64}$B$_{15}$ under moderate tensile stress
   (right). The labels indicate the frequency of the applied field.}
    \label{fig:time-seq}
   \end{figure}

In Fig.~\ref{fig:time-seq}, we show some typical experimental BK data obtained in a
polycrystalline Fe-Si 7.8 wt.\% ribbon and in an amorphous Fe$_{64}$Co$_{21}$B$_{15}$
ribbon, the latter measured under moderate tensile stress. The different sequences,
acquired during the same time interval, correspond to increasing applied field rates.
At low rates, the BK signal clearly comes out as a sequence of distinguishable and well
separated avalanches: at larger rates, this separation is progressively lost and the
noise resembles a continuous sequence of peaks. This feature is common to both
materials, even if this similarity is only seeming, as their statistical properties are
very different and appear to be representative of two types of behavior, the
'universality classes', with different DW dynamics, as we will describe in detail.

To investigate the statistical distribution of BK jumps is necessary to univocally
define the size and the duration of an individual jump. This poses a delicate problem,
given the self-similar fractal nature of the BK signal, and the fact that, in addition,
the background noise limits the detection of smaller avalanches. The common way to
solve this problem is to define a threshold, a 'resolution coefficient'
\cite{LIE-72,WIE-79,MEI-92,BER-94,DUR-95a}, in order to set the temporal limits of a
single avalanche. This procedure has been justified in the context of the fractal
properties of the signal, as it is possible to associate to it a random Cantor dust,
with a fractal dimension $\leq 1$, which can be used to calculate the critical
exponents \cite{DUR-95a}. The determination of the threshold can vary, but in general
we found the results are nearly insensitive to changes in its value \cite{DUR-95a}, as
long as it is not too small or too large. Because the BK signal amplitude $v$ follows a
power law of the type
\begin{equation}\label{eq:pdv-def}
    P(v) = v^{-(1-c)}f(v/v_0),
\end{equation}
where $c$ is proportional to the driving rate, we usually assume the threshold between
5 and 15 \% of $v_0$.

With this definition, the duration $T$ and the size $S$ (the integral within the time
limits) are unambiguously defined. Like the signal amplitude, also the distributions of
$T$ and $S$ follow power laws
\begin{equation}\label{eq:pdT-def}
P(T)=T^{-\alpha}g(T/T_0),
\end{equation}
and
\begin{equation}\label{eq:pds-def}
    P(S) = S^{-\tau}f(S/S_0),
\end{equation}
where $T_0$ and $S_0$ are two cutoffs, and $\alpha$ and $\tau$ the critical exponents.

\begin{table}
  \centering
  \begin{tabular}{|p{.7cm}|p{3cm}|c|p{4.3cm}|c|c|c|c|}
  \hline
  Ref. & Material & Type & Dimensions & $\tau$ & $\alpha$ & $1/\sigma \nu z$ & Rate dep. \\
  \hline
  \cite{LIE-72} & 81\%NiFe & Wire & 50cm x $\Phi$ 1 mm & 1.73 & 2.28 & 1.63 & Yes \\
  \hline
  \cite{SPA-96} & Vitrovax 6025X & Ribbon & 4 cm x 1 cm x 30 $\mu$m & 1.77 & 2.22 & 1.51 & ? \\
  \hline
  \cite{URB-95a} & Perminvar & Strip & 5 cm x 2 cm x 100 $\mu$m & 1.33 & - & - & No \\
  \hline
  \cite{MCM-93} & Annealed steel & Strip & 2.5 cm x 13 cm x 840 $\mu$m & 1.24;1.27 & - & - & No \\
  \hline
  \cite{DUR-00} \cite{DUR-02} & Fe$_{64}$Co$_{21}$B$_{15}$,  Fe$_{21}$Co$_{64}$B$_{15}$
amorph. (tensile stress)
  & Ribbon & 28 cm x 1 cm x 20 $\mu$m & 1.3 & 1.5 & $\sim$1.77 & No \\
  \hline
  \cite{MEH-02} & Fe$_{21}$Co$_{64}$B$_{15}$ amorph. (no stress)& Ribbon & 21 cm x 1 cm x 20 $\mu$m
  & 1.46 & 1.74 & 1.70 & No \\
  \hline
  \cite{DUR-95} & SiFe 1.8\% & Strip
   & 20 cm x 1 cm x 180 $\mu$m & 1.5 & 2 & - & Yes\\
  \hline
  \cite{DUR-00} & SiFe 6.5\%, SiFe 7.8\%, Fe$_{64}$Co$_{21}$B$_{15}$ partially cryst. & Ribbon
   & 28 cm x 0.5 cm x 48 $\mu$m
   30 cm x 0.5 cm x 60 $\mu$m
   30 cm x 1 cm x 20 $\mu$m & 1.5 & 2 & $\sim$2 & Yes\\
  \hline
\end{tabular}
\caption{Experimental critical exponents reported in the literature which can be
considered sufficiently accurate and reliable. Exponents $\tau$, and $\alpha$ are the
critical exponent of the size and duration distributions, respectively. The exponent
$1/\sigma \nu z$ relates the average size of an avalanche to its duration (see text).
"Rate dep." indicates if the exponents $\tau$ and $\alpha$ depend on the applied field
rate.}\label{tab:crit-exp-exp}
\end{table}

\begin{figure}
    \begin{center}
    \begin{tabular}{ccc}
      \includegraphics[width=8cm]{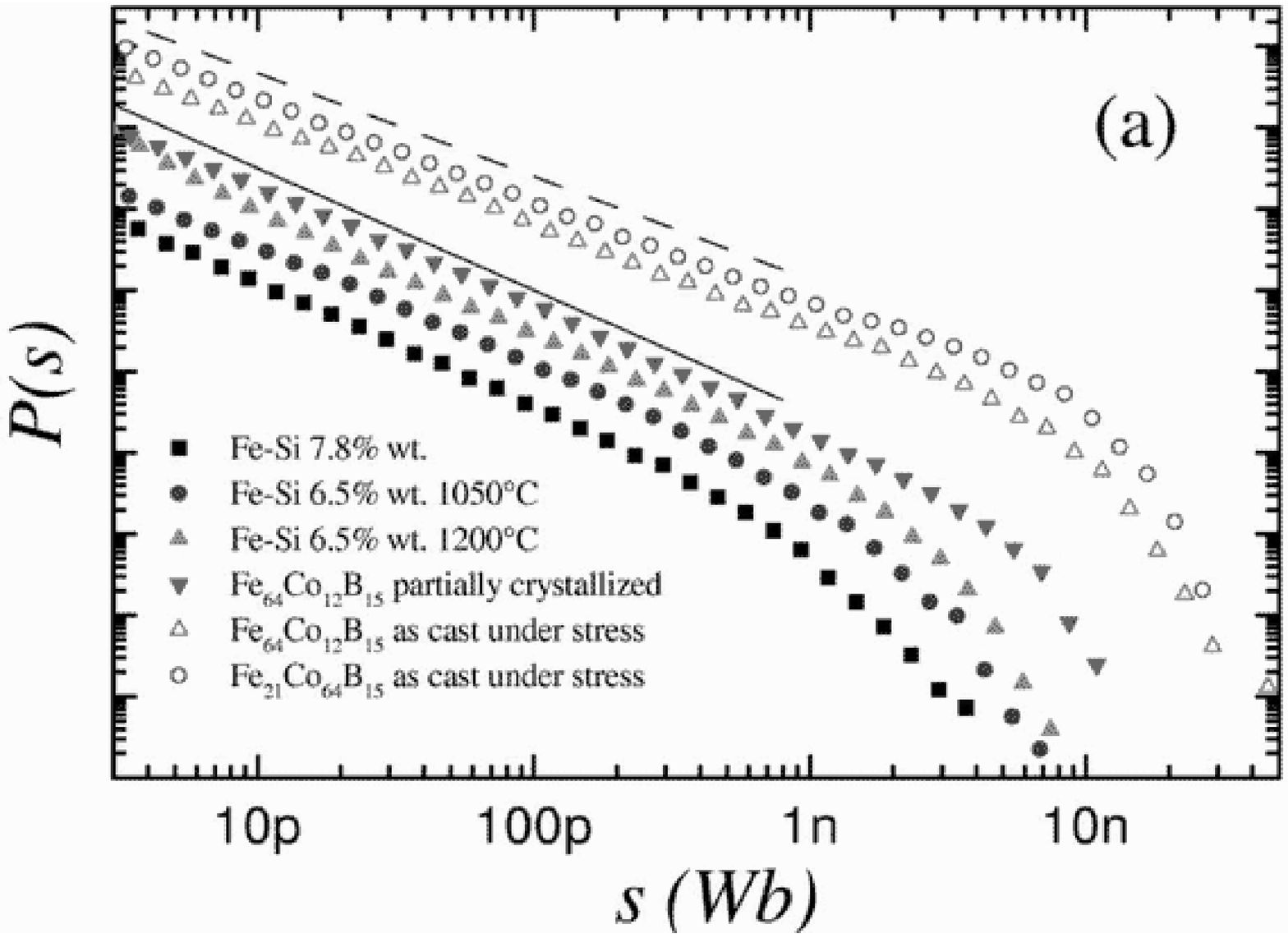} & &
      \includegraphics[width=8cm]{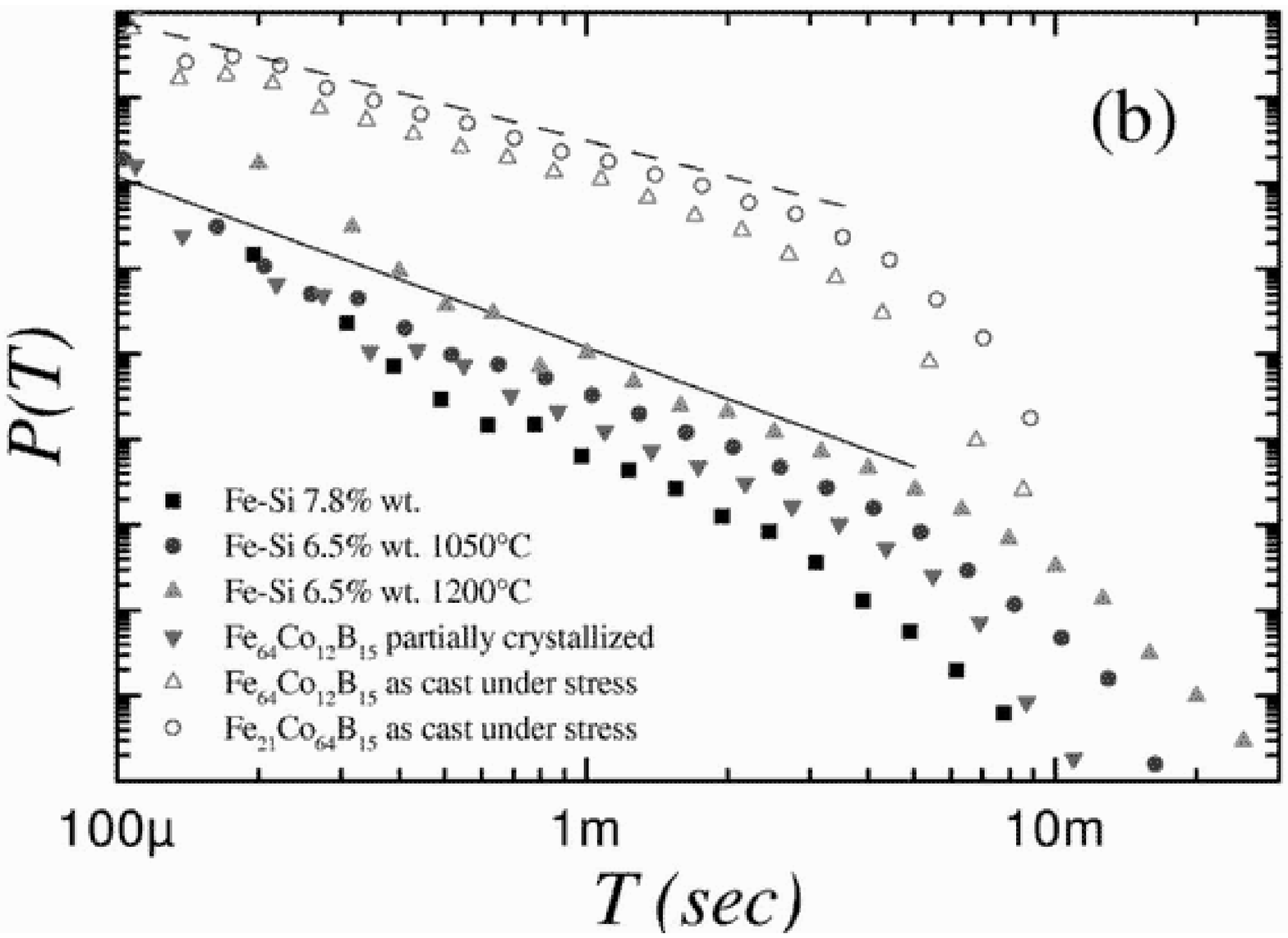}
   \end{tabular}
   \end{center}
   \caption{(Left) Distributions of Barkhausen jump sizes measured in
different materials.  The solid line has a slope $\tau \sim 1.5$ while for the dashed
one $\tau \sim 1.27$.  (Right) Similar plot for duration distributions. The solid line
has a slope $\alpha \sim 2$, while for the dashed one $\alpha \sim 1.5$. [From
\cite{DUR-00}, Fig. 1, p. 4706]}
    \label{fig:dur00f1}
   \end{figure}

\begin{figure}
   \begin{center}
   \begin{tabular}{ccc}
   \includegraphics[height=6cm]{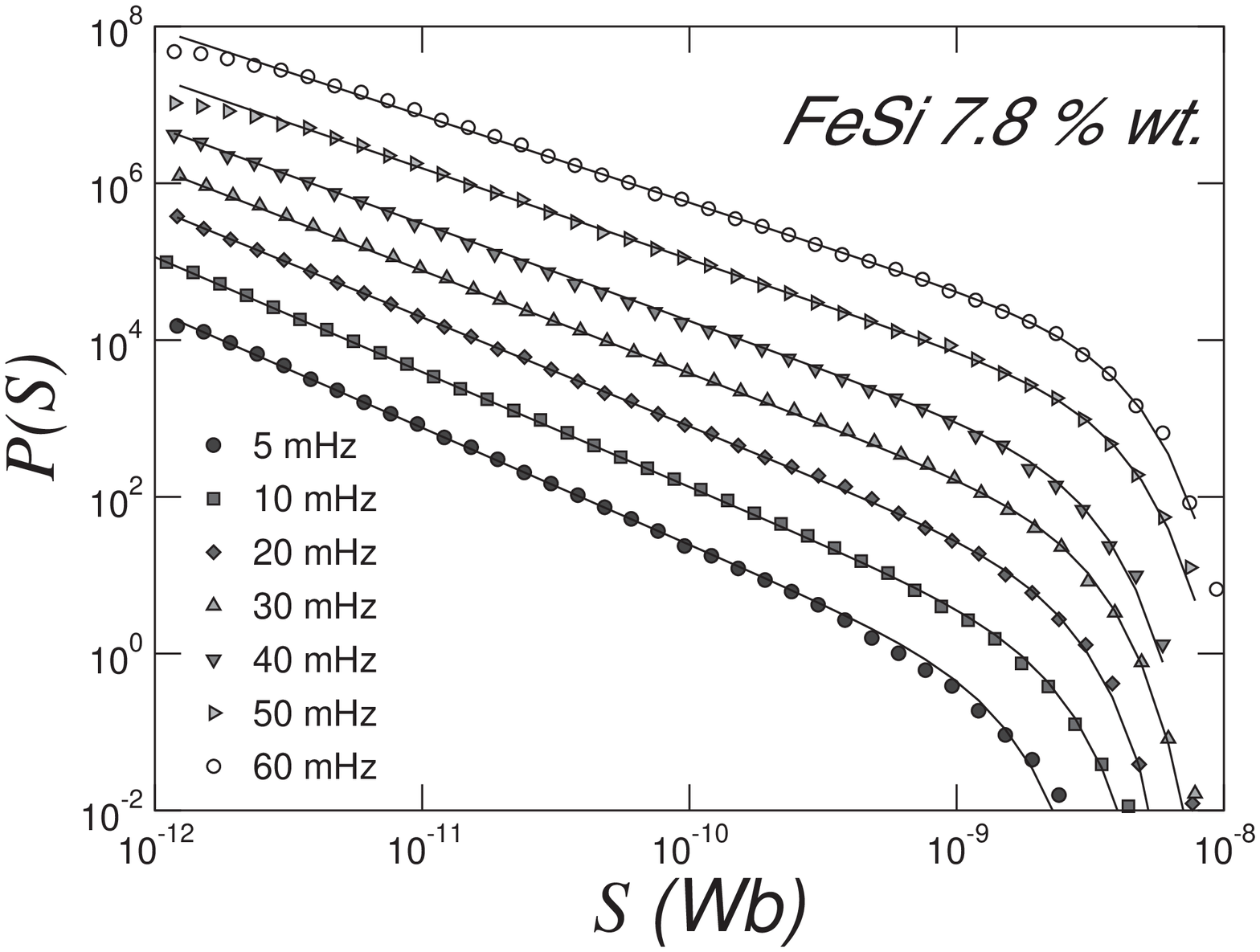} & &
      \includegraphics[height=6cm]{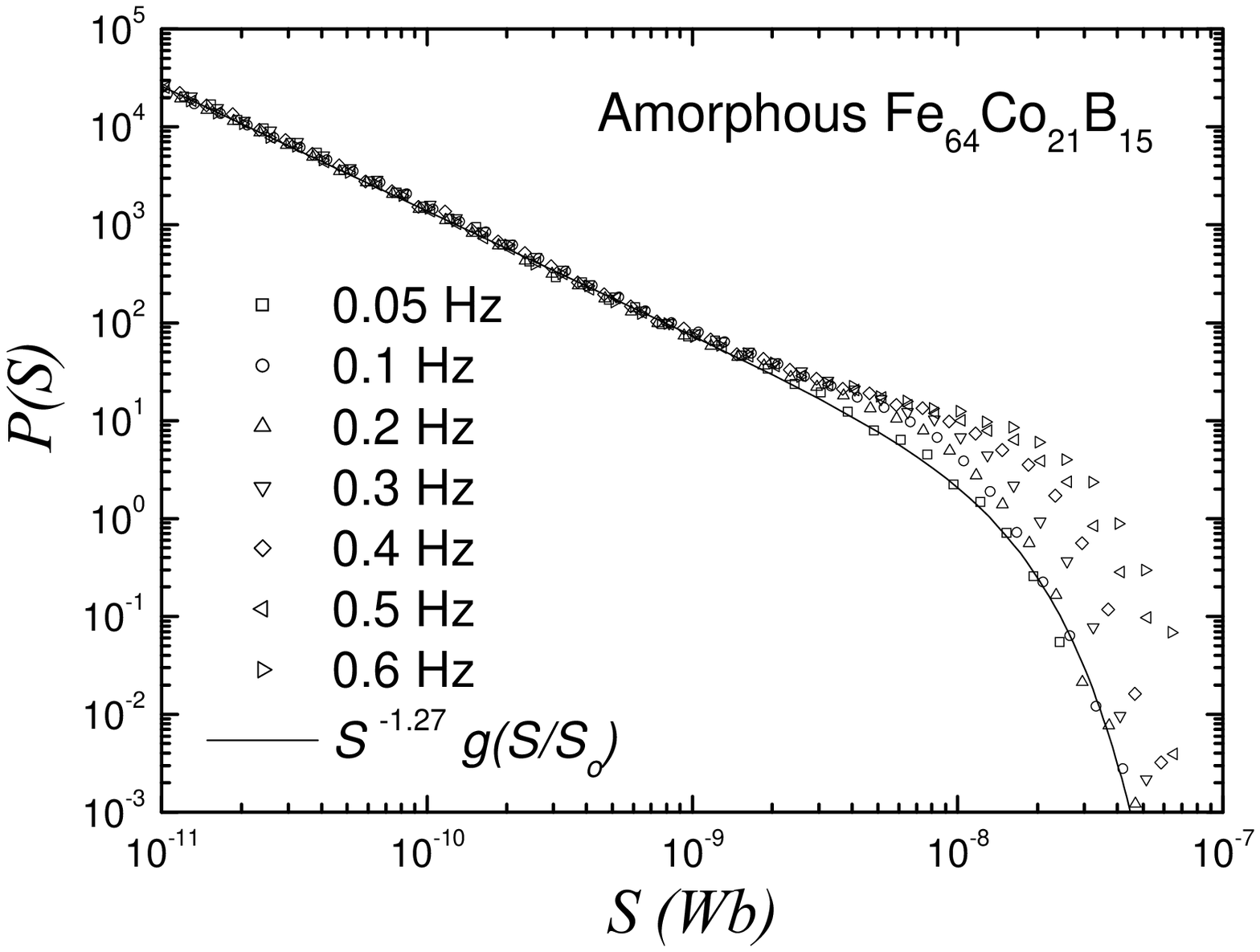}
   \end{tabular}
   \end{center}
   \caption{Distributions of the avalanche sizes $S$ for the
    polycrystalline SiFe (left) and the amorphous (right) material at
    different applied field frequencies. In the polycrystalline
    sample, the critical exponent $\tau$ varies linearly with the
    frequency \cite{DUR-95,DUR-95a} between 1.5 and 1.15, while it is
    independent in the amorphous one $\tau \sim 1.27$.}
    \label{fig:pds-rate-dep}
   \end{figure}

It is interesting to note that, despite the large number of papers regarding the BK
noise, the reliable estimates of these critical exponents are very limited. We report
in Tab.~\ref{tab:crit-exp-exp} the results presented in the literature we believe are
sufficiently accurate to be considered and analyzed. We leave out, for instance, the
results of the quite famous papers of Cote and Meisel \cite{COT-91,MEI-92}, who first
suggested the BK exhibits self-organized criticality \cite{BAK-87}, because they are
based on a very poor statistics. At first sight, the measured exponents span a quite
large range of values, and no general behavior can be easily guessed. But it is worth
to consider in detail each experiment analyzing the particular conditions of the
material and of the measurement to  find a possible explanation for this large
variability. In particular, two key factors appear to affect the results, and should
thus be carefully checked: the applied field rate dependence of the critical exponents,
and if the data are taken in a small bin around the coercive field or along all the
hysteresis loop. After this warning, we proceed to analyze the literature results in
more detail:

\begin{enumerate}
    \item The 81\%NiFe used in Ref.~\onlinecite{LIE-72} is the only
wire which has been measured accurately. The authors use 3 different samples with
different annealing conditions, but do not report any indication about the structural
characteristic of the materials. They also did notice a dependence of the statistics on
the applied field rate, but to simplify matter, they made the measurements at the same
\emph{magnetization rate}, which later has been established to be the correct parameter
to take into account \cite{ALE-90,ALE-90a}. As it is not explicitly mentioned, the
noise has been probably recorded along all the hysteresis loop and it is thus likely
that it is not statistically stationary.

 \item Similar critical exponents have been reported by Spasojevi\'{c}
\textit{et al.}  \cite{SPA-96}, with a completely different material. As mentioned in
Sec.~\ref{sec:exp_setup}, this experiment has some serious drawbacks due to the large
extension of the pickup coil in respect the sample length, which have been shown to
affect the size distributions \cite{MCC-76}. The data are taken in a small field
interval, presumably around the coercive field, but it is not clear from the paper if
the applied field is large enough to saturate the material. The rate dependence of the
critical exponents has not been discussed.

 \item The Perminvar used in \cite{URB-95a} has been subjected to an
annealing at 450 $^\circ$C for 24 h. This prolonged treatment has the effect to induce
the precipitation of micro-crystalline phases, which create a large number of strong
pinning centers for the DW, resulting in a low permeability of about 250. The
estimation of the critical exponent is limited to the size exponent $\tau$, but it is
made with high accuracy.

 \item A comparable size critical exponent has been
measured for a sheet of interstitial-free, low-carbon, annealed steel \cite{MCM-93}.
The external field is applied both parallel and perpendicular to the rolling direction.
The corresponding hysteresis loops show a different permeability, which affects the
value of the cutoff $S_0$, and slightly the critical exponent $\tau$. As matter of
fact, the estimation at the lowest permeability (smaller $S_0$) appear much less
accurate, due to the limited region of the power law dependence.

\item Ref.~\cite{DUR-95} reports the first estimation of the critical exponents in a
polycrystalline SiFe strips (1.8 \% wt.). Within the experimental errors, the critical
exponents show a linear dependence on the applied field rate, expressed by $\alpha = 2
-c$, and $\tau = 1.5 - c/2$, where $c$ is the parameter comparing in
Eq.~\ref{eq:pdv-def}.

\item In Ref. \onlinecite{DUR-00}, we have reported measurements of the critical
exponents in polycrystalline SiFe ribbons (with high Si contents) and amorphous alloys
with composition Fe$_{x}$Co$_{85-x}$B$_{15}$. We also measured a partially crystallized
Fe$_{64}$Co$_{21}$B$_{15}$, obtained after annealing for 30 min at 350 $^\circ$C and
then for 4h at 300 $^\circ$C under an applied tensile stress of 500 MPa. This induces
the formation of $\alpha$-Fe crystals of about 50 nm, with a crystal fraction of $\sim
5\%$ \cite{BAS-96}. The SiFe alloys and the partially crystallized alloy show $\tau
\sim 1.5$ and $\alpha \sim 2$, measured at the lowest possible frequency $f$ = 3-5 mHz
(see Fig.~\ref{fig:dur00f1}-left). They also show a linear dependence on the applied
field rate, as shown in Fig.~\ref{fig:pds-rate-dep}-left \cite{DUR-95,DUR-95a}. The
highly magnetostrictive Fe$_{x}$Co$_{85-x}$B$_{15}$ alloys ($\lambda_s \sim 30-50
\times 10^{-6}$) have been measured under a tensile stress of $\sigma \sim 100$ MPa.
The applied stress is found to enhance the signal-noise ratio, reducing biases in the
distributions, but does not change the exponents \cite{DUR-99}, as long a small stress
is applied (see below). These alloys yield $\tau \sim 1.27$ and $\alpha \sim 1.5$
(Fig.~\ref{fig:dur00f1}- right), independent of the applied field rate, as shown in
Fig.~\ref{fig:pds-rate-dep}-right. In contrast, the shape of the cutoff changes
drastically as the rate increases.

\item The authors of Ref.~\onlinecite{MEH-02} used the same type of amorphous alloy of
Ref. \onlinecite{DUR-00}, but without the application of a tensile stress. The
resulting critical exponents are larger, and the authors have no a simple explanation
of the difference. As they claimed, the residual stress may be the origin of the
anomaly. We can confirm that unstressed amorphous materials have in general hysteresis
loops with restricted regions of linear permeability, thus the determination of the
exponents can be difficult and less accurate.
\end{enumerate}

In Tab.~\ref{tab:crit-exp-exp} we also report the values of another critical exponent
$1/\sigma \nu z$ relating the average size $\langle S \rangle$ to the duration $T$ as
$\langle S \rangle \sim T^{1/\sigma \nu z}$. This exponent has been shown to be in
strict relation with the power spectrum exponent. (see Secs.~\ref{sec:ps} and
\ref{sec:th-jump-distr}).


To understand the origin of the variability of the critical exponents is important to
first focus on the values which has been found to be close in different samples and
experiments. In addition, it is better to consider the data recorded in a small bin
around the coercive field,  that can thus be considered as stationary. Two sets of
exponents emerges: one is measured in polycrystalline alloys or with embedded grains,
characterized by $\tau \approx 1.5$, and $\alpha \approx 2$, with a linear dependence
on the applied field rate. The other is measured in materials where the crystalline
matrix is absent (amorphous samples under tensile stress) or progressively lost by
induction of a strong disorder (like in the Perminvar of Ref.~\cite{URB-95a}). They
show $\tau \approx 1.3$, and $\alpha \approx 1.5$, and no rate dependence is found.
These results support the idea that there are (at least) two kinds of behavior, the
\emph{universality classes}, to which correspond different sets of critical exponents.
We will came back to this matter in more detail in Sec.~\ref{sec:th-jump-distr}, where
we will also try to explain the occurrence of the other experimental exponents.

\begin{figure}
   \begin{center}
   \begin{tabular}{ccc}
   \includegraphics[height=5cm]{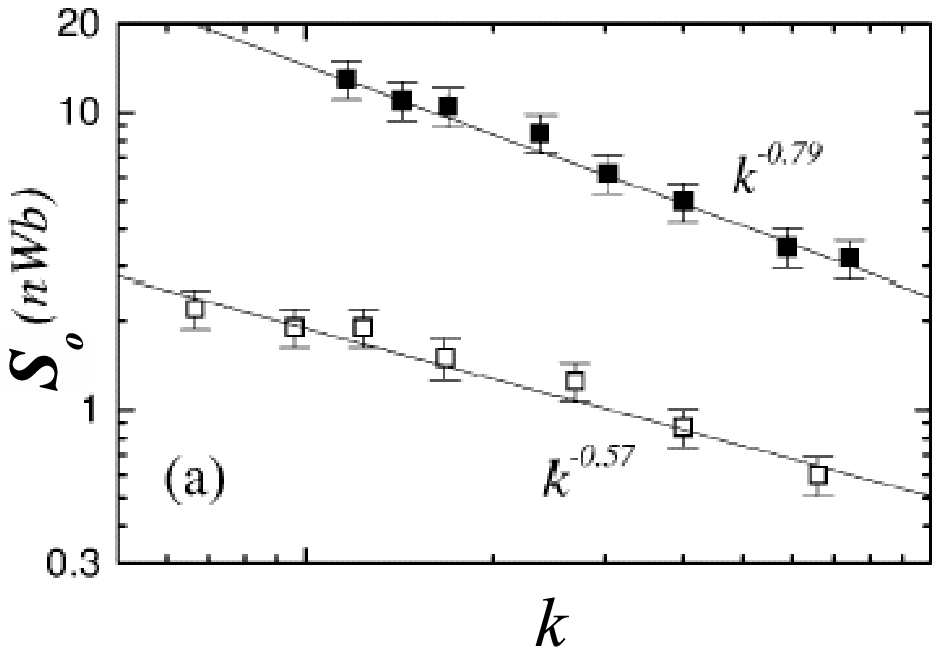} & &
      \includegraphics[height=5cm]{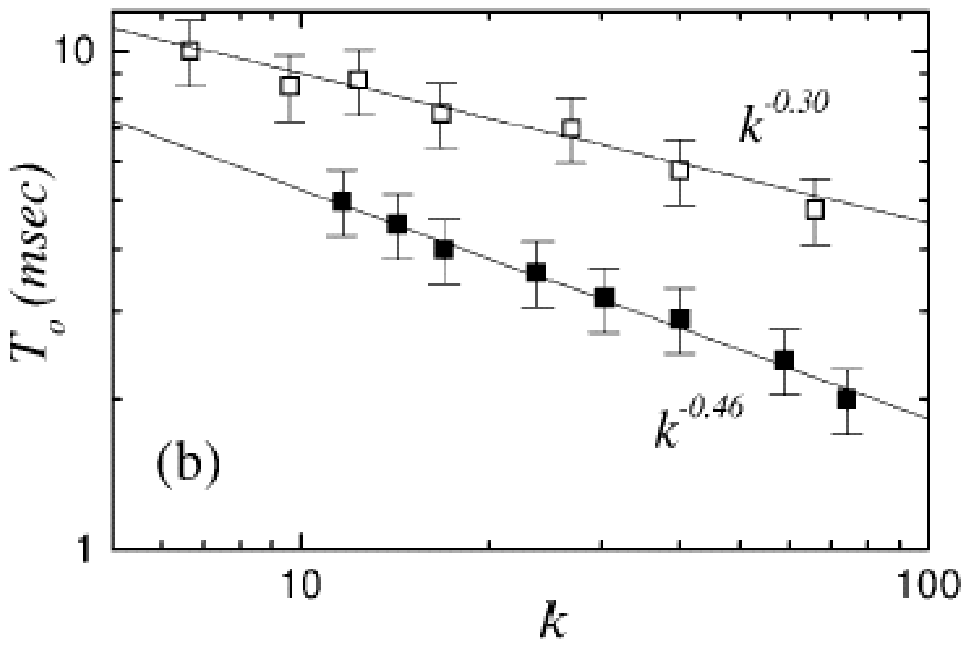}
   \end{tabular}
   \end{center}
   \caption{\label{fig:kdep}The cutoff of the Barkhausen size (a) and
duration (b) distributions as a function of the demagnetizing factor $k$ in Fe-Si 6.5
wt\% alloy (empty symbols) and an amorphous Fe$_{21}$Co$_{64}$B$_{15}$ sample under
constant tensile stress (filled symbols).}
   \end{figure}

Given the variability of the number shown above, it is quite clear that the sole
comparison of the values of the exponents $\tau$ and $\alpha$ can be misleading. In
other words, it necessary to determine other experimental critical exponents which can
confirm the existence of the two universality class. In this respect, an important
suggestion has been given by Perkovi\'{c} \textit{et al.}, who pointed out that "an
explanation for the experiment must involve collective motion of many domains; it must
provide an explanation for the power-law scaling regions, \emph{an it must provide an
explanation for the cutoff} [From \cite{PER-95}, pg. 4528]. We have already stressed
the importance of the demagnetizing effect in the DW dynamics, so that we investigated
in detail the effect of the change of the demagnetizing factor $k$ on the power-law
distributions. We used two samples belonging to the different universality classes, and
measured the avalanche distributions, progressively cut the samples to change the
apparent permeability $\mu_{app}$, and thus calculate the factor $k$ using the well
known expression $\mu_{app}= \mu / (1+k \mu)$, with $\mu$ the intrinsic permeability.
We verified that the cutoffs $S_0$ and $T_0$ strongly depend on $k$, as in
Fig.~\protect\ref{fig:kdep}, and follow the scaling relations $S_0 \sim
k^{-1/\sigma_k}$, and $T_0 \sim k^{-\Delta_k}$. Best fit estimation of the exponents
gives $1/\sigma_k \sim 0.57$, $\Delta_k \sim 0.30$ for the polycrystalline class, and
$1/\sigma_k \sim 0.79$, $\Delta_k \sim 0.46$ for the amorphous one.

\subsection{Power spectra and avalanche shapes}
\label{sec:ps}

If we merely count the huge number of papers in the literature concerning the spectral
properties of the BK noise, the contrast with those considering scaling exponents
results clearly. Since the earlier papers, but especially during the 70s-80s, almost
all the authors publishing about the BK effect considered the problem of the detection
and explanation of the power spectrum. The main reason for this abundant production is
related to the practical need to reduce the spectral noise in applications where soft
materials are used. In addition, it seemed easy to explain all the details the power
spectral shape in terms of a superposition of 'elementary' Barkhausen jumps (see for
instance \cite{MAZ-65}). These elementary units were generally defined \textit{a
priori}, for instance, as exponential or square events: a proper superposition of these
independent jumps can give practically any measured power spectrum shape. Even if
successful to give a general description of the power spectrum shape, this approach was
not able to account for the microscopic magnetization events and the details of the DW
dynamics, so it really did not represent a significant step towards the fundamental
understanding of the BK noise. Remarkably, after the publication of the two seminal
papers by Alessandro \textit{et al.} \cite{ALE-90,ALE-90a}, introducing the ABBM model
(Sec.~\ref{sec:abbm}), the investigation of the power spectrum apparently lost
interest, even though the problem remained basically a puzzle. Only recently the
subject, seen under a completely new point of view, received a new impulse.

As it should be clear from the discussion in the preceding sections, the detection of
stationary or non-stationary Barkhausen signals has a strong effect also on the
estimation of spectral properties. More precisely, traditional power spectral
calculations using FFT must be applied only to stationary data, otherwise they can give
unpredictable results. A non-constant value of the permeability, for instance,
introduces low frequency components unrelated to the BK noise. Thus we must consider
only two types of spectral estimation: i) those taken \textit{with a single DW at
constant velocity}, as in \cite{GRO-77,GRO-77b,POR-79b,VER-81}, where a feedback setup
keeps the flux rate constant, and ii) those taken in a small bin around the coercive
field, as first performed in \cite{WIE-78,BER-81}.

Measurements of power spectrum of type (i) are particularly important. In fact, not
only they are a measure of the time correlation of the BK signal (in the frequency
domain), but more they are a measurement of the spatial fluctuation of local fields
which are needed to unpin the DW. In other words, these measurements describe the
spatial spectrum of the random pinning field (see Sec.~\ref{sec:abbm}). Measurements on
single crystal of SiFe give a $f^{-2}$ power spectrum, and thus a $k^{-2}$ spectrum of
the pinning field \cite{GRO-77,VER-81}, because of the linear relation between $f$ and
$k$ at constant $v$. It is curious to note that measurements on a toroidal Metglass
sample gave a $k^{-1.7}$ spectrum \cite{GRO-77b}. Even if a single measurement cannot
be taken as conclusive, this result can be regarded as a manifestation of a different
universality class, similar to what reported for the avalanche critical exponents.

As anticipated, the most accurate report of power spectra in a polycrystalline material
has been presented in Ref.~\cite{ALE-90a}. A strip of non-oriented 3\% SiFe, having a
grain size of about 100 $\mu$m is measured under a small tensile stress ($\sim$ 5 MPa),
and at variable apparent permeability obtained varying the air gap with a mumetal yoke
(see Fig.~\ref{fig:ale90af1}). The effect of the grain size on similar materials is
reported in Ref.~\cite{BER-90b}. Successively, a few other papers have reported the
power spectra of single crystal and polycrystalline SiFe materials \cite{DUR-97}, as
well as of amorphous materials \cite{DUR-96,PET-98a}.

\begin{figure}
   \begin{center}
   \begin{tabular}{ccc}
   \includegraphics[height=8cm]{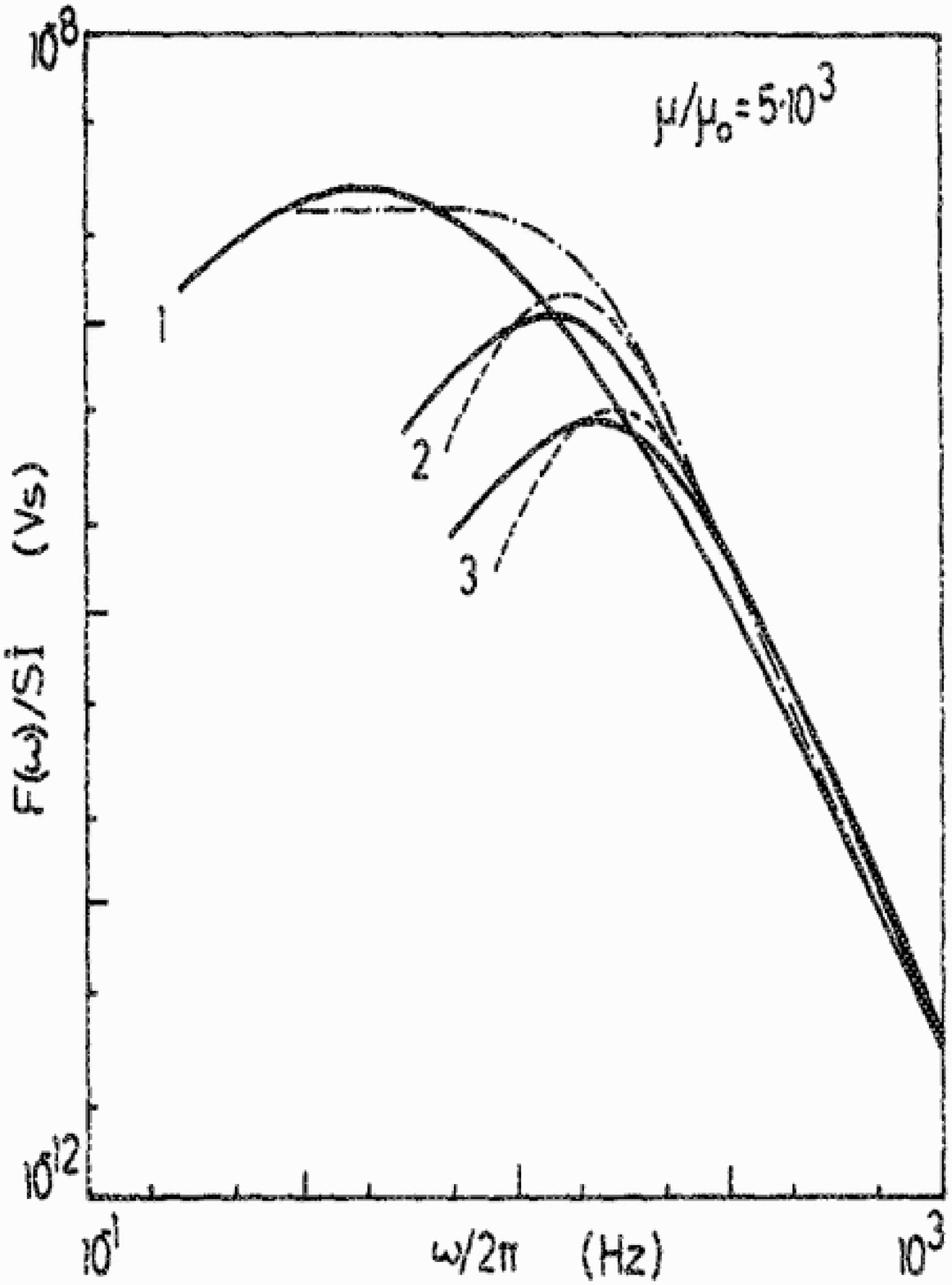} & &
      \includegraphics[height=8cm]{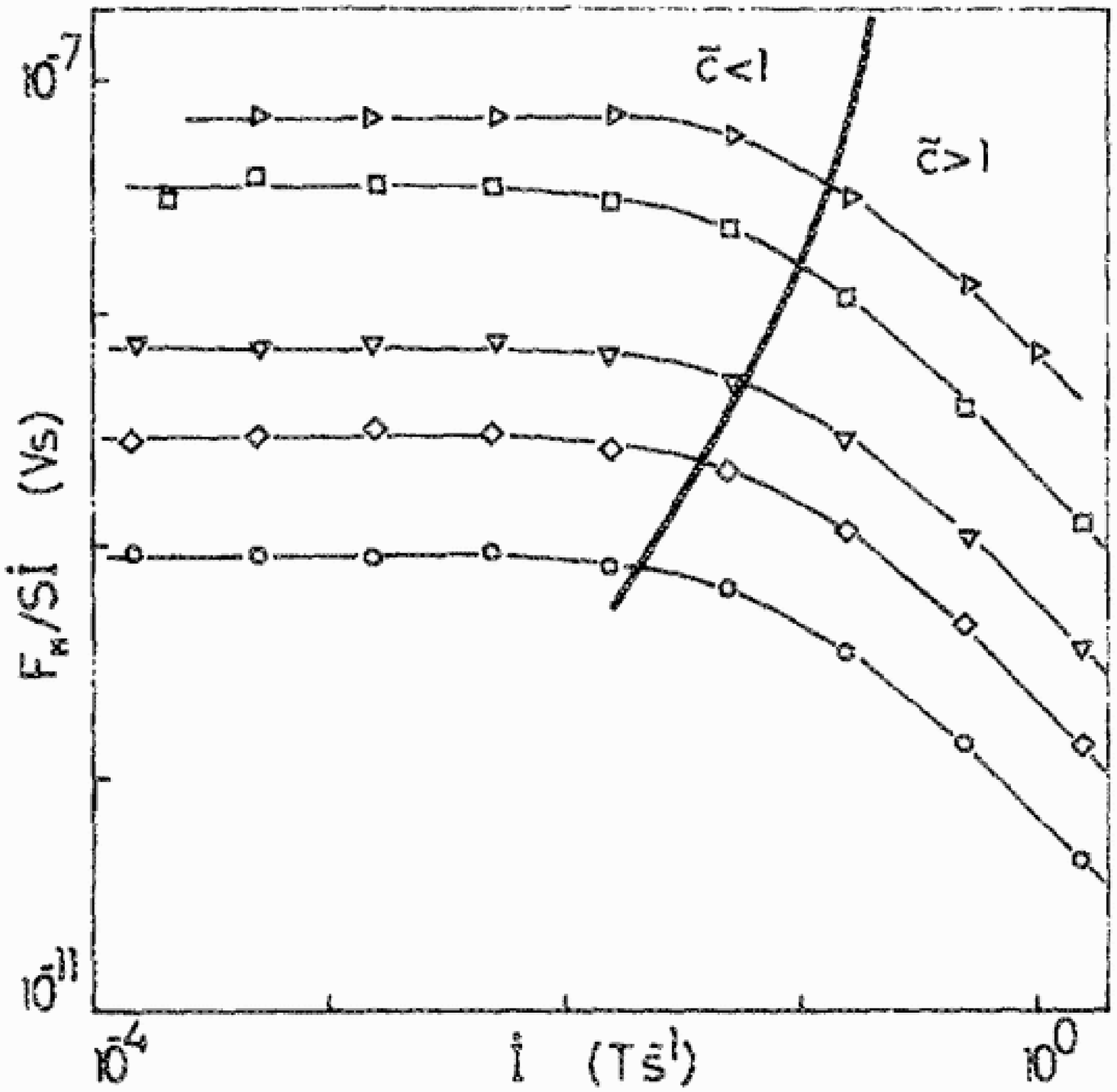}
   \end{tabular}
   \end{center}
   \caption{(Left) Normalized power spectra $F(\omega)/S\dot{I}$ at
    fixed permeability of a polycrystalline 3\% SiFe strip. The solid lines
    are the experimental results while the dotted lines are the
    prediction of the ABBM model (Sec.~\ref{sec:abbm}); (right)
    Normalized maximum of the power spectra $F_M/S\dot{I}$ for
    permeability ranging from 2500 ($\circ$) up to 60,000
    ($\triangleright$). Continuous lines are guide to eyes only. The
    bold line is the boundary between the intermittent ($\tilde{c}<1$)
    and the continuous ($\tilde{c}>1$) DW motion. Here $\tilde{c}$ is
    equivalent of the exponent $c$ of eq.~\ref{eq:pdv-def}. [From
    Ref.~\cite{ALE-90a}, figs. 7 and 10, pg.  2912-2913]}.
    \label{fig:ale90a}
   \end{figure}

There some general common characteristics that can be summarized as follows:

\begin{enumerate}
    \item The spectrum $F(\omega)$ has a typical $1/f^\vartheta$ shape at
high frequency, with $\vartheta = 1.7 \div 2$, and scales linearly with the average
magnetization rate $\dot{I}$, so that the spectra normalized with the average flux
$F(\omega)/S\dot{I}$ coincide at high frequency (Fig.~\ref{fig:ale90a});

 \item At lower frequency, the spectrum shows a marked peek at a
 frequency roughly proportional to $\dot{I}^{1/2}$, with an amplitude
 which takes a constant value at low rates, as shown in
 Fig.~\ref{fig:ale90a}.  This amplitude roughly scales with the
 permeability, as $\mu^{1.5}$ \cite{ALE-90a};

 \item At frequencies lower than the peek, the spectrum scales as
   $f^\psi$, with $\psi \sim 0.6$, or $\sim 1$.

\end{enumerate}

\begin{figure}
   \begin{center}
   \begin{tabular}{ccc}
   \includegraphics[height=8cm]{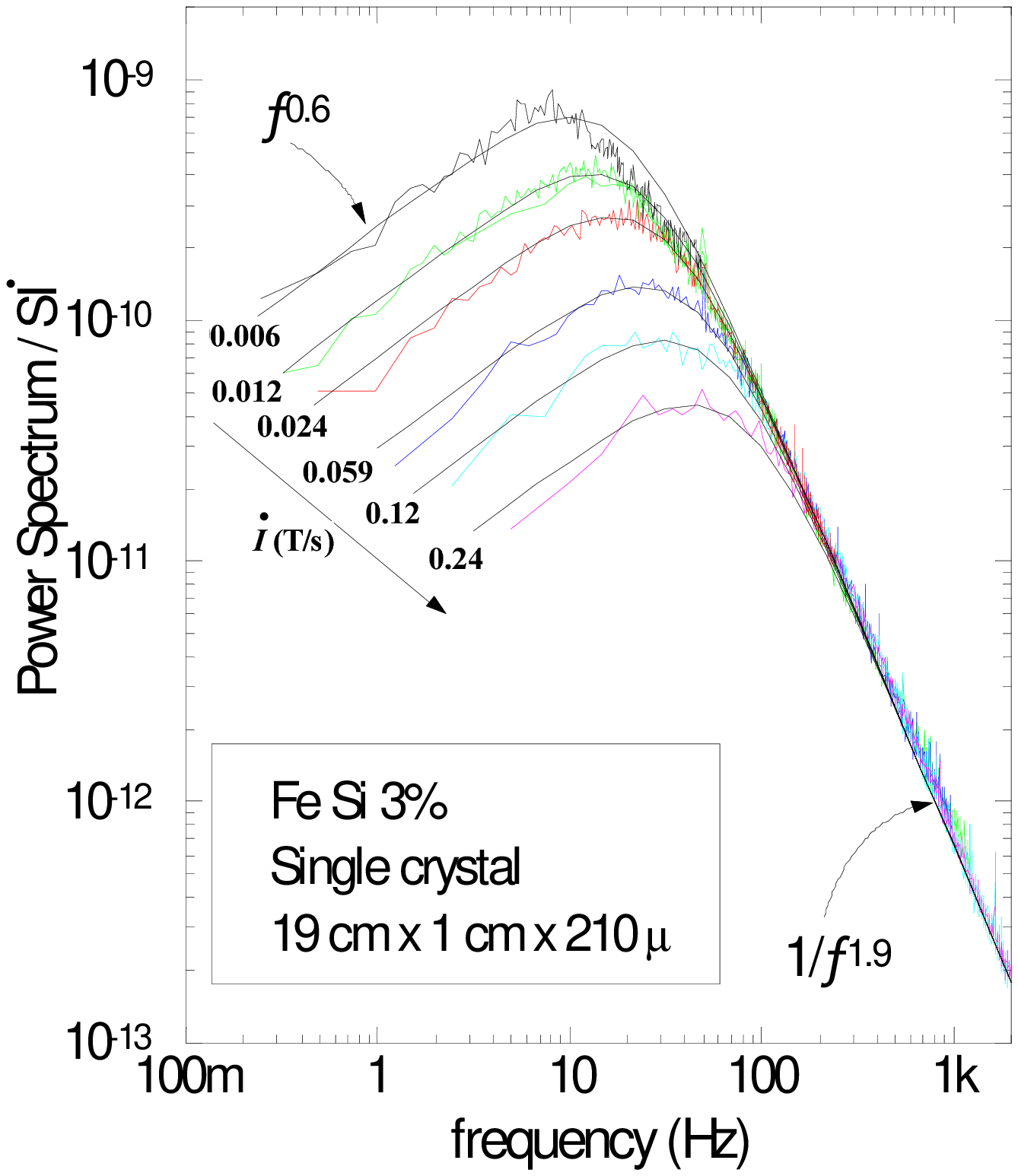} & &
      \includegraphics[height=8cm]{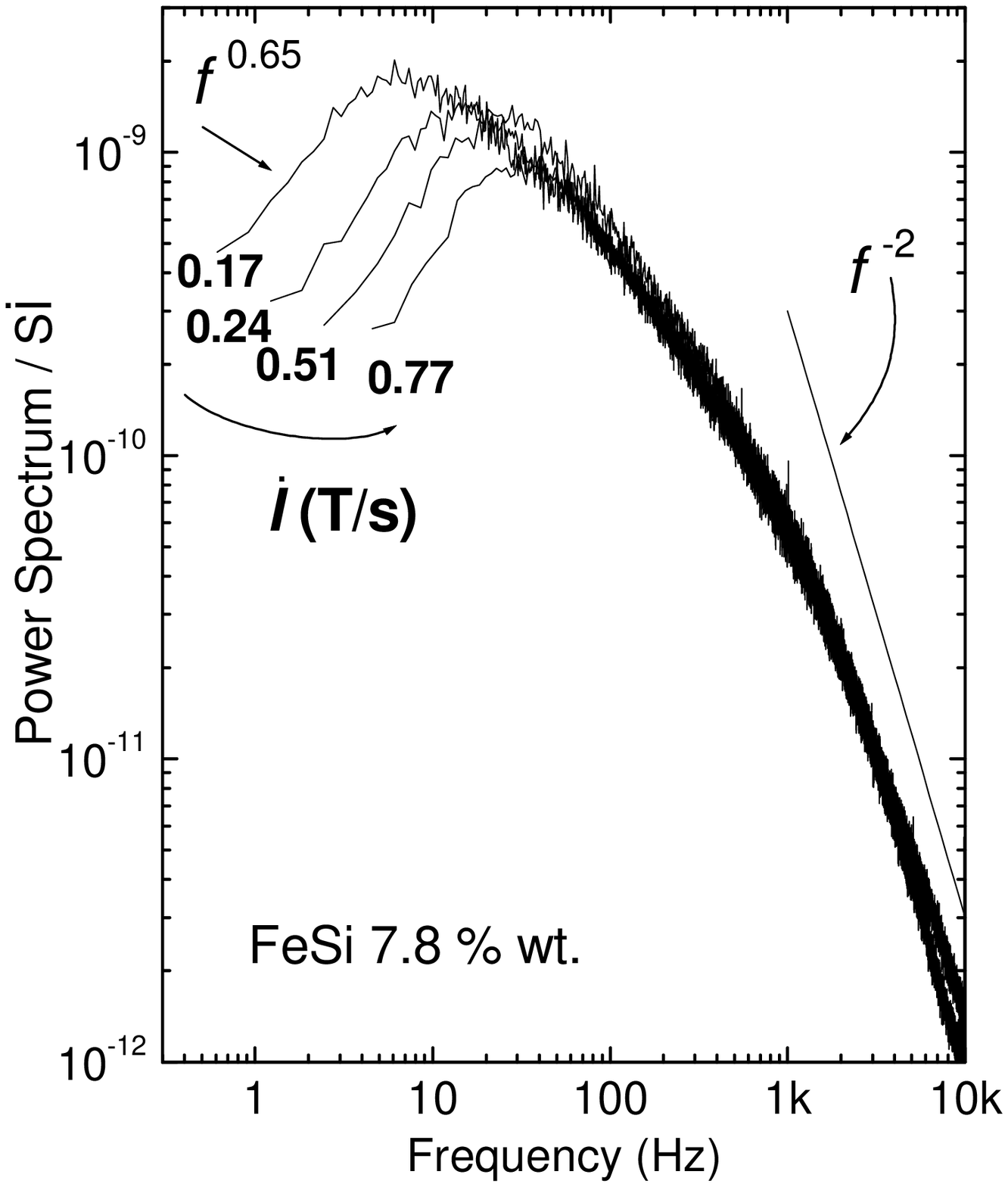}
   \end{tabular}
   \end{center}
   \caption{Normalized power spectra of a single crystal 3\% SiFe strip
    (left, reproduced from \cite{DUR-97}, Fig. 3, pg. 580], and a
    polycrystalline 7.8 \% SiFe ribbon (right) as a function of the
    magnetization rate $\dot{I}$.}
    \label{fig:spectra-sife}
   \end{figure}

\begin{figure}
   \begin{center}
   \begin{tabular}{ccc}
   \includegraphics[height=8cm]{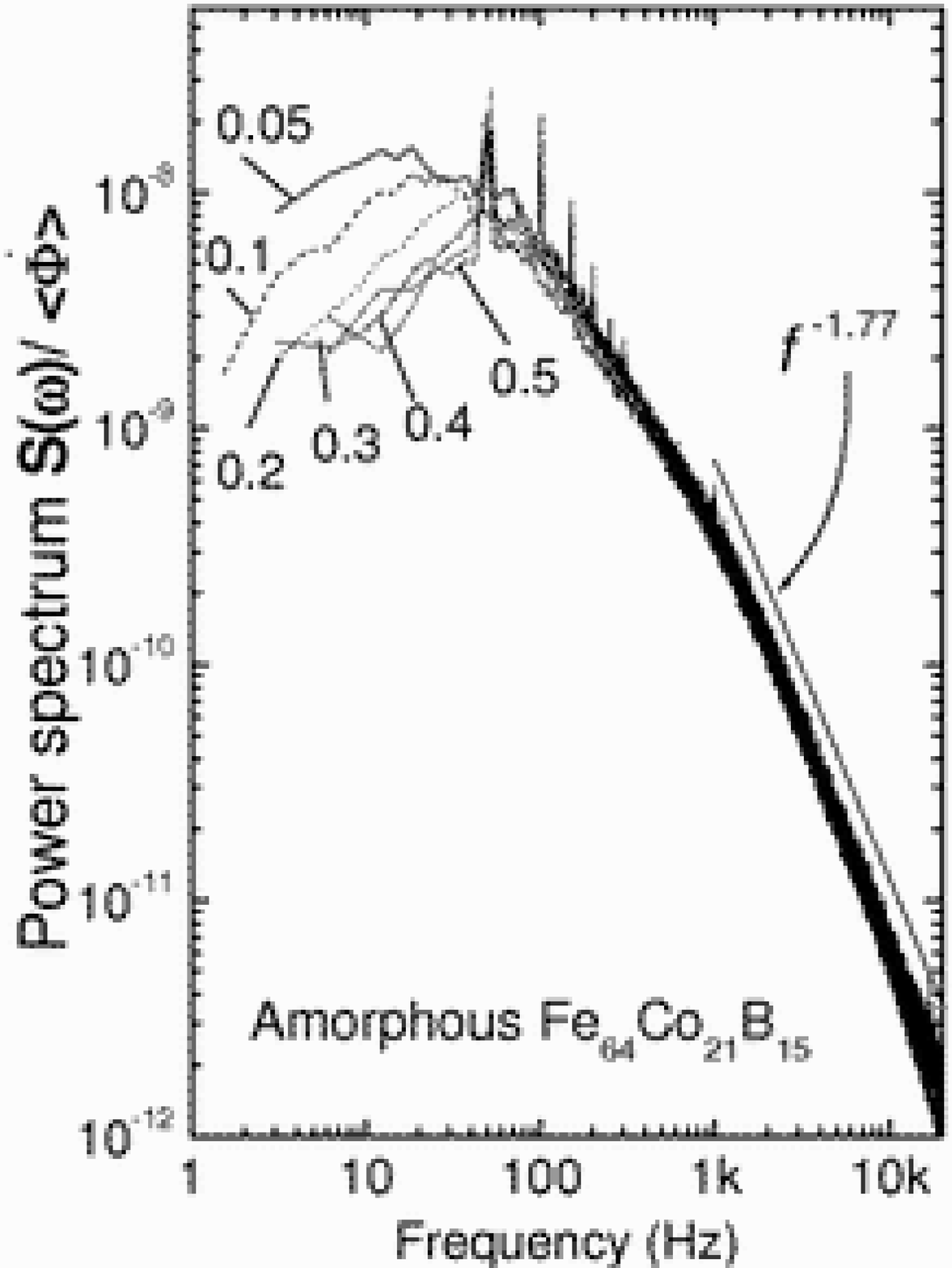} & &
      \includegraphics[height=8cm]{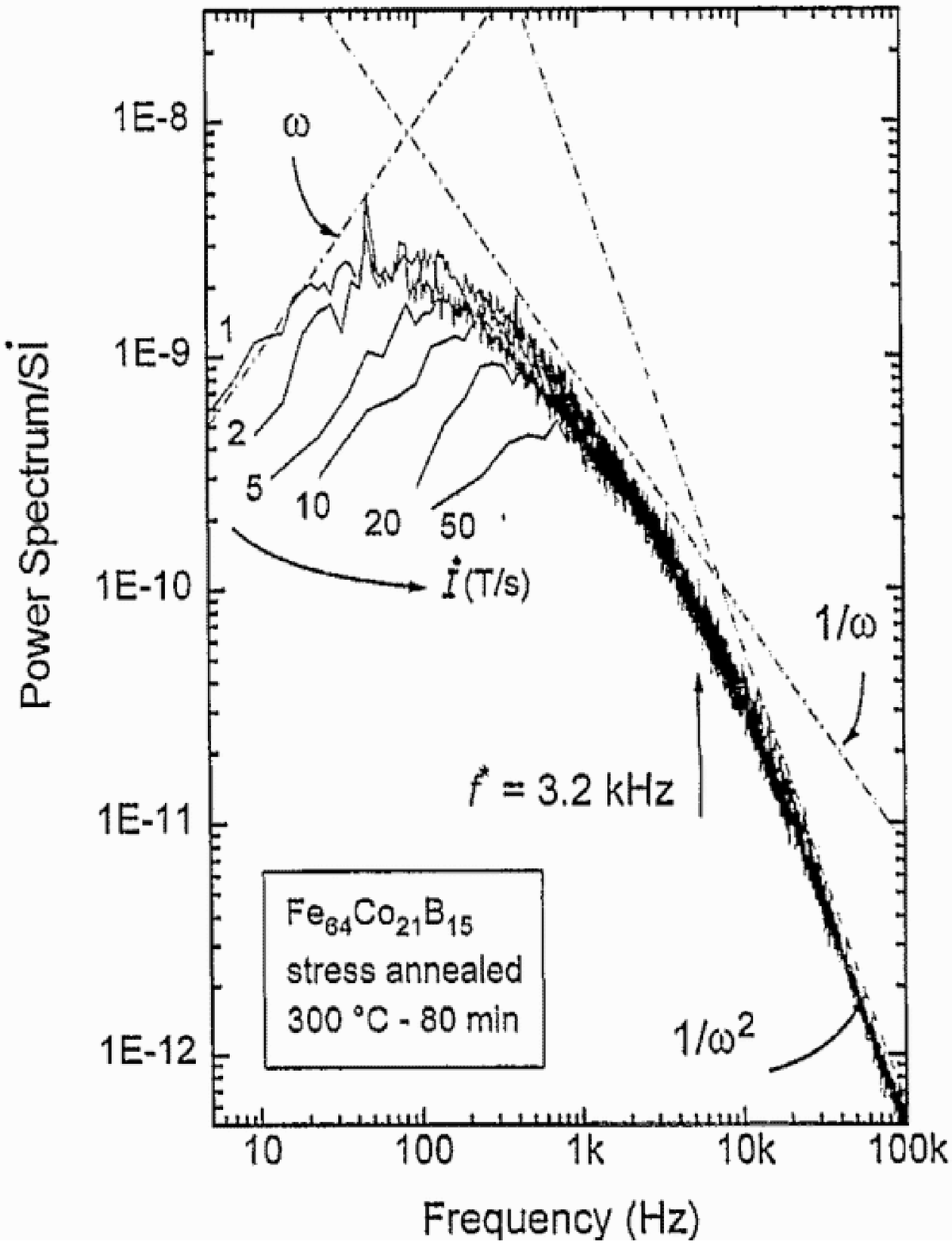}
   \end{tabular}
   \end{center}
   \caption{Normalized power spectra of the same amorphous material of
    Fig.~\ref{fig:pds-rate-dep} (left), and of a partially crystallized
    (16\%) sample induced by annealing (right, reproduced from
    \cite{DUR-96}, Fig. 2, pg. 300) as a function of the magnetization
    rate $\dot{I}$. In the left Fig., $f^*$ is roughly the frequency
    where the shape changes from $f^{-1}$ to $f^{-2}$}.
    \label{fig:spectra-amorph}
   \end{figure}

These general features describe only qualitatively the observed results. More
precisely, there are at least four different typical shapes, as shown in
Figs.~\ref{fig:spectra-sife}-\ref{fig:spectra-amorph}.

Single crystal and polycrystalline SiFe strips show the simplest power spectrum, with
$\vartheta \sim 2$, and $\psi \sim 0.6$. The polycrystalline SiFe ribbons at high Si
content keep the same exponents, but present a large region at intermediate frequencies
where the spectrum strongly deviates from the simple $1/f^2$ law. Amorphous materials
under tensile stress show instead $\vartheta = 1.7 \div 1.8$, and $\psi \sim 1$, as
shown in Fig.~\ref{fig:spectra-amorph}. Similar data are found in the unstressed case,
$\vartheta = 1.73 \pm 0.08$ and $\psi \sim 1$, for the same amorphous alloy
\cite{MEH-02} (private communication). Amorphous materials, where a proper annealing
induces a partial crystallization, show a sort of intermediate case: the high frequency
part is more similar to the polycrystalline SiFe samples with $\vartheta \sim 2$, while
$\psi \sim 1$ keeps the values of a pure amorphous material. The former property is not
a merely coincidence, as we observed a corresponding change in the critical exponents
of size and duration distributions towards the values measured in SiFe, as shown in
Tab.~\ref{tab:crit-exp-exp}. Curiously, all the spectra, despite these differences,
show the same dependence of the peek frequency on $\dot{I}^{1/2}$.

\begin{figure}
   \begin{center}
   \includegraphics[height=8cm]{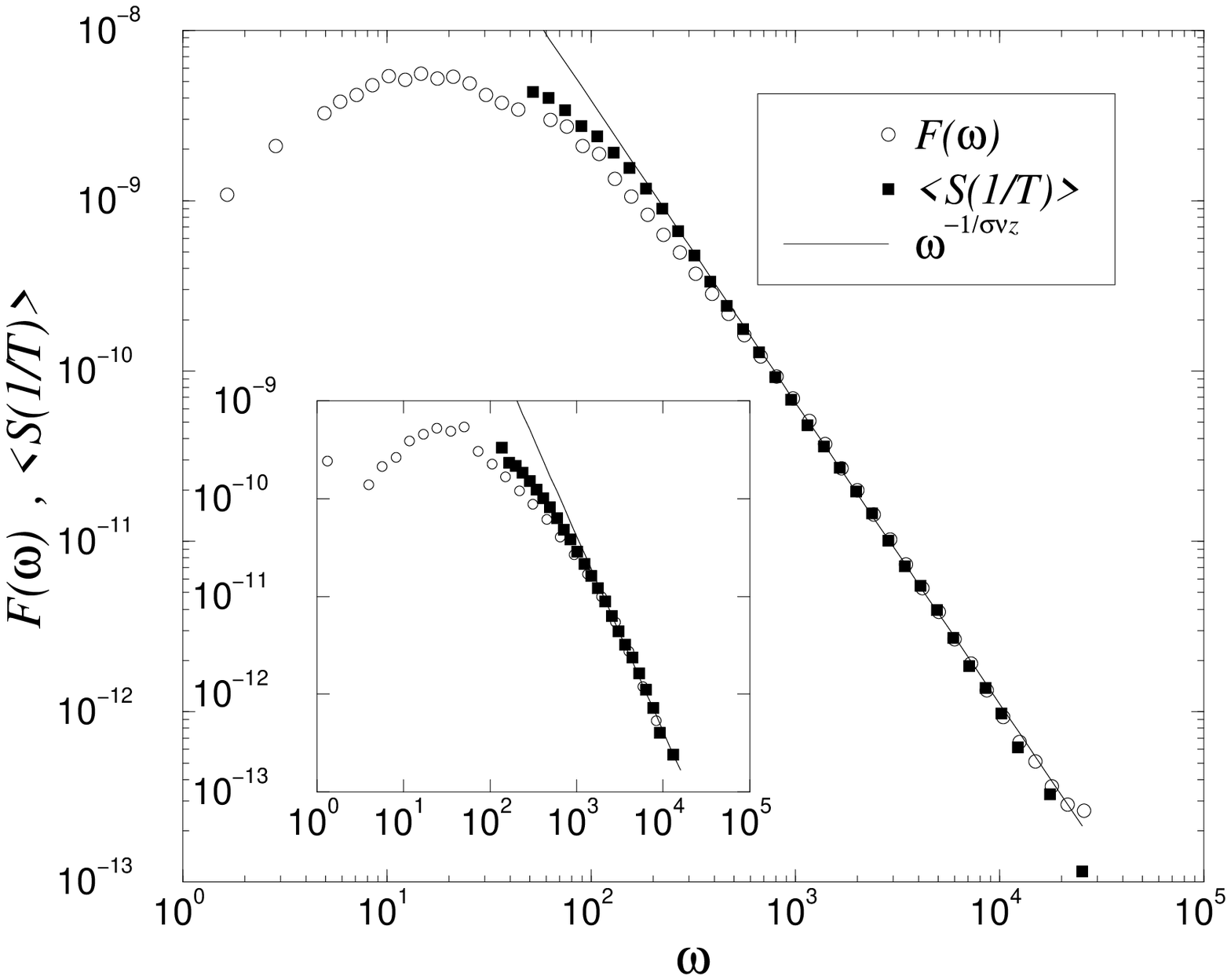}
   \end{center}
   \caption{Comparison of the power spectrum $F(\omega)$ with the average size $\langle S \rangle$ as a
        function of the inverse of duration $T$, for the amorphous material and
        the polycrystalline SiFe ribbon (7.8 \%, inset) of Fig.~\ref{fig:dur00f1}. The theoretical prediction
        of Ref.~\cite{KUN-00} is also shown, with $1/\sigma \nu z$ equal to 1.77 and 2
        [From \cite{DUR-02}, fig. 1, pg. 1086]}.
    \label{fig:spectra-vs-snuzzu}
   \end{figure}

The interpretation of all these complicated details still represents a challenging
puzzle, although some progress has been recently made \cite{KUN-00,DUR-02}. Earlier
estimates of the exponent $\vartheta$ (\cite{LIE-72}, and similarly in
\cite{PER-95,SPA-96,DAH-96}) led to the conclusion that "the asymptotic decay of the
spectrum is not given by the shape of the pulses, but is exclusively determined by the
distribution of the pulse duration and the correlation between the size and the
duration" [Adapted from \cite{LIE-72}, pg. 15]. In our notation, $\vartheta = 1 +
2\sigma \nu z - \alpha$, or analogously $(3-\tau)/\sigma \nu z$ (see also
Sec.~\ref{sec:th-ps}). Using the data of Tab.~\ref{tab:crit-exp-exp}, this implies
$\vartheta \sim 3$, much larger than the experiments. This result was successively
found to be valid only for $\tau > 2$, while for $\tau < 2$, Kuntz and Sethna
\cite{KUN-00} calculated $\vartheta = 1/ \sigma \nu z$, which is the same critical
exponent relating the average avalanche size to the duration
(Tab.~\ref{tab:crit-exp-exp}). This prediction agrees pretty well with the experiments,
as we can see in Fig.~\ref{fig:spectra-vs-snuzzu}. In particular, it works for the
amorphous material in an extended frequency range; the same is not valid for the
polycrystalline SiFe ribbon, and the reason will be discussed after introducing the
hypothesis at the base of the calculations by Kuntz and Sethna.

\subsubsection{\label{sec:shape}Avalanche shapes}

The simple result $\vartheta = 1/\sigma \nu z$ obtained by Kuntz and Sethna \cite{KUN-00}
is based on the existence of a certain number of scaling relations and on a few
hypothesis, like the complete separation of the avalanches in time. We leave the
discussion of these important aspects to a further section (see Sec.~\ref{sec:th-ps}),
focusing here on the new proposed scaling relations which can be tested experimentally.
These relations involve universal scaling functions whose \emph{shape} could provide a
critical test for model, as "a sharper tool for discriminating between different
universality classes than critical exponents" \cite{SET-01,MEH-02}.

The first relation states that the average avalanche shape scales in a universal way so
that
\begin{equation}
\label{eq:vt-shape-exp} v(t,T)=T^{1/\sigma \nu z-1}f_{shape}(t/T),
\end{equation}
where $v$ is the BK signal, $t$ is the time and $f_{shape}(t/T)$ is the universal
scaling function. Similar to this relation, we can consider the voltage as a function
of magnetization $s = \int ^T _0 v dt$ to get
\begin{equation}
v(s,S)=S^{1-\sigma \nu z}g_{shape}(s/S) \label{eq:vs-shape-exp}
\end{equation}

A third relation deals with the fluctuations of avalanche sizes, and considers the
probability $P(v|S)$ that a voltage $v$ occurs in an avalanche of size $S$. This
probability scales as:
\begin{equation}
  P(v|S)=v^{-1} f_{voltage}(vS^{\sigma \nu z-1})
    \label{eq:PVS-exp}
\end{equation}
and, again, $f_{voltage}$ is another universal scaling function.

\begin{figure}
   \begin{center}
   \begin{tabular}{ccc}
   \includegraphics[height=8cm]{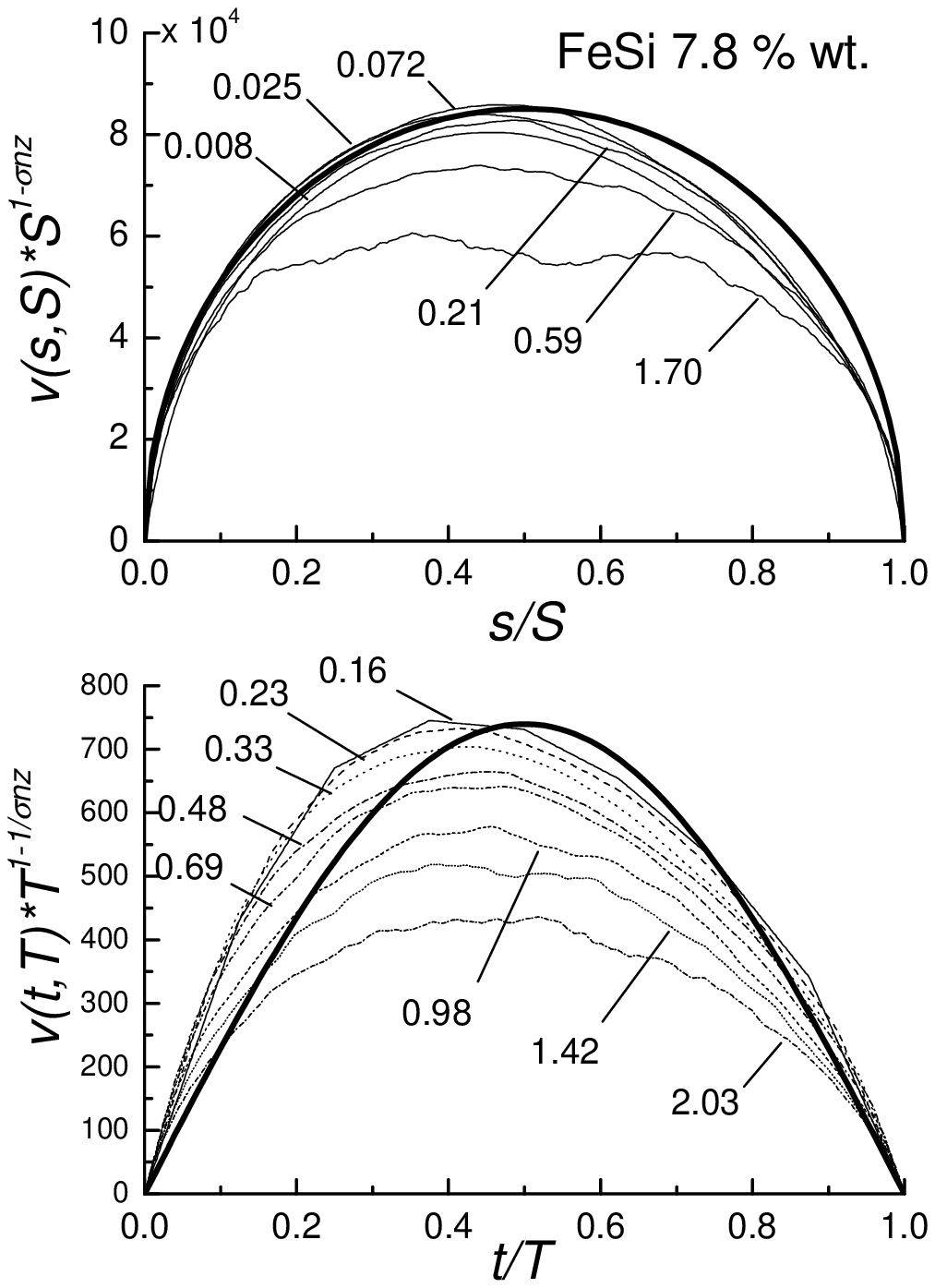} & &
      \includegraphics[height=8cm]{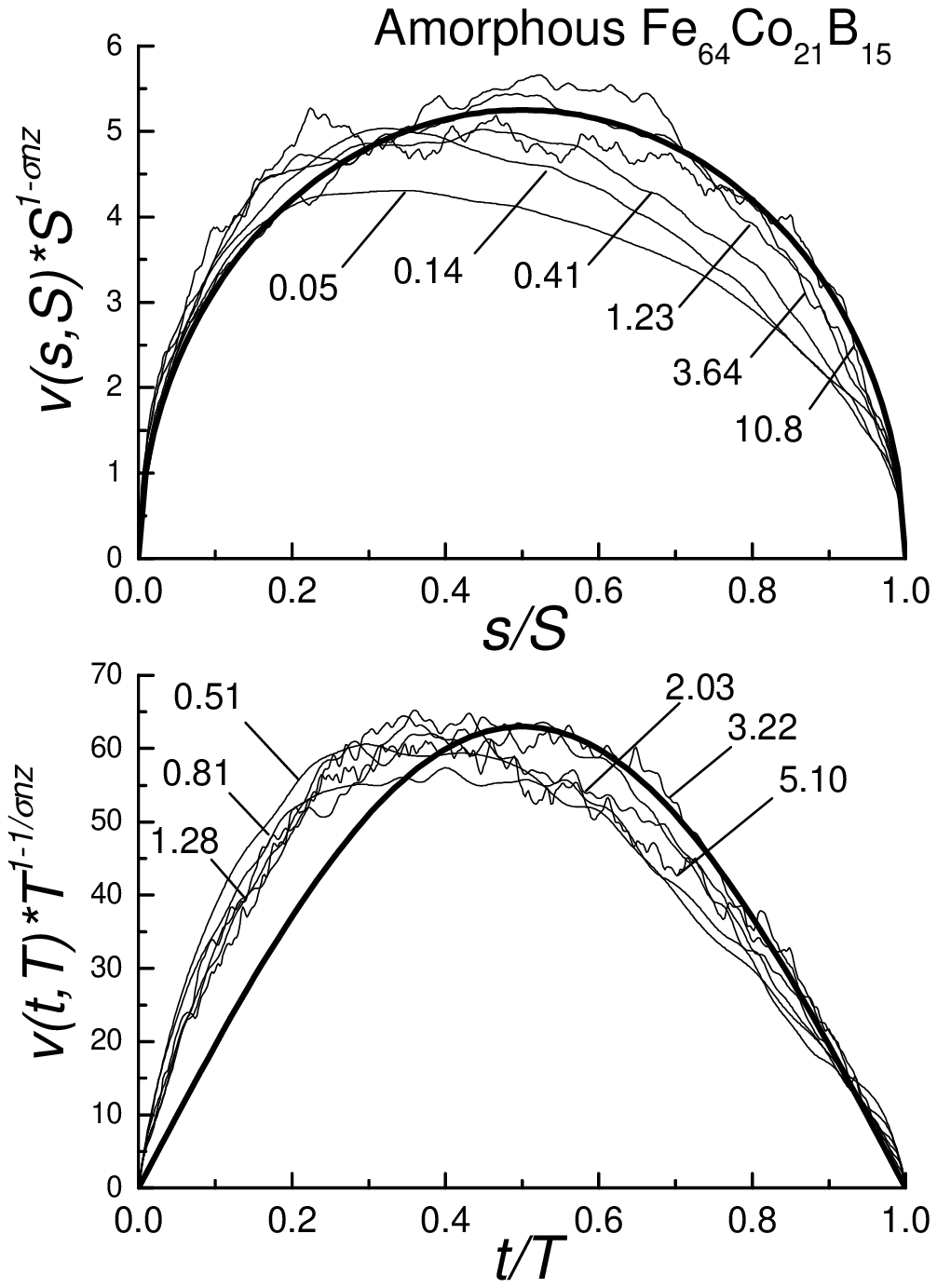}
   \end{tabular}
   \end{center}
   \caption {\label{fig:semi}Average avalanche shapes for the
   7.8 \% SiFe ribbon (left) and the amorphous sample (right) of
   Fig.~\ref{fig:time-seq}. Time signal $v(t,T)$ and magnetization
   signal $v(s,S)$ are scaled according to
   Eqs.~\ref{eq:vt-shape-exp}-\ref{eq:vs-shape-exp}.  Bold lines are
   the theoretical predictions of
   Eqs.~\ref{eq:Vtnorm}-\ref{eq:Vsnorm}, a semicircle and an arc of
   sinusoid, respectively (see Sec.~\ref{sec:th-shape}).  Numbers in
   the graphs denote avalanche size (in nWb, upper figures) and
   duration (in milliseconds, lower figures).}
   \end{figure}

A partial experimental verification of these relations has been reported in a couple of
papers \cite{DUR-02,MEH-02}. In Fig.~\ref{fig:semi}, we show the complete universal
scaling laws of Eqs.~\ref{eq:vt-shape-exp} and  \ref{eq:vs-shape-exp} for the same 7.8
\% SiFe ribbon and the amorphous sample of Fig.~\ref{fig:time-seq}, scaled using the
theoretical estimation of the critical exponent $\sigma \nu z$. The experimental data
are compared with the theoretical average shape predicted using the ABBM model (see
Eqs.~\ref{eq:Vtnorm}-\ref{eq:Vsnorm}). In the case of the amorphous material, the
scaling is pretty good, except for the magnetization signal at small sizes. In
particular, both time signals show a marked time asymmetry, similar to what found in
\cite{SPA-96} where the average of avalanches is made without any scaling. The
asymmetry is also visible in the magnetization signal, even if less marked. Here, the
shape calculated using the ABBM (see Eq.~\ref{eq:Vsnorm} and Sec.~\ref{sec:th-shape})
reproduces pretty well the experiments, except for the small sizes in the amorphous
material. It is interesting to note that all the models presented in the literature
predict symmetric time and magnetization signals. In Sec.~\ref{sec:th-shape}, we will
discuss the relevance of the shape of the scaling function as indicator of universality class.

\begin{figure}
   \begin{center}
   \begin{tabular}{ccc}
   \includegraphics[width=8cm]{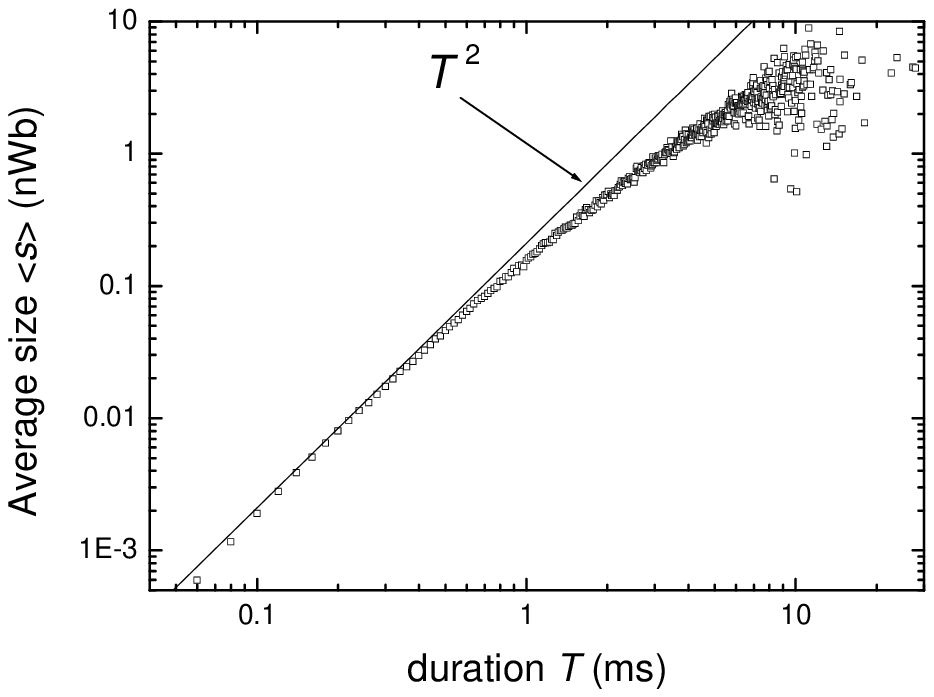} & &
      \includegraphics[width=8cm]{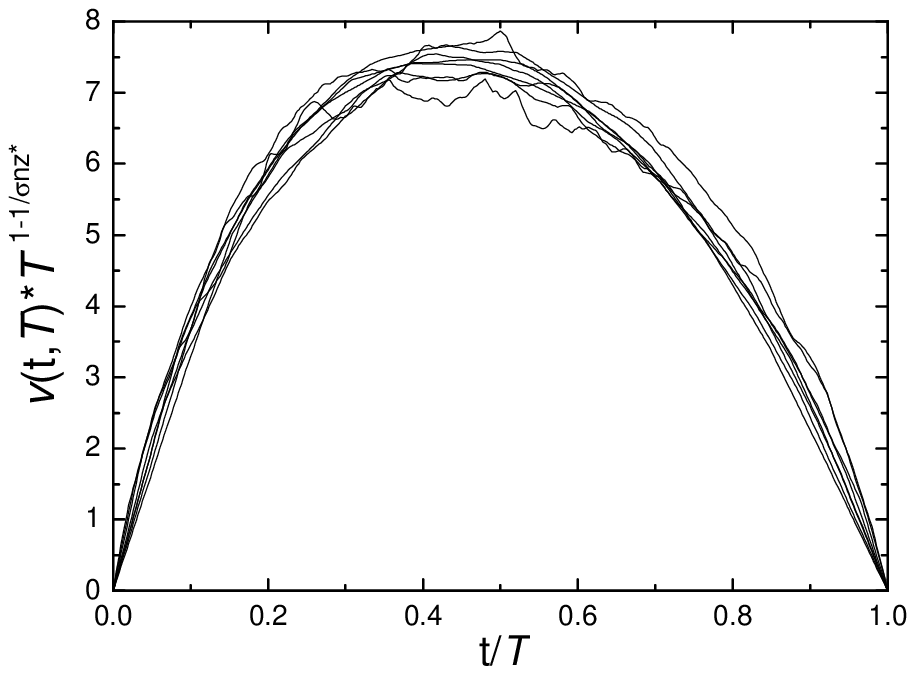}
   \end{tabular}
   \end{center}
   \caption {\label{fig:snuzzu-SiFe} Average avalanche size as a
    function of the duration for the 7.8 \% SiFe ribbon (left).
    Time signals are rescaled as in Eq.~\ref{eq:vt-shape-exp} using a
    variable exponent $1/\sigma \nu z^*$, deduced by the plot on the
    left.}
   \end{figure}

The scaling of time signal data of 7.8 \% SiFe ribbon using the theoretical exponent
$1/\sigma \nu z = 2$ is clearly incorrect. The reason is illustrated in
Fig.~\ref{fig:snuzzu-SiFe}, where we plot the average avalanche size $\langle S
\rangle$ as a function of duration $T$. The exponent $1/\sigma \nu z$ (the slope of the
curve in the log-log plot) slightly varies from $\sim$ 2 down to about 1.75. If we use
this 'variable' exponent $1/\sigma \nu z*$, the rescaling of the time signal is highly
satisfying.


\begin{figure}
   \begin{center}
      \includegraphics[width=10cm]{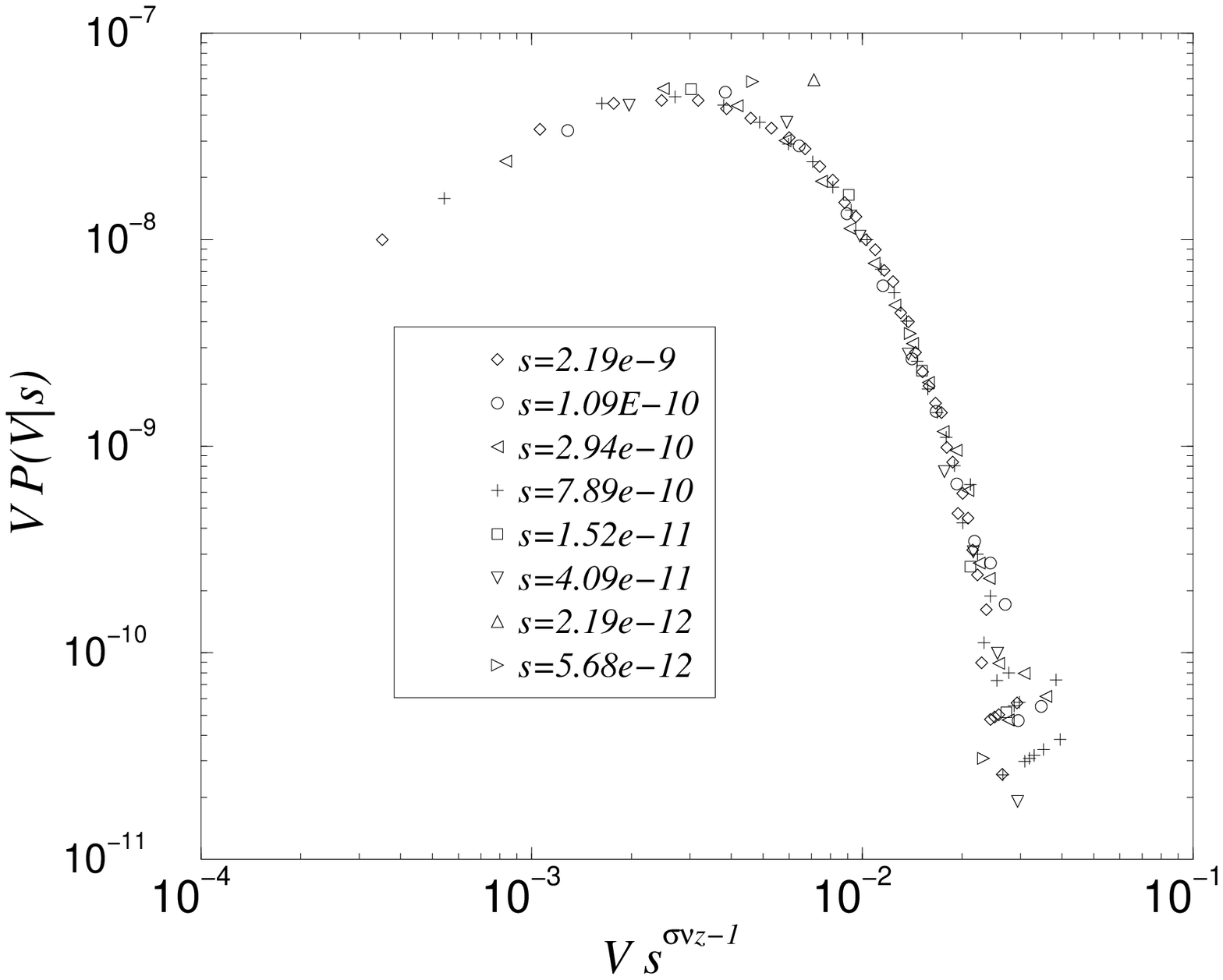}
   \end{center}
   \caption {\label{fig:pvs-amorph} Distribution of time signal at
    fixed size as predicted by Eq.~\ref{eq:PVS-exp} for the amorphous
    material using the theoretical scaling exponent $1/\sigma \nu z =
    1.77$. [From \cite{DUR-02}, Fig. 3, pg. 1087]}
   \end{figure}

The third universal scaling function (Eq.~\ref{eq:PVS-exp}) has been reported in
\cite{DUR-02} for an amorphous material, as shown in Fig.~\ref{fig:pvs-amorph}. This
confirms the validity of scaling relations proposed in \cite{KUN-00}, at least for the
amorphous material. In fact, the relation in Eq.~\ref{eq:PVS-exp} dos not hold so well
for the 7.8 \% SiFe ribbon.

\subsubsection{\label{sec:second_spectra}High order power spectra}

In Ref.~\cite{OBR-94}, O'Brien and Weissman introduced a further statistical
characterization of BK signal calculating power spectra of order higher than those
shown in the previous section. This analysis is motivated by the possibility to prove
the BK as a real example of SOC \cite{BAK-87}, lacking of any inertial effect unlike,
for instance, grain falling in sandpiles. In fact, the mere "existence of pulses with a
range of sizes and durations should not be taken as conclusive evidence for SOC" [From
\cite{OBR-94}, pg. 3446]. Thus it is important to understand if the observed scaling
properties reflect a real self-organization involving "cooperative avalanche processes
invoked by SOC", or simply reflect the scaling properties of the pinning field, or,
alternatively, the vague proximity of a disorder-driven critical transition
\cite{PER-95}.

\begin{figure}
   \begin{center}
   \begin{tabular}{ccc}
   \includegraphics[height=8cm]{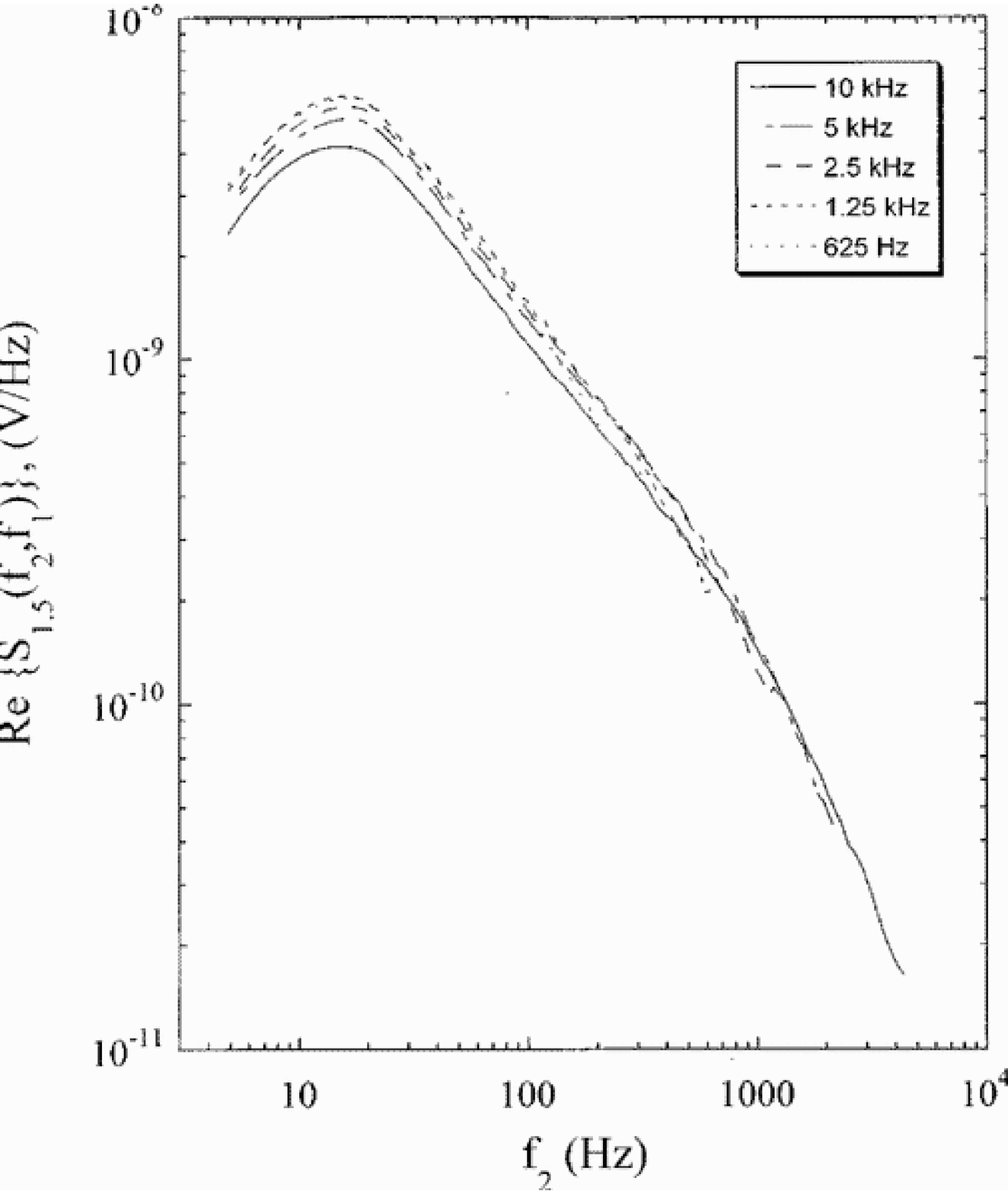} & &
      \includegraphics[height=8cm]{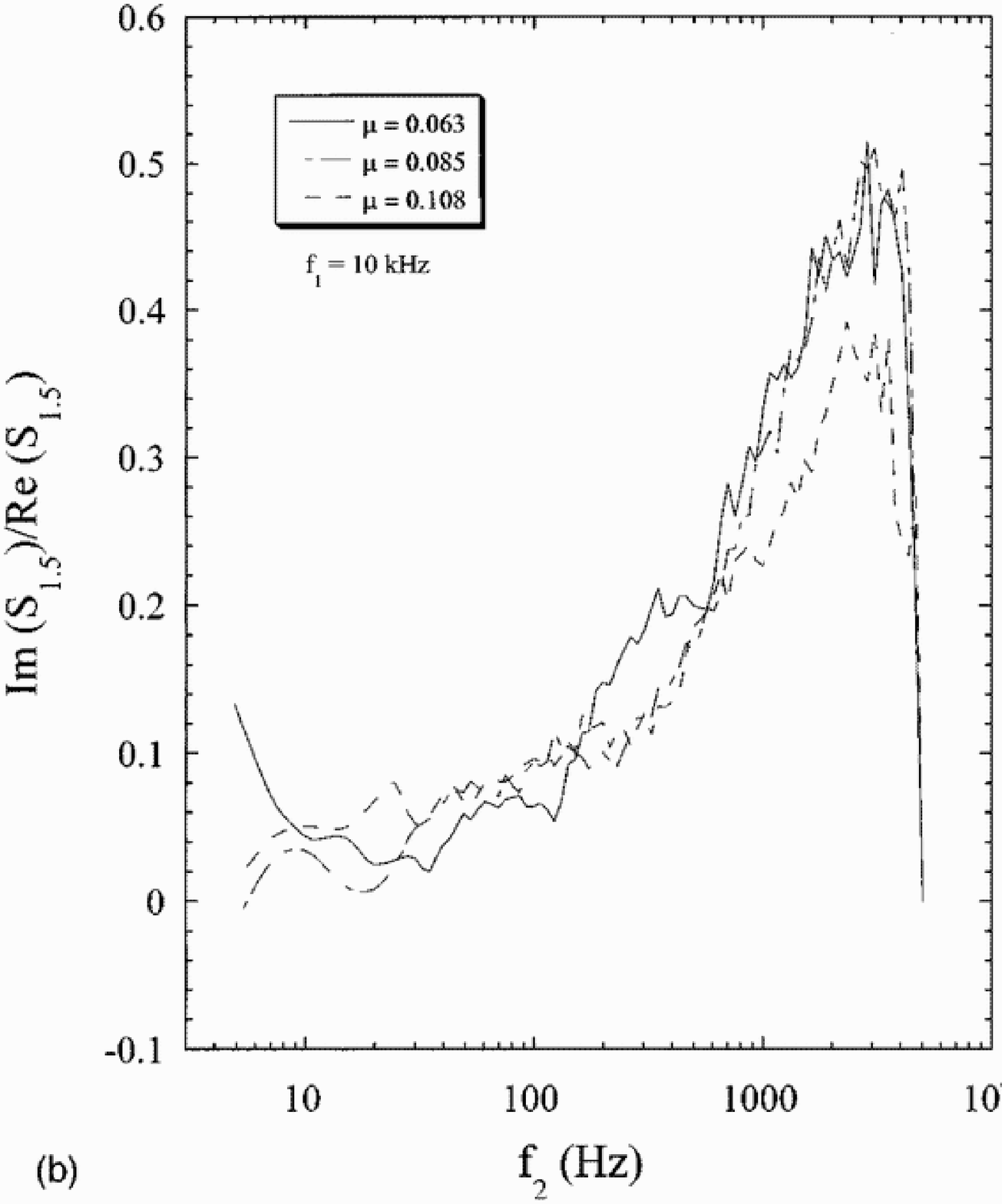}
   \end{tabular}
   \end{center}
   \caption {The 1.5 power spectrum defined in Eq.~\ref{eq:1.5ps} for
   a Fe$_{21}$Co$_{64}$B$_{15}$ amorphous ribbon in the unstressed
   state, at 5 different frequencies $f_1$. [From \cite{PET-98a},
   figs. 6 and 9, pg. 6366-6367]
  \label{fig:second-spectra}}
   \end{figure}

The high order spectra involve correlations between different temporal scales, so for
instance, could clarify if events on a certain scale systematically precede event on
another scale, as in the case of systems showing precursors such as earthquakes.
Another important aspect, more specific of BK noise, is the possibility to distinguish
whether the high frequency spectrum $1/f^\vartheta$ primarily comes from short independent
pulses, as often assumed, or conversely from the fine structure of longer pulses
\cite{PET-98a}. This is addressed considering the 1.5 power spectrum
$S_{1.5}(f_2,f_1)$, involving a third moment of time signal $v(t)$, defined as:
\begin{equation}\label{eq:1.5ps}
  S_{1.5}(f_2,f_1) = \frac{F\{v(t)\} F\{H(t,f_1)\}^\ast}{<H(t,f_1)>}
\end{equation}
where $H(t,f_1)$ is the time-dependent Haar power at the frequency $f_1$, $F\{\cdot\}$
is the usual Fourier transform those components are calculated at frequency $f_2$
\cite{PET-98a}. Similarly, the second spectrum $S_{2}(f_2,f_1)$ is defined as:
\begin{equation}\label{eq:2ndps}
  S_{2}(f_2,f_1) = \frac{F\{H(t,f_1)\} F\{H(t,f_1)\}^\ast}{<H(t,f_1)>^2}
\end{equation}
A complete presentation of these spectra have been reported in Ref.~\cite{PET-98a} for
a Fe$_{21}$Co$_{64}$B$_{15}$ amorphous ribbon (in the unstressed state). In
Fig.~\ref{fig:second-spectra} we show the 1.5 power spectrum. The weak independence of
$Re\{S_{1.5}(f_2,f_1)\}$ on $f_1$ indicates that the high frequency spectrum
$1/f^\vartheta$ contains a large contribution from pulses with duration much longer
than $1/f_1$. This definitely demonstrates that, in this case, the power spectrum does
not come from independent short pulses. The imaginary part, instead, is sensitive to
the asymmetry of the time signal. The positive sign of
$Im\{S_{1.5}(f_2,f_1)\}/Re\{S_{1.5}(f_2,f_1)\}$ suggests that, on average, high
frequency components precede the low frequency ones. This is in accord with the time
asymmetry of the average avalanche shape of Fig.~\ref{fig:semi}.

\begin{figure}
   \begin{center}
   \includegraphics[width=8cm]{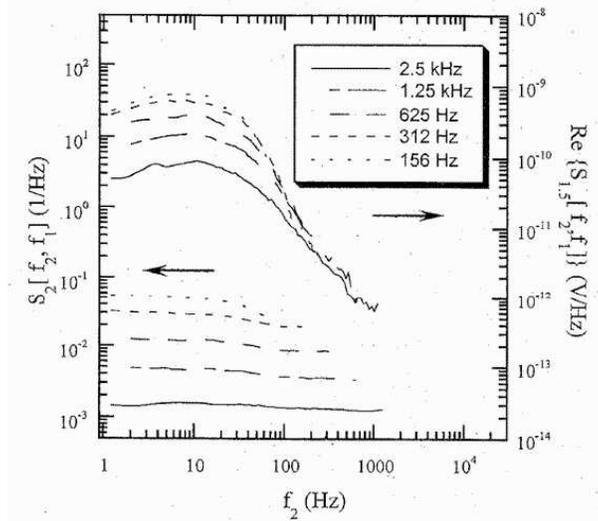}
   \end{center}
   \caption {Second spectra $S_2$ and 1.5 power spectra $S_{1.5}$ for
     a 1.8 \% SiFe single crystal [From \cite{PET-98}, Fig. 3,
       pg. 1172]. \label{fig:sec-sp-sing-crys}}
   \end{figure}

Interestingly, high order power spectra of 1.8 \% SiFe single crystals have a
completely different behavior \cite{PET-98}, as shown in
Fig.~\ref{fig:sec-sp-sing-crys}. Here $S_{2}(f_2,f_1)$ is nearly flat as a function of
$f_2$ and falls off as $f_1^{-1.2}$, which indicates that most of the power at $f_1$
comes from separated pulses of duration $1/f_1$. The 1.5 power spectrum confirms this
hypothesis, as it strongly depends on $f_1$, in contrast with the weak dependence in
the amorphous ribbon. In this case, it is appropriate to assume the power spectrum as a
superposition of individual independent pulses.

\subsection{\label{sec:films}Thin films}

All the results presented so far are related to materials which are essentially
three-dimensional, with the sole exception of the wire reported in
Ref.~\onlinecite{LIE-72} (see Tab.~\ref{tab:crit-exp-exp}). The behavior of the BK
noise in 2D thin films is much less known, and the number of papers reporting a
complete set of experimental results is very limited
\cite{WIE-77,WIE-78,WIE-79,PUP-00a,KIM-03}. Many questions remain to be solved, both on
the experimental and the theoretical side, so that the topic represents an interesting
and promising field of study for the near future.

Contrary to bulk three dimensional systems with a relatively simple magnetic structure
of nearly parallel DWs, thin films show richer and often more complicated DW patterns.
The film thickness plays a fundamental role because of the increasing importance of
stray fields in the direction perpendicular to the sample, as first pointed out by
N\`{e}el \cite{NEE-55}. In addition to usual Bloch DWs, other different types exist,
such as symmetric and non-symmetric N\`{e}el walls, and the cross-tie walls, a
complicated pattern of 90$^\circ$ N\`{e}el walls. A complete description of the
existing configurations together with many experimental images can be found in the
excellent book of Hubert and Sch\"{a}fer \cite{Hubert}. To have an idea of the
complexity, it is worth reporting their very words about the role of film thickness:
"The typical DW in very thin films, i.e. the symmetric N\`{e}el wall, needs basically a
one-dimensional description of its magnetization structure. The stray field of this
wall reaches out into the space above and below the film, however, and cannot be
treated one-dimensionally. Cross-tie walls in thin films need at least a
two-dimensional description for the magnetization and a three-dimensional treatment of
the stray field. Bloch walls in thick films and their relatives are two-dimensional
both in the magnetization and in the stray field. Most complex are cross-tie walls in
thicker films (~100 nm for Permalloy), which need a three-dimensional description of
both magnetization and field and which have never been analysed theoretically up to
now" [From Ref.~\cite{Hubert}, pg. 241].

It is clear that the experimental BK data can be understood only through a precise
knowledge of the active DWs, and their role in the magnetization process. Considering
that the present research is still at a preliminary stage, the task will surely require
some time to be fulfilled. Nevertheless, it is worth to discuss in detail the main
results presented in the literature. A quite impressing work has been done in the Ph.D
thesis of N.J. Wiegman in 1979 \cite{WIE-79}, who explored the properties of Permalloy
thin films using a inductive setup. A selected number of results had been previously
reported in a couple of papers \cite{WIE-77,WIE-78}.

\begin{figure}
   \begin{center}
   \includegraphics[width=12cm]{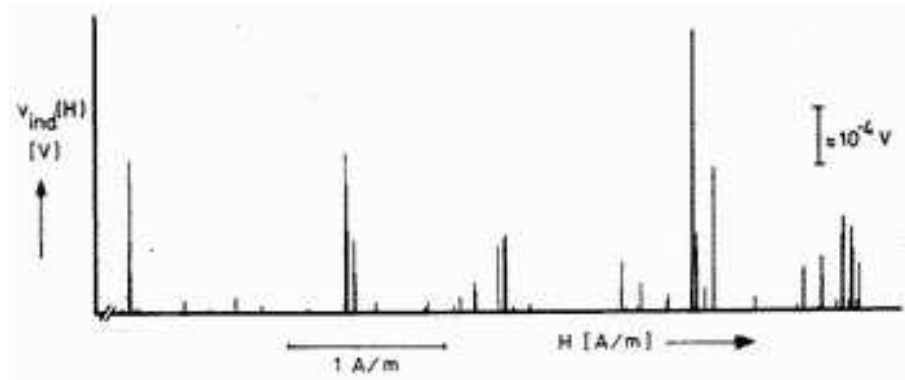}
   \end{center}
   \caption  {A typical time signal for a permalloy thin film
    [From \cite{WIE-79}, Fig. 3.2, pg. 44]. \label{fig:time-seq-film}}
   \end{figure}

\begin{figure}
   \begin{center}
   \begin{tabular}{ccc}
   \includegraphics[height=4cm]{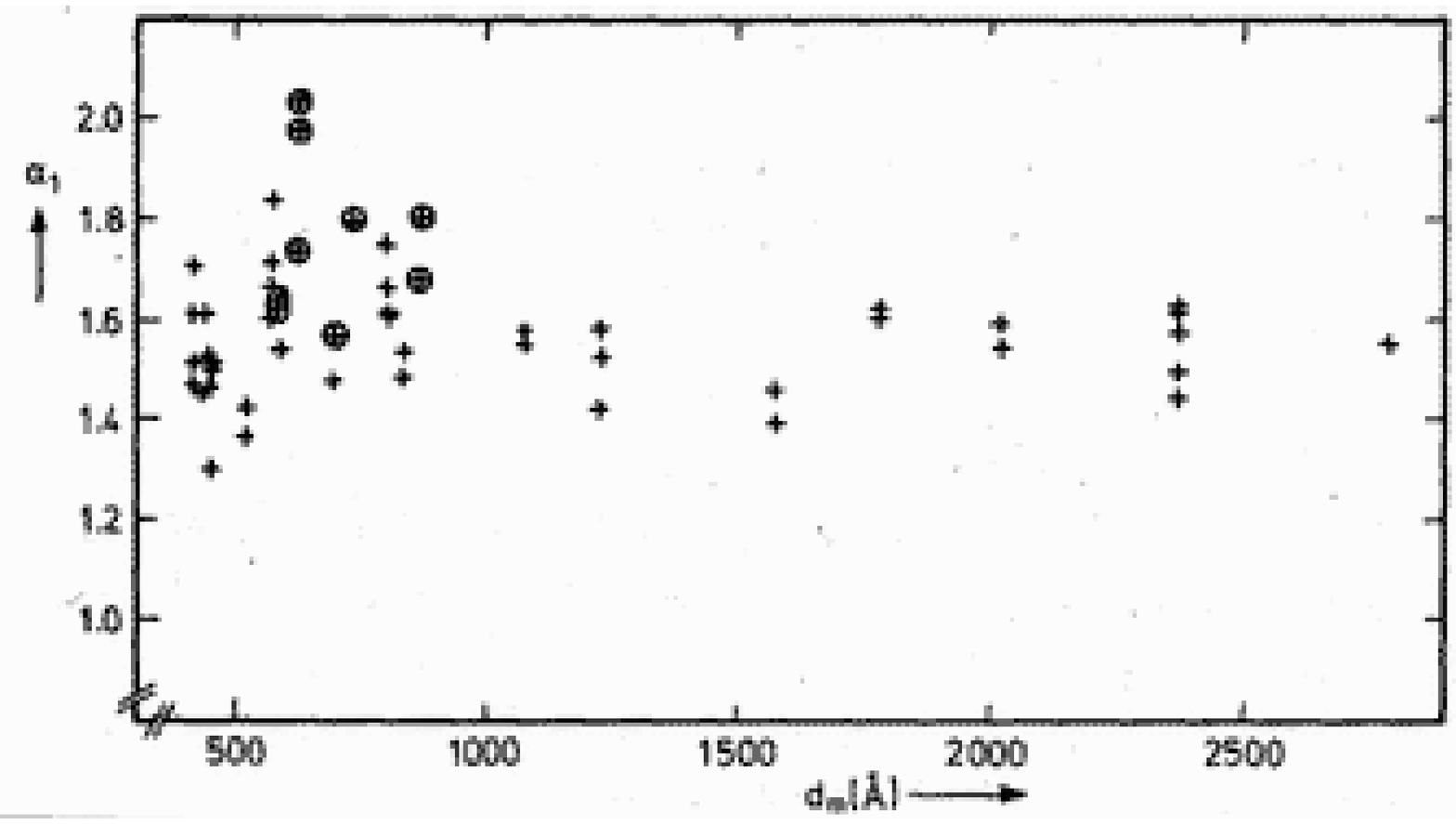} & &
      \includegraphics[height=4cm]{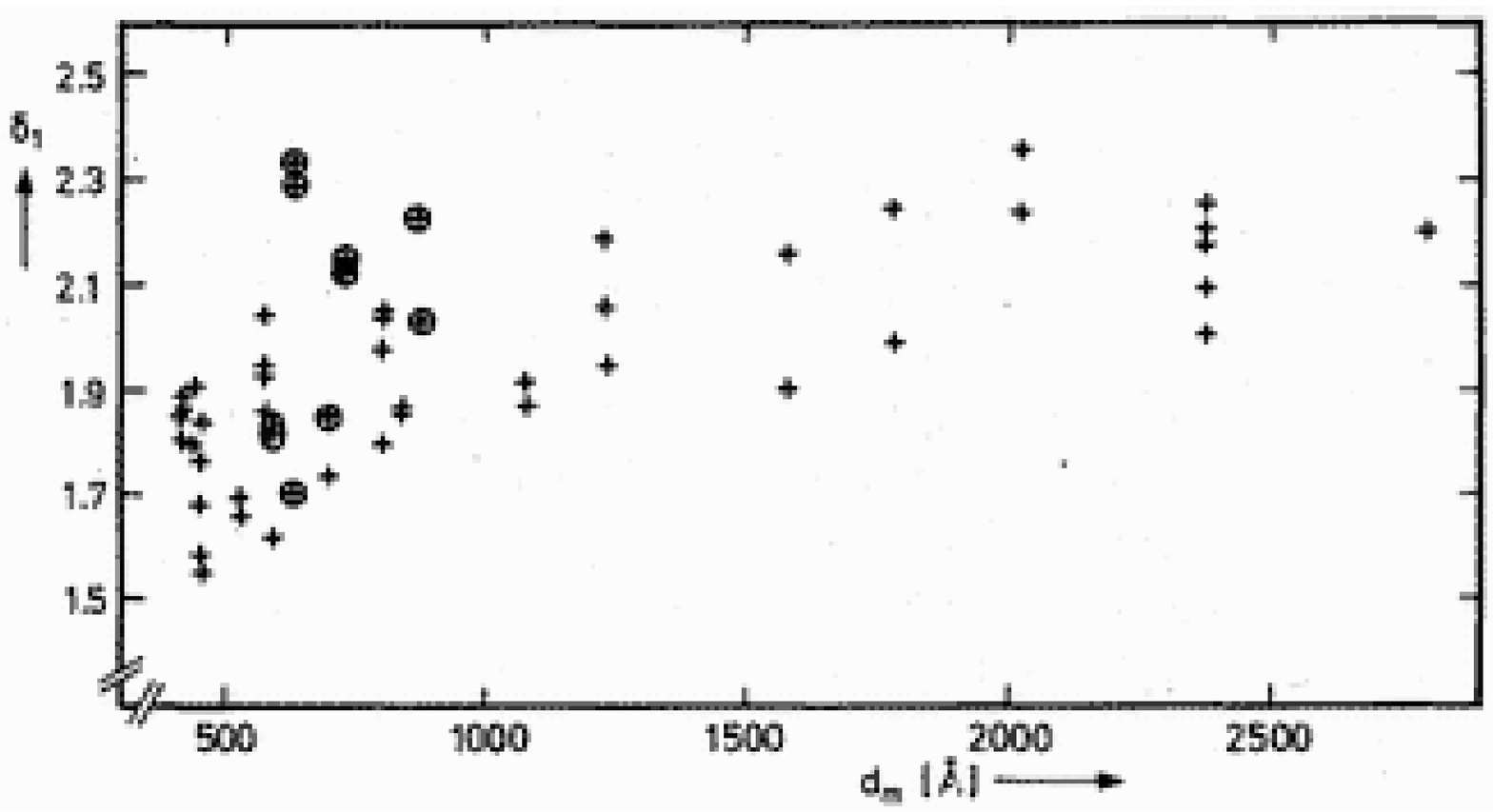}
   \end{tabular}
   \end{center}
   \caption {Values of the critical exponents $\tau$ ($\alpha_1$ in
    Wiegman's notation, left) and $\alpha$ ($\delta_1$, right) for
    permalloy films as a function of the thickness.  [From
    \cite{WIE-79}, Figs. 3.10 and 3.14, pg. 57;62] }
    \label{fig:wie-exp}
   \end{figure}

\begin{figure}
   \begin{center}
   \includegraphics[width=10cm]{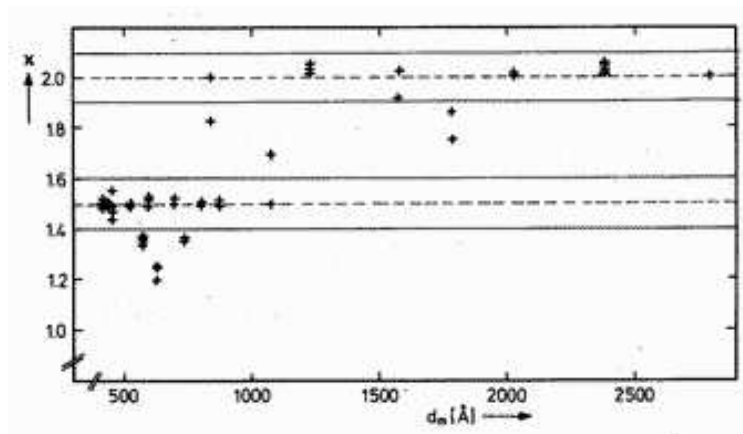}
   \end{center}
   \caption {The critical exponent $1/\sigma \nu z$ ($\chi$ in
    Wiegman's notation) for permalloy films as a function of the
    thickness.  [From \cite{WIE-79}, Fig. 3.7, pg. 53].}
    \label{fig:wie-snuzzu}
   \end{figure}

The simple visual comparison of a typical noise data of Permalloy films with those of
SiFe or amorphous samples (see Fig.~\ref{fig:time-seq}) reveals striking differences.
As shown in Fig.~\ref{fig:time-seq-film}, the BK avalanches appear as sharp peaks, well
separated in time, with nearly no inner structure. This behavior is confirmed by the
data of the size and duration distributions, where the maximum duration is nearly two
orders of magnitude less than the corresponding value in ribbons, as noted before,
while the same does not occur for the maximum size. In other words, the avalanches in
these films are rather sharp peaks of short duration, reversing a magnetization volume
comparable to the value observed in ribbons. The behavior of the critical exponents
$\tau$ ($\alpha_1$ in Wiegman's notation) and $\alpha$ ($\delta_1$), shown in
Fig.~\ref{fig:wie-exp}, as a function of the film thickness is quite intriguing. The
exponents do not show any strong thickness dependence, except for the increasing
fluctuations below 100 nm, which, by the way, is the region where complex cross-tie
walls are observed. In the same region, the critical exponent  $1/\sigma \nu z$
($\chi$) displays a sharp transition of from 2 to 1.5, as clearly shown in
Fig.~\ref{fig:wie-snuzzu}.

  \begin{figure}
   \begin{center}
   \includegraphics[width=8cm]{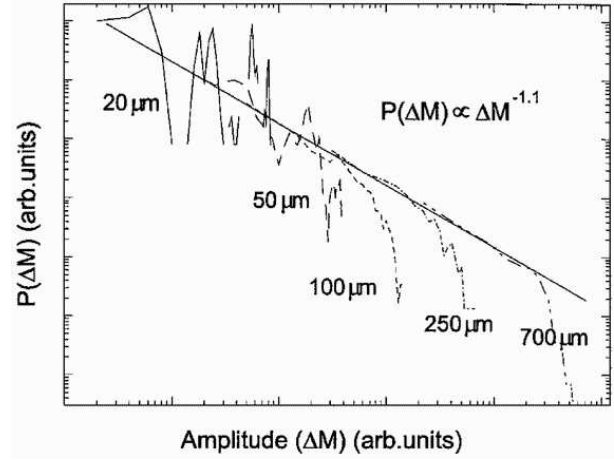}
   \end{center}
   \caption {\protect\label{fig:pup}
    Distribution of magnetization steps $\Delta M$ in a 90 nm
    Fe film. Data taken at different laser spots are properly
    rescaled to show a single distribution [From \protect\cite{PUP-00a},
    Fig. 5, pg. 5418].}

   \end{figure}

  \begin{figure}
   \begin{center}
   \includegraphics[width=12cm]{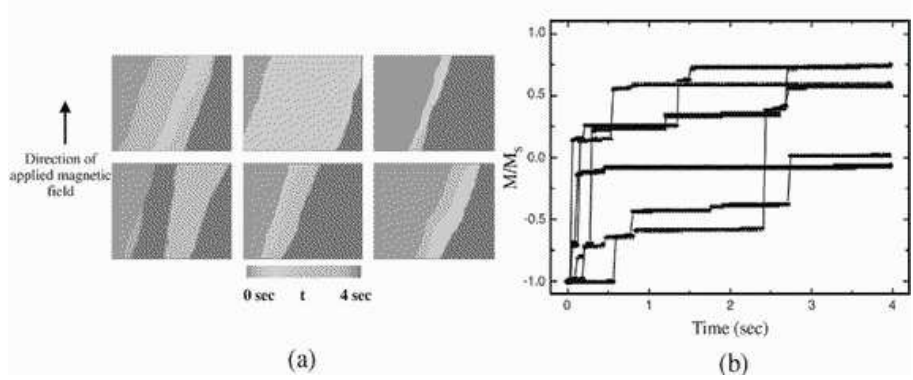}
   \end{center}
   \caption {\protect\label{fig:kim}
    (a) Domain images of same 400 x 320 $\mu m^2$ area for a
    25-nm Co film between 0 and 4 sec. The sample was saturated
    downward first, and then a constant field was applied upward,
    denoted by the solid arrow, during observation. (b) Magnetization
    reversal curves obtained from the corresponding domain patterns of
    (a)[From \protect\cite{KIM-03}, Fig. 2, pg. 087203-2].}

   \end{figure}

These interesting results should be considered as the basis for further investigations.
In fact, the critical exponents have the same values observed in polycrystalline
ribbons ($\tau \sim 1.5, \alpha \sim 2, 1/\sigma \nu z \sim 2$) for thicknesses down to
100 nm, suggesting a typical three-dimensional behavior with predominant long-range
interactions. The situation at smaller thicknesses is more complicated and the role of
DW pattern, such as cross-tie walls, is not clear. While an accurate estimate of the
critical exponents is desirable, it is also necessary to perform an extended scaling
analysis, not restricted to a single exponent (i.e. $\tau$ as in most recent studies).

In Ref.~\cite{PUP-00a}, the critical exponent $\tau$ is estimated in a Fe film of 90 nm
using the MOKE setup (Sec.~\ref{sec:exp_setup}), and considering different laser spot
sizes from 20 to 700 $\mu$m. A single distribution $P(\Delta M)$ is obtained properly
rescaling the size of the avalanche $\Delta M$ and the distribution $P$, as shown in
Fig.~\ref{fig:pup}. Unfortunately, data at small spot sizes are not averaged over
different positions, so that the overall distribution appears quite unreliable under
100 $\mu$m. Given that, the estimated exponent is $\tau = 1.1 \pm 0.05$. A successive
estimation using a spot of 20 $\mu$m averaged over 10 different locations yields $\tau
= 1.14$ \cite{PUP-00b}. It worth noting that these values are much smaller than the
ones observed in permalloy, a fact we will comment very soon. The other critical
exponents are not considered, even if their estimation appears possible.

The paper of Kim \textit{et al.} \cite{KIM-03} reports an estimate of the critical
exponent $\tau$ in Co films, using a different MOKE setup able to visualize directly
the DWs. As shown in Fig.~\ref{fig:kim}, both ends of the DWs are usually outside the
frame, so that the authors deliberately neglect those jumps having one of the ends
inside the frame. The resulting power-law distribution has an exponent close to 1.33,
averaged over 4 different thicknesses, from 5 to 50 $\mu$m. As we explain more
extensively in Sec.~\ref{sec:th-2dim}, this exponent is different from $\tau$, which
should be evaluated, in principle, considering only the jumps fully inside the frame.
As a matter of fact, this appears quite difficult to perform, given the limited number
of such jumps.

It worth noting that a similar comment could be appropriate also for the previous
experiment of Puppin. In that case, it is not possible to distinguish between jumps
fully inside the laser spot and the others. Moreover, the relative probability of the
two types of events is not known, so that in principle we cannot be sure if the
exponent $\tau$ is correctly estimated. With many others, this could be one of the
reasons to get the lower value in respect to the permalloy. But, as discussed above,
all these matters have be confirmed by further experiments.

\section{Models and theories of magnetization dynamics \label{sec:theory}}

A theory of the Barkhausen effect should explain the statistical features of the noise
such as the power spectrum, the avalanche distributions and the pulse shape.  In
particular, the theory should be able to predict the values of the scaling exponents
characterizing the distributions and possibly link the noise properties to
microstructural details of the material under study. It is thus important to establish
a controlled pathway from the microscopic reversal processes and the resulting
macroscopic noise response.

We can list here a set of crucial questions that need to be addressed by the theory:
\begin{enumerate}
\item an explanation for the occurrence of scaling and quantitative estimates for the
exponents. \item the origin of cutoff in the scaling regime. \item the effect of
microstructural parameters (lattice structure, anisotropies, spin interactions) on the
noise statistics.
\end{enumerate}
Clearly these questions are particularly broad and it is dubious that a single answer
will be valid for all materials and experimental situations. The observation of scaling
laws, however, can signify that relatively simple models could capture and reproduce a
wide class of experiments. This is what is expected in critical phenomena, where
universality implies that only the main symmetries and conservation laws influence the
behavior of the system, while many other quantitative details are irrelevant. Thus, if
the Barkhausen effect reflects the presence of an underlying critical phenomenon, one
expects relatively broad universality classes spanning different materials, as it is
indeed observed.

Discussing the main theoretical approaches proposed to understand the Barkhausen
effect, we will pay particular attention to the way each theory addresses the main
questions reported above. In order to put the models into context, we first discuss the
general properties of ferromagnetic materials in terms of the micromagnetic free
energy. This represents a natural microscopic starting point to build up mesoscopic
models and theories amenable to an analytic treatment and large scale numerical
simulations.

\subsection{General properties of ferromagnetic systems}
\label{sec:genprop} A ferromagnetic materials can be represented as an ensemble of
localized magnetic moments or spins, interacting between each other and with the
external magnetic field $\vec{H}$. The macroscopic properties of the material, such as
the Barkhausen noise and the hysteresis loop, are due to microscopic rotations of the
spins and could in principle be described by a microscopic theory. As a first step in
this direction, we will discuss the interactions governing the dynamics of the local
magnetization and the associated energetic contributions. In particular, we can write
the energy of a ferromagnetic material as a sum of different terms
\begin{equation}
E = E_{ex} + E_{m} + E_{an} + E_{dis},
\end{equation}
where $E_{ex}$ represents exchange interactions, $E_{m}$ the magnetostatic energy,
$E_{an}$ the anisotropy and $E_{dis}$ the disorder. The detailed form of these terms,
as a function of the local magnetization $\vec{M}(\vec{r})$ will be discussed in the
following sections.

\subsubsection{Exchange energy}

The most important energetic contribution comes from exchange interactions, which are
typically short-ranged and tend to align spins. For a set of spins $\vec{s}(\vec{r}_i)$
on the lattice, the exchange energy can be written as
\begin{equation}
E_{ex}=\sum_{ij}J(|\vec{r}_i-\vec{r}_j|) \vec{s}(\vec{r}_i)\cdot\vec{s}(\vec{r}_j),
\label{eq:ex}
\end{equation}
where $J(x)$ decays rapidly for large $x$ and the sum is over all atom pairs.
Eq.~\ref{eq:ex} can be approximated in the continuum limit, through an elastic
approximation which involves the removal of all small scale details. The set of spins
$\vec{s}(\vec{r}_i)$ are replaced by a continuous field $\vec{M}(\vec{r})$ and the
exchange energy can be written as
\begin{equation}
E_{ex}= A\int d^3r \sum_{\alpha=1}^{3} (\vec{\nabla}M_\alpha(\vec{r}))^2,
\label{eq:ex2}
\end{equation}
where $A$ is the exchange coupling, which can be derived from $J(x)$.

\subsubsection{Magnetostatic energy}
\label{sec:magnetostatic} The magnetostatic energy is due to the interactions between
the spins and the external field and to dipole-dipole interactions between different
spins. For a uniformly magnetized sample the contribution of the external field to the
magnetostatic energy is simply given by
\begin{equation}
E_m = -\frac{\mu_0}{8\pi}V\vec{M}\cdot\vec{H}, \label{eq:hm}
\end{equation}
where $\vec{M}$ is the magnetization and $V$ is the volume.

In addition, we should  consider the energy due to the demagnetizing field
$\vec{H}_{dm}$, the magnetostatic field generated by the local magnetization. In order
to compute $H_{dm}$, it is useful to define ``magnetic charges'' associated to the
discontinuities of the normal component of the magnetization. For a surface separating
two regions of magnetizations $\vec{M}_1$ and $\vec{M}_2$, the surface charge density
is given by
\begin{equation}
\sigma=\hat{n}\cdot(\vec{M}_1-\vec{M}_2) \label{eq:surfq}
\end{equation}
where $\hat{n}$ is the vector normal to the surface. For instance, at the boundary of
the sample, where the magnetization varies abruptly from $M_s$ to zero, the charge
density is given by $M_s\cos\theta$, where $\theta$ is the angle between the
magnetization direction and $\hat{n}$. It is important to remark that magnetic charges
are not physical charges, but are a convenient tool to compute the demagnetizing field
and magnetostatic energy of magnetized bodies. Everything proceeds as in electrostatics
provided we replace electric charge with magnetic charges and the electric field with
the demagnetizing field which is thus given by
\begin{equation}
\vec{H}_{dm}(\vec{r})=-\vec{\nabla}\cdot \int \frac{dS^\prime \;
\sigma}{|\vec{r}-\vec{r}^{\;\prime}|}, \label{eq:hdem}
\end{equation}
where the integral is over the surface separating two regions of constant
magnetization. For a uniformly magnetized ellipsoid $\vec{H}_{dm}$ is constant and
proportional to magnetization vector
\begin{equation}
\vec{H}_{dm}=-k\vec{M},
\end{equation}
where $k$ is a geometry dependent demagnetizing factor. In this case the
total magnetostatic energy is simply obtained by replacing $\vec{H}$ in Eq.~\ref{eq:hm}
by $\vec{H}+\vec{H}_{dm}$.

In general the demagnetizing field is not constant and Eq.~\ref{eq:hm} has to be
replaced by an integral
\begin{equation}
E_m = -\frac{\mu_0}{8\pi}\int d^3r \vec{M}\cdot(\vec{H}+\vec{H}_{dm}(\vec{r})),
\label{eq:hm2}
\end{equation}
where $H_{dm}$ is given by the contribution of surface charges (i.e. Eq.~\ref{eq:hdem})
summed to the contribution due to the bulk variation of the magnetization. The bulk
dipolar energy can be written explicitly as
\begin{equation}
E_m = -\frac{\mu_0}{8\pi}\int d^3 r d^3 r^\prime \sum_{\alpha,\beta=1}^{3}
\left(\frac{\delta_{\alpha\beta}}{|\vec{r}-\vec{r}^{\;\prime}|^3} -
\frac{3(r_\alpha-r^\prime_\alpha)
(r_\beta-r_\beta^\prime)}{|\vec{r}-\vec{r}^{\;\prime}|^5}\right)
M_\alpha(\vec{r})M_\beta(\vec{r}~{\;\prime}). \label{eq:dip}
\end{equation}
As in the case of the  surface term  Eq.~\ref{eq:dip}, can also be rewritten in terms
of the density of magnetic volume charges
$\rho(\vec{r})=\vec{\nabla}\cdot\vec{M}(\vec{r})$ as
\begin{equation}
E_m = -\frac{\mu_0}{8\pi}\int \frac{d^3r\;d^3r^\prime
\rho(\vec{r})\rho(\vec{r}^\prime)}{|\vec{r}-\vec{r}^{\;\prime}|}.
\end{equation}

\subsubsection{Magnetocrystalline anisotropy and magnetoelastic energies}

The magnetization in a ferromagnetic material typically has preferential directions
corresponding to the crystallographic axis of the material. Therefore, it is easier to
magnetize the sample along the easy directions. This observation can also be expressed
in terms of the energy of magnetocrystalline anisotropy
\begin{equation}
E_{an} = \int d^3r \sum_{\alpha,\beta}K_{\alpha\beta} M_\alpha M_\beta, \label{eq:anis}
\end{equation}
where $M_\alpha$ is the $\alpha$ component of the vector $\vec{M}$ and $K_{ij}$ is a
symmetric tensor, describing the anisotropy of the material. In the simplest case of a
uniaxial crystal, Eq.~\ref{eq:anis} reduces to
\begin{equation}
E_{an}=\int d^3r K_0 (\vec{M}\cdot \hat{e})^2=\int d^3r K_0 M^2 \sin^2 \phi
\label{eq:an}
\end{equation}
where $\phi$ is the angle between the easy axis $\hat{e}$ and the magnetization vector
and $K_0$ is the uniaxial anisotropy constant.

The variations of the magnetization inside a ferromagnetic sample can cause deformation
in the lattice structure, a phenomenon known as magnetostriction. Conversely, when an
external mechanical stress is applied to the sample the magnetic structure can in
principle be modified. To describe this effect it is useful to introduce the
magnetoelastic energy, which in the most general form can be written as
\begin{equation}
E_{an} = \int d^3r
\sum_{\alpha,\beta,\gamma,\delta}\lambda_{\alpha\beta\gamma\delta}\sigma_{\alpha\beta}
M_\gamma M_\delta,
\end{equation}
where $\sigma_{\alpha\beta}$ is the stress tensor and
$\lambda_{\alpha\beta\gamma\delta}$ is the magnetoelastic tensor. For a crystal with
isotropic magnetostriction, under a uniaxial stress $\sigma$ the anisotropy energy
takes the simple form of Eq.~\ref{eq:an} with $K_0$ replaced by $K_0+3/2 \lambda
\sigma$, where $\lambda$ is the uniaxial magnetostriction constant.

\subsubsection{Disorder}

In the previous discussion we have considered only an homogenous system, in which the
interactions are globally defined and do not depend on position. In general, however,
different sources of inhomogeneities are found in virtually all ferromagnetic
materials. The presence of structural disorder is essential to understand the
fluctuations in the Barkhausen noise, which would be strongly suppressed in a perfectly
ordered system. The nature of the disorder can be inferred from the microscopic
structure of the material under study.

We can thus distinguish several contributions to the magnetic free energy due to the
disorder: in a crystalline materials disorder is due to the presence of vacancies,
dislocations or non-magnetic impurities. In polycrystalline materials we should add to
these defects the presence of grain boundaries and variations of the anisotropy axis in
different grains. Finally in amorphous alloys disorder is primarily due to internal
stresses and the random arrangement of the atoms. It is important to notice that in the
following we will consider the disorder as quenched (or ``frozen''): it does not evolve
on the timescales of the magnetization processes under study. It is not always simple
to quantify the energetic contributions of the different sources of disorder but we can
highlight here the main effects.

The presence of randomly distributed non-magnetic inclusions give rise to a
magnetostatic contribution, due to the magnetic charges that form at the boundaries of
the inclusions \cite{NEE-46}. The energetic contribution due to a collection of those
inclusions can be expressed as a local fluctuation of the dipolar coupling. A similar
discussion can be repeated in the case of exchange interactions, leading to a
fluctuating exchange coupling. The typical fluctuations of the coupling depend on the
volume fraction $v$ of the non magnetic inclusions: the typical strength of the
(exchange or dipolar) coupling $g(\vec{r})$ will be of the order of $\bar{g}\simeq
(1-v)g_0$ and the fluctuations $\langle (g-\langle g \rangle)^2\rangle \simeq vg_0^2$,
where $g_0$ is the coupling without impurities. This type of disorder is conventionally
called of random-bond type and should also be present in amorphous alloys, due to the
random arrangements of the atoms.

In polycrystalline samples, each grain has a different crystalline anisotropy. In
particular, the direction of the anisotropy axes will fluctuate in space and in the
simple case of uniaxial anisotropies the anisotropy energy will be given by
$E_{an}=\int d^3r K (\vec{M}\cdot \hat{e}(\vec{r}))^2,$ where $\hat{e}$ is a random
function of position reflecting the grain structure of the material under study.

In magnetostrictive samples internal stresses play a similar role than anisotropies, as
we discussed in the previous section. In particular, a random distribution of internal
stresses produces an random energy of the type of Eq.~\ref{eq:an} with a random
anisotropy constant $K(\vec{r}) \propto K_0+3/2\lambda\sigma(\vec{r})$. While internal
stresses can have multiple origins, it is possible to compute explicitly their
distribution in some particular cases, such as in a crystal with a random distribution
of parallel dislocations. The stress in $\vec{r}$ due to a dislocation in the origin is
given in cylindrical coordinates by $\sigma_{\alpha\beta}=b\mu
C_{\alpha\beta}(\theta)/r$, where $b$ is the Burgers vector, $\mu$ is the shear modulus
and $C_{\alpha\beta}$ is an angular function depending on the dislocation type
\cite{Hirth}. The distribution of the internal stresses is formally given by
\begin{equation}
P(\tilde{\sigma}_{\alpha\beta})=\int d^{2N}r D(\vec{r}_1,...\vec{r}_N)
\delta(\tilde{\sigma}_{\alpha\beta} - \sum_k \sigma_{\alpha\beta}(\vec{r}-\vec{r}_k)),
\end{equation}
where $D$ is the distribution of the position of the dislocations. The internal stress
distribution has been evaluated in the case of a random short-range correlated
dislocation arrangement: $P(\sigma)$ is Gaussian for small stresses, with variance
$\langle \Delta\sigma^2 \rangle \propto\rho$, where $\rho$ is the dislocation density
and at larger stresses displays a power law tail, scaling as $\sigma^{-3}$
\cite{GRO-98}. It is interesting to remark that $P(\sigma)$ is an experimentally
accessible quantity, since it is directly related to the X-ray spectrum \cite{SZE-00}.

We have discussed two common contributions to the disorder energy in ferromagnetic
materials: random bonds and random anisotropies. To conclude the discussion, we remark
that some diluted antiferromagnets in a field can be effectively described by a
random-field type disorder, with energy given by
\begin{equation}
E_{dis}=\int d^3r \vec{h}(\vec{r})\cdot \vec{M}
\end{equation}
where $\vec{h}$ is randomly distributed in space. Despite the fact that random field
disorder is not present in ferromagnetic materials, it provides a useful theoretical
ingredient to understand the role of disorder in hysteresis, as we will discuss in the
following.

\subsubsection{Micromagnetic equations}

In the previous sections we have discussed the different energetic contributions to
ferromagnetic materials. Collecting all the terms, considering for simplicity a
uniaxial material, we can write the energy
\begin{equation}
E=\sum_{\alpha=1}^{3}\int d^3r  [A(\vec{\nabla}M_\alpha)^2+K(M_\alpha e_\alpha)^2-
(H_\alpha+H_{dem}^{(\alpha)})M_\alpha], \label{eq:totmm}
\end{equation}
$A$, $K$ and $e_\alpha$ in principle depend on position and $H_{dem}^{(\alpha)}$ is the
component $\alpha$ of the demagnetizing field, discussed in section
\ref{sec:magnetostatic}.

The energy function reported in Eq.~\ref{eq:totmm} can in principle be used to compute
the equilibrium properties of a ferromagnetic materials. This problem is in general
very complex and we do not attempt to treat it here. In fact, to describe the
Barkhausen effect we are interested in the evolution of the magnetization in response
of an increasing external field. The equation of motion for the magnetization can
readily be obtained from Eq.~\ref{eq:totmm}
\begin{equation}
\frac{\partial \vec{M}}{\partial t}=\gamma \vec{M}\times \vec{H}_{eff}
\label{eq:micromag}
\end{equation} where $\gamma$ is the charge to mass ratio and
$\vec{H}_{eff}\equiv-\delta E/\delta \vec{M}$. Eq.~\ref{eq:micromag} does not include
dissipation mechanism and would predict an indefinite precession of the magnetization
vector. This problem can be overcome introducing phenomenological laws for the
dissipation.  This results for instance in the Landau-Lifshitz-Gilbert equation whose
numerical integration for different microstructures and boundary conditions is the
subject of micromagnetism.

It is tempting to use a micromagnetic approach describe collective effects arising in
the Barkhausen effect from first principles.  An attempt on this direction was made in
Refs.~\cite{GON-97,CHU-98,CHU-98a} , where a Montecarlo version of the micromagnetic
equation was used to simulate the Barkhausen effect in polycrystals.

The model consists in a chain of $N$ magnetic moments with orientations defined by two
angles $\phi_i$ and $\theta_i$. The total energy of the system is given by
\begin{equation} E = \frac{1}{2}\sum_i (
\sin^2\alpha_i+2H\cos\theta_i+m\sin^2\theta_i\sin^2\phi_i -2a \cos \beta_{i,i+1}),
\end{equation} where $\alpha_i$ is the angle between the magnetic moment and the local
(random) easy axis, $H$ is the magnetic field, $a$ is the exchange-anisotropy energy
ratio and $m$ controls the magnetostatic energy. The last term takes into account the
interaction between magnetic moments and $\beta_{i,j}$ is the angle between the moments
in $i$ and $j$.

The evolution of the magnetization is dictated by a Montecarlo algorithm. As the
external field is slowly increased the magnetic moments rotate in bursts which are
associated with the Barkhausen jumps.  The results obtained for the jump distributions
are in some range of parameters reminiscent of the experiments (i.e. a power law with a
cutoff) but a quantitative comparison with experiments would be questionable.  The
model treats very accurately some aspects of ferromagnetic system, namely spin
rotations, nucleation and random anisotropies, but completely neglects the fundamental
aspects of the problem: dimensionality (the model is one dimensional), the range of
interactions (only nearest-neighbor interactions are present, while in reality dipolar
interactions give rise to long-range terms decaying as $1/r^3$) and demagnetizing
effects.  For these reasons, this model is not able to reproduce quantitatively the
scaling properties of the Barkhausen effect.  To this end, it would be necessary to
study three dimensional models with long-range interactions, a task that promises to be
computationally very demanding.

\subsection{Random energy models}
\label{sec:abbm} The difficulties arising in the direct analysis of the micromagnetic
equations naturally lead to search for simplified models that could capture the
essential mechanism underlying the Barkhausen noise without excessive complications.
The main ingredient of any model for the Barkhausen effect is the presence of disorder,
thus the simplest approach is to neglect most of the details of the micromagnetic
energy, keeping only a uniform magnetostatic term $E_m$ and replacing the rest with a
random function of the magnetization $m$
\begin{equation}
E=F(m)-mH_{eff},
\end{equation}
where $H_{eff}=H+H_{dem}$. The rationale behind this approximation can be understood
considering a single rigid domain wall dividing the sample in two domains. In this case
the magnetization is proportional to the wall position $x$: $m= M_s (2x/L-1)$, where
$M_s$ is the saturation magnetization and $L$ is the sample width. The sample is
magnetized as the applied magnetic field pushes the domain wall across a disordered
landscape represented by the random potential $F(m)$. In order to define completely the
problem, one needs to specify the statistical features of $F(m)$, such as its
distribution and correlations.

\begin{figure}[h]
\centerline{\psfig{file=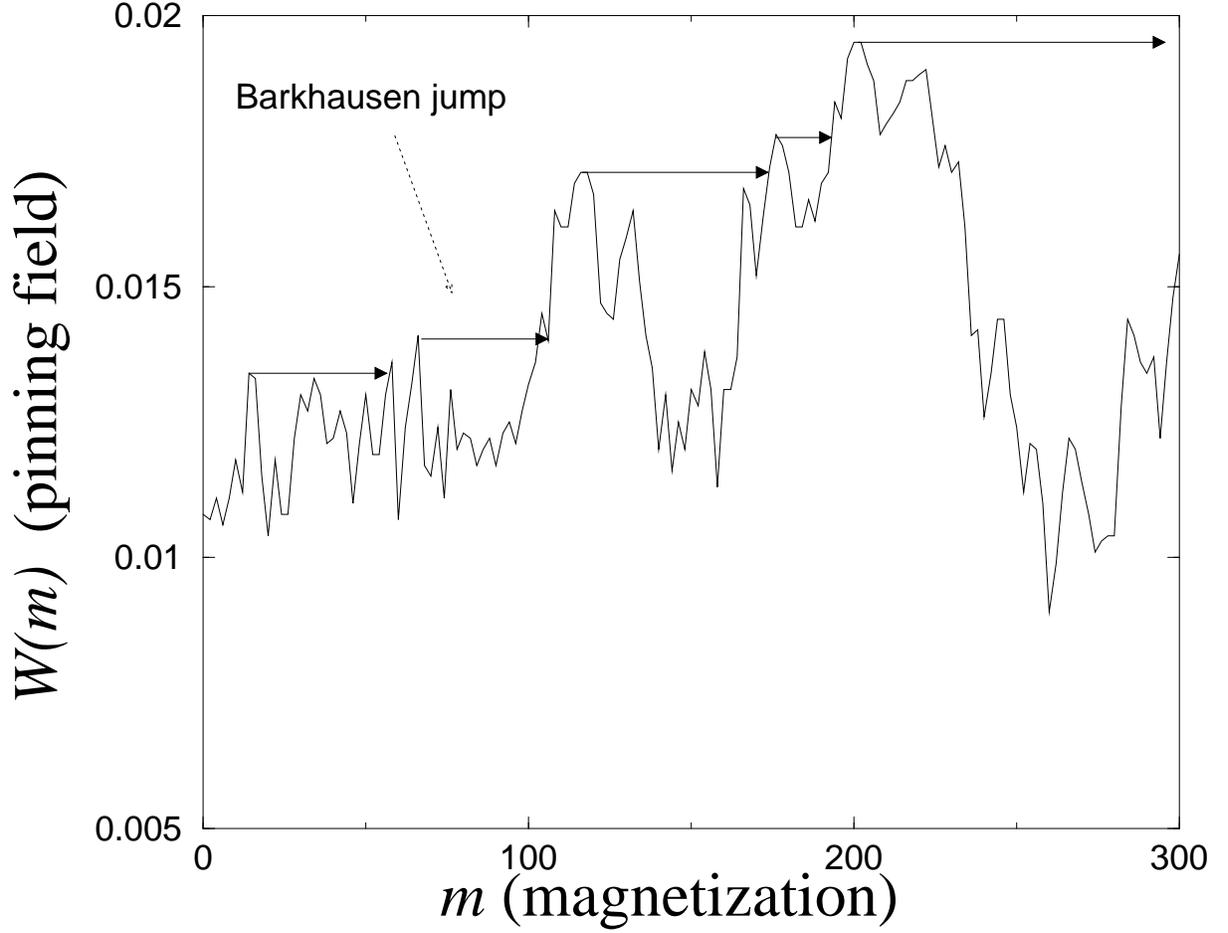,width=16cm,clip=!}} \caption{The Barkhausen jumps
(denoted by arrows) from the perspective of the random energy model. }
\label{fig:randomenergy}
\end{figure}

Ne\'el was the probably the first to use a random energy model to study the hysteresis
loop properties at low fields in his theory of the Rayleigh loops \cite{NEE-42,NEE-43}.
In the N\'eel model the random function was schematized as a series of parabolas with
randomly distributed curvatures. The Barkhausen noise can be constructed from the
random energy model by a visual inspection of the pinning field $W(m)\equiv-\frac{d
F}{dm}$ as shown in Fig.~\ref{fig:randomenergy}. As the effective field is increased,
the domain wall is pinned as long as $H_{eff}<W(m)$, when $H_{eff}$ reaches a local
maximum of $W(m)$ the domain wall jumps forward until the condition $H_{eff}<W(m)$ is
met. The Barkhausen sizes $S$ are just the changes in magnetization $\Delta m$
occurring after these jumps or, in more mathematical terms, the first returns of the
stochastic process described by $W(m)$.  The random energy model in the form proposed
by N\'eel, however, can not describe correctly the Barkhausen noise because the pinning
field is essentially uncorrelated. Thus the first return distributions decays
exponentially, while the experiments indicate a power law distribution.

In order to obtain a power law distribution of returns the pinning field has to be
long-range correlated. This point was first realized by Bertotti \cite{BER-86,BER-87},
who included a Brownian pinning field in a random energy model. The model, widely known
as ABBM \cite{ALE-90,ALE-90a}, yields a quite accurate description of the Barkhausen
noise statistics. In the ABBM model the magnetization evolves according to an
overdamped equation of motion \beq \frac{dm}{dt}= ct-k m +W(m), \label{eq:abbm} \eeq
where the external field increases at constant rate $H=ct$, $k$ is the demagnetizing
factor, and the damping coefficient has been set to unity rescaling the time units. The
pinning field is a Brownian process where correlations grow as \beq \langle
(W(m)-W(m'))^2 \rangle = D|m-m'|. \eeq

The main predictions of the ABBM model can be obtained deriving Eq.~(\ref{eq:abbm})
with respect to time and defining $v \equiv d m/ dt$ \beq \frac{d v}{dt}=c-k v + v
f(m), \label{eq:dtabbm} \eeq where $f(m)\equiv dW/dm$ is an uncorrelated random field
with variance $D$.

Expressing Eq.~\ref{eq:dtabbm} as a function of $v$ and $m$ only \beq \frac{d
v}{dm}=\frac{c}{v}-k + f(m), \label{eq:abbm-log} \eeq we obtain a Langevin equation for
a random walk in a confining potential, given by $U(v) = kv- c\log(v)$. Asymptotically,
the statistics of $v$ is ruled by the Boltzmann distribution \beq P(v, m\to \infty)
\sim \exp(-U(v)/D)=v^{c/D} \exp(-kv/D). \eeq

The distribution in the time domain is obtained by a simple transformation and it is
given by \cite{ALE-90} \beq P(v)\equiv P(v, t\to \infty)=\frac{k^{c/D}v^{c/D-1} \exp(-k
v/D)}{D^{c/D}\Gamma(c/D)}. \label{eq:pv-abbm} \eeq A consequence of
Eq.~(\ref{eq:pv-abbm}) is that the domain wall average velocity is given by $\langle v
\rangle = c/k$. For $c/D<1$ the velocity distribution in Eq.~\ref{eq:pv-abbm} is a
power law with an {\em upper} cutoff that diverges as $k \to 0$. In this regime, the
domain wall moves in avalanches whose size and durations are also distributed as power
laws. For $c/D>1$ the motion is smoother with fluctuations that decrease as $c/D$
increases.

The avalanche size distribution equals the distribution of first return times of a
random walk in the confining potential $U(v)$. In the limit $k\to 0$, we can directly
apply the exact result reported in Ref.~\cite{BRA-00} for the first return times of a
random walk in a logarithmic potential, which implies that the avalanche exponents
depends on $c$ and are given by \beq \tau=3/2-c/2D~~~~~\alpha =2-c/D. \label{eq:exp-c}
\eeq See also Ref.~\cite{DUR-95,DUR-95a} for alternative but not rigorous derivations
of these results.

The scaling of the cutoff of the avalanche distributions in the ABBM model can be
obtained similarly in the limit $c\to 0$ solving for the first return probability of a
biased random walk and is given by \beq S_0 \sim k^{-2}. \label{eq:schi} \eeq Using
similar arguments one can also show that the cutoff of avalanche durations scales as
$T_0 \sim k^{-1}$ \cite{ZAP-98}.

\subsection{Spin models}

In this section we discuss the attempts to understand the statistical properties of the
Barkhausen effect by models of interacting spins, that can either be simulated on a
computer or in some cases analyzed theoretically.  This approach is conceptually simple
and theoretically appealing, since it tries to derive directly macroscopic properties
from a microscopic model, replacing the vectorial micromagnetic equations by simplified
rules for the evolution of integer valued spins. The basis of this kind of approach is
the notion of universality which is expected for critical phenomena: the values of the
critical exponent and the shape of the scaling functions are independent on the
microscopic details of the system, provided that the relevant symmetries of the problem
are correctly taken into account.  For this reason, we can expect to obtain the correct
large scale behavior, such as the scaling properties of the Barkhausen effect, from
relatively simple models.

\subsubsection{The random field Ising model}

Sethna et. al \cite{SET-93} have proposed the driven random-field Ising model as a
prototype for hysteresis and avalanches at first-order phase transitions. The model was
studied extensively analytically and numerically and the resulting scaling behavior was
compared with the experimental results on the Barkhausen effect
\cite{DAH-93,PER-95,DAH-96,PER-99,SET-01}.

In this models, a spin $s_i = \pm 1$ is assigned to each site $i$ of a $d-$dimensional
lattice. The spins are coupled to their nearest-neighbors spins by a ferromagnetic
interaction of strength $J$ and to the external field $H$.  In addition, to each site
of the lattice it is associated a random field $h_i$ taken from a Gaussian probability
distribution with variance $R$, $P(h)=\exp(-h^2/2R^2)/\sqrt{2\pi}R$.  The Hamiltonian
thus reads
\begin{equation} E = -\sum_{\langle i,j \rangle}Js_i s_j
-\sum_i(H+h_i)s_i, \label{eq:rfim}
\end{equation}
where the first sum is restricted to nearest-neighbors pairs. In the zero temperature
dynamics used by Sethna et al. \cite{SET-93}, the external field is ramped from
$-\infty$ to $\infty$ and the spins align with the local field \cite{BER-90}
\begin{equation}
s_i = \mbox{sign}(J\sum_j s_j  + h_i +H).
\end{equation}
In this way a single spin flip can lead the neighboring spins to flip, eventually
trigger an avalanche.

For small values of the disorder $R$ the first spin flips are likely to generate a big
avalanche whose size is comparable to the system size, leading to a discontinuous
magnetization reversal. On the other hand, a large disorder prevents the formation of
large avalanches and the magnetization reversal is smooth. In fact, these two regimes
are separated by a critical disorder $R_c$, where the avalanches are distributed as a
power law. The behavior of the model close to $R_c$ is very similar to that of
equilibrium systems at the critical points and can thus be studied by standard methods,
like mean-field theory and the renormalization group.

A first qualitative picture of the behavior of the RFIM can be obtained by mean-field
theory \cite{SET-93,DAH-96}. To this end, we consider Eq.~\ref{eq:rfim} extending the
sums to all the pairs of sites and obtain
\begin{equation}
E = -\sum_{i}(JM+H+h_i)s_i, \label{eq:rfim.mf}
\end{equation}
where $M=\sum_i s_i$. The magnetization can then be obtained self-consistently as
\begin{equation}
M=1-2 P(JM+H+h_i<0)=1-2\int_{-\infty}^{-JM-H} \rho(h)dh.
\end{equation}
This equation has a single valued solution for $R>R_c=\sqrt(2/\pi)J$, while the
solution becomes multivalued for $R<R_c$. In the second case, we have an hysteresis
loop with a jump at $H_c$ and a diverging slope $dM/dH$. For high disorder there is no
hysteresis, but this is an artifact of mean-field theory that is not observed in low
dimensions. At the critical point $R=R_c$ the magnetization has a diverging slope, but
no finite jump.  In analogy with critical phenomena it is possible to obtain scaling
laws describing the singularities occurring at the transition. In particular the
magnetization obeys the scaling law
\begin{equation}
M=|r|^\beta g_\pm(h/|r|^{\beta\delta}),
\end{equation}
where $r\equiv (R-R_c)/R_c$ and $h \equiv (H-H_c)/H_c$ are the reduced control
parameters, and $\beta=1/2$, $\delta=3$. The scaling function $g_\pm$ is obtained as
the root of a cubic equation.

The distribution of avalanche sizes near the critical point can be computed
analytically in the framework of mean-field theory. The result can be summarized in the
scaling form
\begin{equation}
P(S,r,h)= S^{-\tau} {\mathcal P}_{\pm} (S/|r|^{-1/\sigma},h/|r|^{\beta\delta}),
\label{eq:psrh}
\end{equation}
where the exponents take the mean-field values $\tau=3/2$, $\sigma=1/2$. The scaling
function ${\mathcal P}_{\pm}$ can be evaluated exactly in terms of $g_\pm$ and is given
by
\begin{equation}
{\mathcal P}_{\pm}(x,y)=(1/\sqrt{2\pi}) \exp (-x(1\mp \pi g_\pm(y)^2/4)^2/2).
\end{equation}
In addition to the size distribution one can also define the distribution of durations
$T$, which obeys an analogous scaling form
\begin{equation}
D(T,r,h)= T^{-\alpha} {\mathcal D}_{\pm} (T/|r|^{-1/(\sigma\nu
z)},h/|r|^{\beta\delta}), \label{eq:rh}
\end{equation}
where $\alpha=2$, $\nu$ is the correlation length exponent ($\xi \sim r^{-\nu}$) and
$z$ is the dynamic exponent relating the correlation length to the characteristic time
($\xi \sim T_0^z$).

Mean-field theory has the advantage of being easily tractable and provides a
qualitative picture of the behavior of the model, but the numerical values of the
exponents are typically inaccurate in dimensions lower than the upper critical
dimension, that for the RFIM it is equal to $d_c=6$. To overcome this problem, one can
perform a renormalization group analysis with an expansion in $\epsilon=6-d$ as in
equilibrium critical phenomena. The renormalization group is quite involved and we do
not discuss it here in detail, but just quote the main results (for a complete
discussion the reader is referred to Ref.~\cite{DAH-96}). The exponent determining the
scaling of the order parameter have been computed to first order in $\epsilon$ and are
estimated to be $\beta=1/2-\epsilon/6$ and $\delta = 3 + \epsilon$. The avalanche
exponent $\tau$ displays only corrections to order $\epsilon^2$ and the cutoff exponent
is estimated as $\sigma=1/2-\epsilon/12$.

\begin{figure}
\centerline{\psfig{file=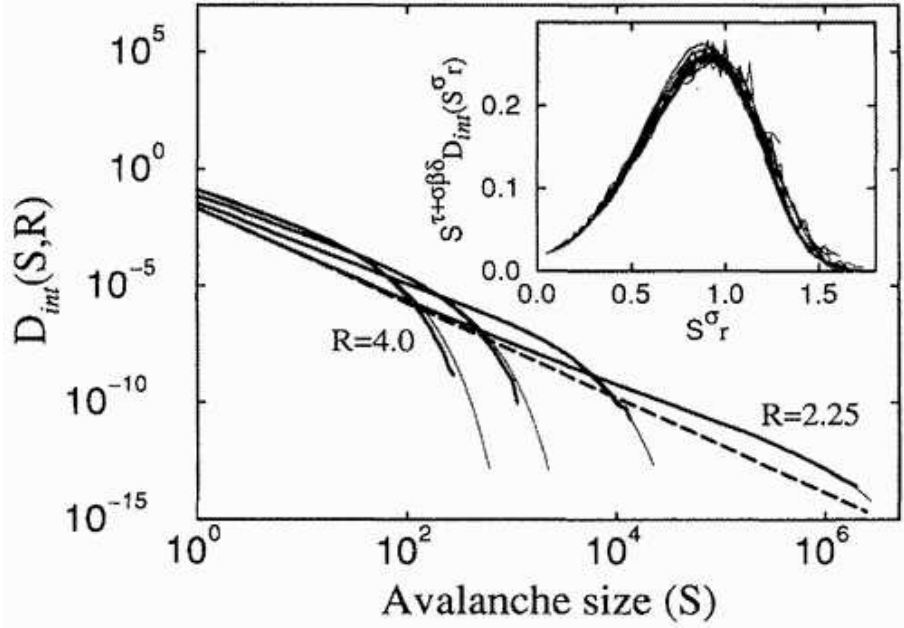,width=12cm,clip=!}} \caption{The integrated
avalanche distribution in the RFIM in $d=3$ simulated in a  lattice of $320^3$ spins.
The inset shows the data collapse. [From \protect\cite{PER-95}, Fig.~1, pg. 4528]}
\label{fig:per95}
\end{figure}

Large scale numerical simulations have been performed, to obtain reliable estimate for
the critical exponents in three dimensions. The avalanche size distribution is most
naturally computed integrating the avalanches over the entire hysteresis loop. The
exponent $\tau_{int}$ measured in this way can be related to the other exponents by a
scaling relation $\tau_{int}=\tau+\sigma\beta\delta$. Typically the distribution
displays an initial power law behavior and a cutoff that depends on the disorder $R$
(see Fig.~\ref{fig:per95}). The distribution for different values of $R$ can then be
collapsed using a scaling form of the type of Eq.~\ref{eq:psrh} \cite{PER-95,PER-99}.
The best numerical estimate for the exponents (or exponent combinations) in three
dimensions is $\tau =1.60\pm 0.06$, $\tau_{int}=2.03\pm 0.03$, $1/\sigma=4.2 \pm 0.3$,
$\sigma\nu z = 0.57 \pm 0.03$, $\alpha =2.02$, $\beta=0.035 \pm 0.03$, $1/\nu=0.71$
\cite{PER-99}.

Looking at the results above one may be tempted to perform a direct comparison with
experiments. However, this should be done with a grain of salt. The RFIM model in fact
predicts scaling only close to a critical disorder $R_c$ while in general there is no
reason to believe that this condition is fulfilled in experiments. To relate the
Barkhausen scaling behavior observed in experiments to a disorder induced phase
transition, such as the one discussed here, it would be necessary to show that the
cutoff to the scaling behavior scales with the disorder strength. Unfortunately such a
measure has not been performed yet.

\subsubsection{Random bonds, random anisotropies and other effects}

The RFIM represents probably the simplest model to study the effect of disorder in
ferromagnetic hysteresis and Barkhausen noise. The model displays a disorder induced
transition with associated critical exponents, which has been recently observed in
experiments. While random fields can be a useful theoretical idealization, they are not
present in ferromagnetic materials. Thus an important question to address is whether
the basic phenomenology of the RFIM carries over to more realistic types of disorder.
While the answer to this question appears to be positive, it is still controversial
whether the disorder induced phase transition is universal with respect to the type of
disorder. A second related question is whether the inclusion of more realistic
ingredients in the RFIM results or not in a change of the scaling exponents.

The first question was addressed by numerical simulations, considering random bonds,
site dilution, and random (infinite) anisotropies.  In the random bond Ising model
(RBIM), the disorder enters into the exchange coupling $J_{ij}$ which is randomly
distributed with mean $J_0$ and standard deviation $R$. Hysteresis in the
RBIM was first studied in the spin glass limit ($J_0=0$)
\cite{BER-90}. In analogy with the RFIM, the
RBIM displays a disorder induced phase transition in the hysteresis loop.  Some of the
scaling exponents, including those ruling the avalanche distributions, have been
evaluated and their values differ slightly from those measured in the RFIM. For
instance, some reported values are $\tau_{int}\simeq 2$, $z\simeq 1.6$ and $\nu \simeq
1$ \cite{VIV-94}. One should beware, however, that the lattice sizes used in this study
are rather limited (up to $40^3$) \cite{VIV-94} compared to those used to estimate
exponents in the RFIM (up to $320^3$) \cite{PER-99}.

In the site diluted Ising model (SDIM), the presence of non-magnetic impurities is
treated by removing a fraction of spins in the lattice. The Hamiltonian is thus simply
given by
\begin{equation}
E= -j\sum_{ij} J_{ij} c_i c_j s_i s_j - \sum_i H c_i\sigma_i
\end{equation}
where $c_i=0,1$ indicates weather the spin is magnetic or not. Typically, the sum in
the exchange interaction is not restricted to the nearest neighbors but extends to a
larger range. Simulations of the SDIM in $d=3$ yield exponents compatible with those of
the RFIM, namely $\tau_{int}=2.0 \pm 0.2$ \cite{VIV-00}.

The random (infinite) anisotropy model is obtained from a vectorial model (a lattice
version of Eq.~(\ref{eq:tot})
\begin{equation}
E=-\sum_{\langle ij \rangle }J \vec{s}_i\cdot\vec{s}_j- \sum_{i} K(\vec{s}_i \cdot
\hat{n}_i)^2+\vec{H}\cdot \vec{s}_i,
 \end{equation}
in the limit of $K\to \infty$. Thus to minimize the anisotropy energy, in each site the
spin should be parallel to the quenched random direction $\hat{n}_i$.  The original
vectorial model becomes then effectively of Ising type as can be seen by introducing a
fictions spin $\sigma_i =\pm 1$ defining the sign of $\vec{s}_i$ along its random axis
(i.e. $\vec{s}_i=\hat{n}_i \sigma_i$). In terms of the new spin variables the
Hamiltonian can be rewritten as
\begin{equation}
E=-\sum_{\langle ij \rangle } J_{ij} \sigma_i \sigma_j - \sum_i H_i\sigma_i,
\end{equation}
with effective random bonds $J_{ij}\equiv J \hat{n}_i\cdot\hat{n}_j$ and random fields
$H_i = n_z$. The model above was simulated numerically with a zero temperature dynamics
analogous to the one employed in the RFIM and with two different distributions for the
anisotropy axis. The measured exponents are in good agreement with those measured in
the RFIM. In particular for the avalanche distribution, the reported values are
$\tau_{int}=2.06\pm 0.05$ and $\tau_{int}=2.10\pm 0.05$ depending on the distribution
\cite{VIV-01}. A recently proposed analysis, based on the renormalization group, of the
vectorial random anisotropy model shows that the model should be described generically
by the RFIM fixed point \cite{DAS-03}, thus supporting the conclusions of the
simulations.

{}From the discussion above, we conclude that the disorder induced critical point in the
RFIM represents a broad universality class describing different sorts of spin models.
This fact, if confirmed by larger scale simulations and renormalization group
calculation, is particularly interesting because it allows to concentrate on a
relatively simple model (the RFIM) to understand quantitatively a more generic scaling
behavior. There is, however, an important point still to be considered: the effect of
long-range dipolar interactions. As we discussed in section \ref{sec:magnetostatic} the
effect is twofold, interactions between spins in the bulk and surface charges at the
boundary of the sample. The bulk interaction has not been studied deeply in the
framework of spin models, apart from some numerical simulations in two dimensions
\cite{MAG-99,MAG-00}. From simple power counting considerations, we would expect that
dipolar forces are relevant in the renormalization group sense, thus modifying the
numerical values of the critical exponents. Therefore, unless one can neglect dipolar
forces (this could be the case in hard rare earth alloys with strong anisotropies), the
numerical results obtained in the RFIM would not apply to real materials.

\begin{figure}
\centerline{\psfig{file=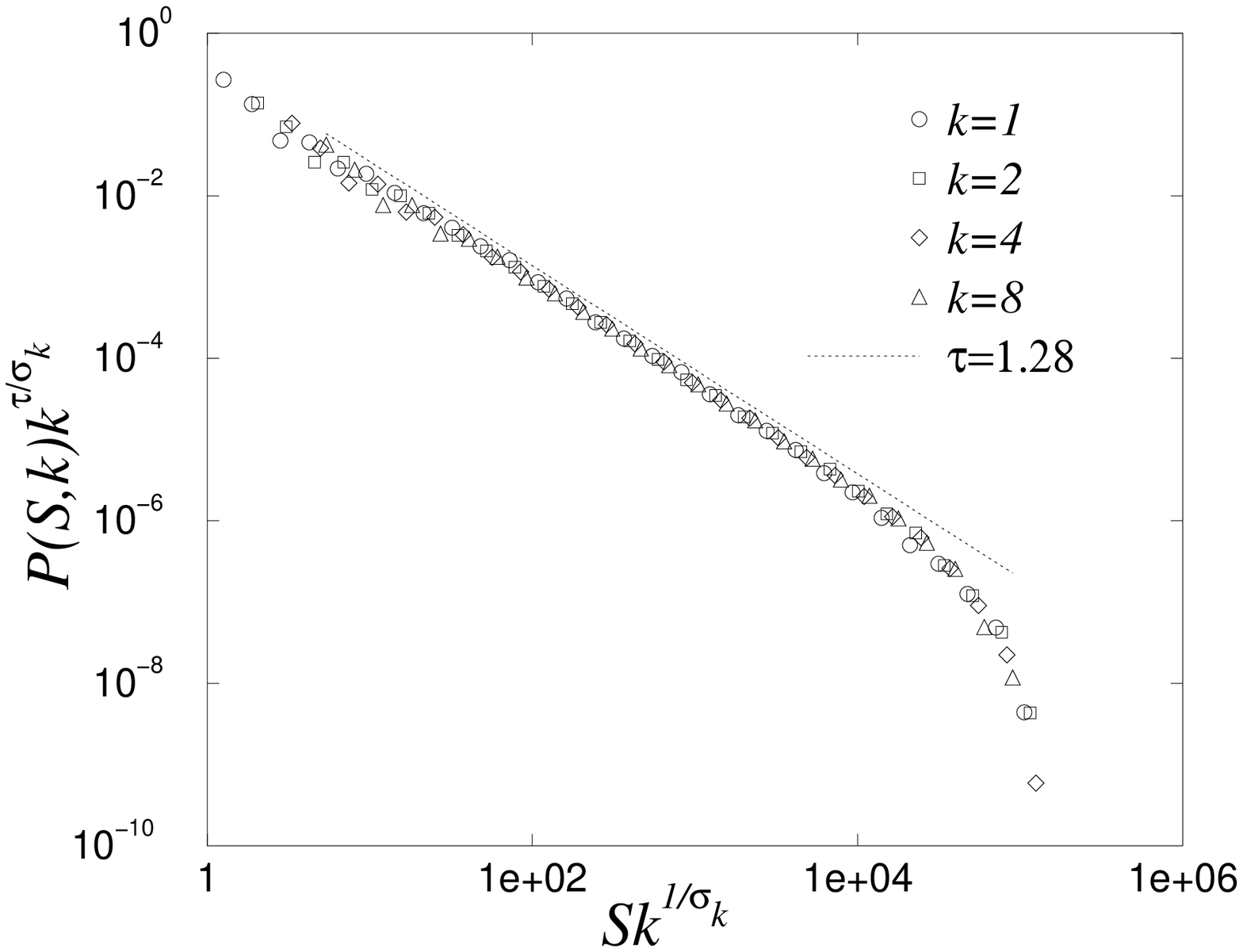,width=12cm,clip=!}} \caption{Data collapse of avalanche
size distribution $P(S) \sim S^{-\tau} f(S/S_0)$ simulated in the RFIM for $R<R_c$ in
presence of a demagnetizing field, using different values of the demagnetizing factor
$k$. The critical exponent $\tau$ is given by 1.28, while $S_0\sim k^{-1/\sigma_k}$
with $1/\sigma_k \sim 0.7$ [From \protect\cite{DUR-02a}, Fig. 1, pg. 233].}
\label{fig:psk_rfim}
\end{figure}

Surface effects have an even more drastic effect on the critical behavior discussed
above. The presence of a demagnetizing field generates an effective correlation length,
destroying scale invariance even at the critical point $R=R_c$. Nevertheless, this
effect is small, since the demagnetizing factor tends to vanish for very large system
sizes.  Simulations of the RFIM in presence a demagnetizing field show remarkable
features in the low disorder phase, where otherwise one expects the magnetization to
change abruptly with a large jump. The demagnetizing field hinders the reversal process
which splits in a series of avalanches of different sizes \cite{DUR-02a,CAR-03}.  The
distribution of size is again a power law distribution and the cutoff is controlled by
the demagnetizing factor (see Fig.~\ref{fig:psk_rfim}) \cite{DUR-02a}.  It is important
to observe that this power law scaling is not related with the disorder induced
critical point, since this behavior is observed generically for $R<R_c$, but can be
explained considering the dynamics of a domain wall.  Indeed for $R<R_c$, at the
coercive field a domain is nucleated and starts to grow. When the domain is large
enough the demagnetizing field reduces the effective field and eventually the growth
stops. At this moment the spins that will flip when the external field is increased are
likely to be those at the domain boundary.  To understand this process in detailed one
should thus analyze domain wall dynamics, as will be discussed in the next section.

\begin{figure}
\centerline{\psfig{file=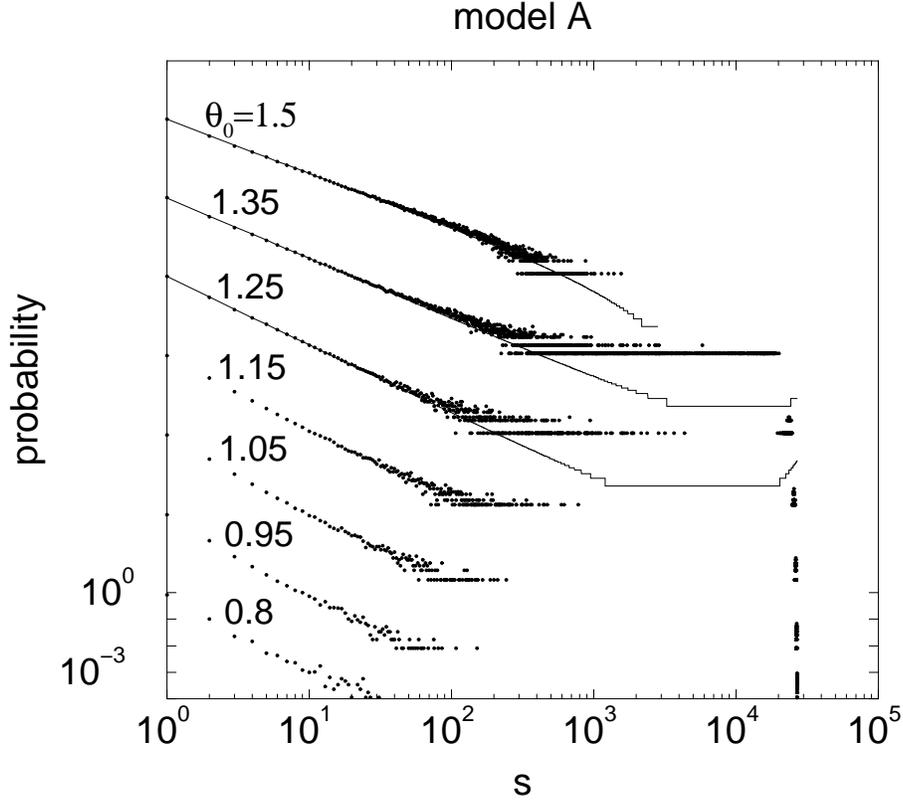,width=12cm,clip=!}} \caption{The integrated
avalanche distribution in the random infinite anisotropy model in $d=3$ simulated in a
lattice of $30^3$ spins. The different curves correspond to different values of
$\theta_0$, corresponding to the average angle between the field axis and the
anisotropy axis [From Ref.~\protect\cite{VIV-01}, Fig.~8, pg. 134431-6].}
\label{fig:viv01}
\end{figure}

\subsection{Domain wall models}

In order to minimize the exchange energy, a ferromagnetic material at sufficiently low
temperature and zero field would display a uniform magnetization, oriented along an
easy axis to lower the magnetocrystalline anisotropy energy. In most cases, however,
this arrangement is hindered by a very high cost in magnetostatic energy, due to the
discontinuities of the magnetization at the sample edges. Thus it is energetically
favorable to create domains, which in the simplest situation span the sample from end
to end. In this condition, the magnetization process occurs prevalently by domain wall
motion, which can then be studied to recover the statistical properties of the
Barkhausen noise \cite{URB-95a,NAR-96,CIZ-97,ZAP-98,BAH-99,QUE-01}.

\subsubsection{Energetics of a domain wall}

\begin{figure}
\centerline{\psfig{file=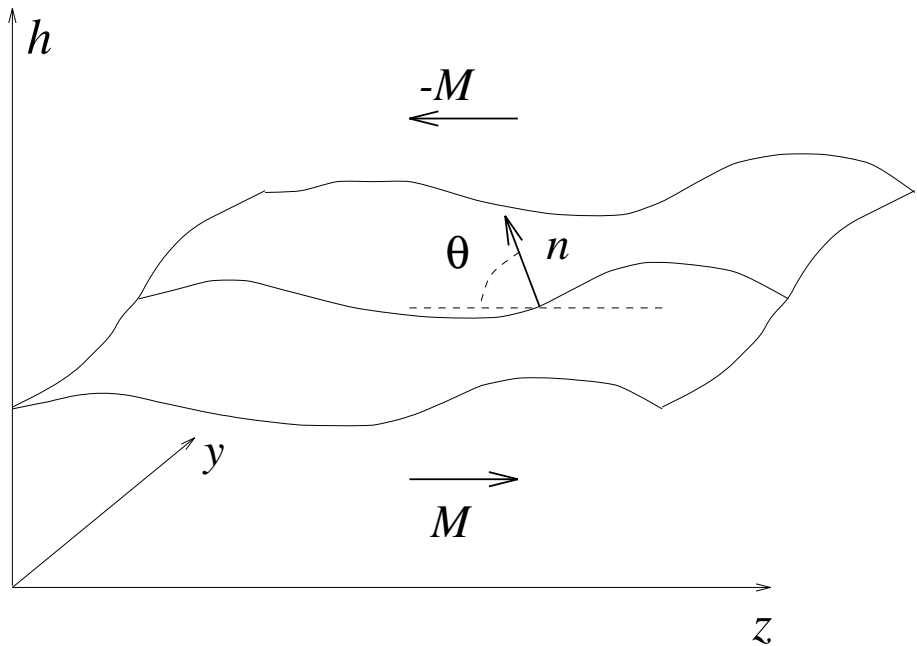,width=12cm,clip=!}} \caption{A domain wall separating
two regions of opposite magnetization. The discontinuities of the normal component of
the magnetization across the domain wall produce magnetic charges.} \label{fig:tela}
\end{figure}

We consider a flexible 180$^\circ$ domain wall separating two regions with opposite
magnetization directed along the $z$ axis.  In absence of disorder the domain wall
would be flat, but in general the wall will bend to accommodate pinning forces. If the
surface has no overhangs, we can describe the position of the domain wall by a function
$h(\vec{r},t)$ of space and time (see Fig.~\ref{fig:tela}). The energy associated with
such a deformation field can be computed expanding the micromagnetic free energy
discussed in Sec.~\ref{sec:genprop}, and we can thus split the energy into the sum of
different contributions due to exchange and magnetocrystalline interactions,
magnetostatic and dipolar fields, and disorder \cite{ZAP-98}.

The contribution from the magnetostatic energy, if the external field $\vec{H}$ is
applied along the $z$ axis is given by
\begin{equation}
E_m = -2 \mu_0 (H+H_{dm}) M_s\int d^2 r\; h(\vec{r},t),
\end{equation}
where we have included the demagnetizing field $H_{dm}$ which is in general a
complicated function of $h(\vec{r},t)$, depending also from the sample shape. In the
simplest approximation, however, the intensity of the demagnetizing field will be
proportional to the total magnetization. Considering the field constant through the
sample, it can thus be written as
\begin{equation}
H_{dm}= -\frac{kM_s}{V}\int d^2r~h(\vec{r},t)
\end{equation}
where the demagnetizing factor $k$ takes into account the geometry of the
domain structure and the shape of the sample and $V$ is a volume factor.

The interplay between magnetocrystalline anisotropy and exchange interactions is
responsible for the microscopic structure of the domain wall. While a very sharp change
of the spin orientation has a high cost in exchange energy, a very smooth rotation of
the spins between two domains is prevented by the magnetocrystalline anisotropy. The
balance between these two contributions determines the width of the domain wall and its
surface energy. The total energy due to these contributions is proportional to the area
of the domain wall
\begin{equation}
E_{dw} = \gamma_w \int d^2r \sqrt{1+|\nabla h(\vec{r},t)|^2},
\end{equation}
where $\gamma_w\simeq 2\sqrt{A K_0}$ is the domain wall surface energy. Expanding this
term for small gradients we obtain
\begin{equation}
E_{dw} = \gamma_w S_{dw}+ \frac{\gamma_w}{2} \int d^2r|\nabla h(\vec{r},t)|^2,
\label{eq:dw}
\end{equation}
where $S_{dw}$ is the area of the undeformed wall. Thus the domain wall energy energy
represents an elastic interaction that tends to keep the domain wall flat.

An additional elastic interaction is due to dipolar forces, since the local distortions
of the domain wall are associated to discontinuities in the normal component of the
magnetization. As discussed in Sec.~\ref{sec:magnetostatic}, this effect can be treated
introducing magnetic charges on the domain wall surface (see Eq.~\ref{eq:surfq}). The
surface charge density is zero when the magnetization is parallel to the wall and for
small distortion it can be expressed as
\begin{equation}
\sigma(\vec{r}) = 2M_{s} \cos\theta \simeq 2M_s\frac{\partial h(\vec{r},t)}{\partial x}
\end{equation}
where $\theta$ is the local angle between the vector normal to the surface and the
magnetization. The energy associated with this distribution of charges is given by
\begin{equation}
E_d = \int d^2r
 d^2r^{\prime}\frac{\mu_0M_s^2}{2\pi|\vec{r}-\vec{r}^{\;\prime}|}
 \frac{\partial h(\vec{r},t)}{\partial z} \frac{\partial
 h(\vec{r}^{\;\prime},t)} {\partial z^{\prime}}
\label{eq:edj2}
\end{equation}
which integrating by part can also be written as
\begin{equation}
E_d = \int d^2r d^2r^{\prime}
 K(\vec{r}-\vec{r}^{\;\prime}) (h(\vec{r},t)-h(\vec{r}',t))^2
\label{eq:edj}
\end{equation}
where the non local-kernel is given by
\begin{equation}
K(\vec{r}-\vec{r}^{\;\prime})=
\frac{\mu_0M_s^2}{2\pi|\vec{r}-\vec{r}^{\;\prime}|^3}\left(1-
\frac{3(z-z^\prime)^2}{|\vec{r}-\vec{r}^{\;\prime}|^2}\right). \label{eq:ker}
\end{equation}
The interaction is long range and anisotropic, as can be seen by considering the
Fourier transform
\begin{equation}
K(p,q) = \frac{\mu_0 M_s^2}{4\pi^2} \frac{p^2}{\sqrt{p^2+q^2}},
\end{equation}
where $p$ and $q$ are the two components of the Fourier vector along $z$ and $y$. In
the preceding derivation we have implicitly assumed an infinitely strong anisotropy, so
that the magnetization never deviates from the easy axis. In practice, however, the
magnetization will rotate slightly from the easy axis, producing additional volume
charges as discussed in section~\ref{sec:magnetostatic}. This effect leads to a slight
modification of the interaction kernel in Eq.~(\ref{eq:edj})
\begin{equation}
\tilde{K}(p,q)\sim\frac{1}{\sqrt{Q}}\frac{p^2}{\sqrt{p^2+Q q^2}}, \label{eq:kernel_an}
\end{equation}
where $Q\equiv 1 + 2\mu_0 M_s^2/K$ is a material dependent constant \cite{NEE-46}.

The disorder present in the material in the form of non-magnetic impurities, lattice
dislocations or residual stresses is responsible for the deformation and the pinning of
the domain wall.  In general, disorder can be modelled introducing a random potential
$V(\vec{r},h)$, whose derivative gives the local pinning field $\eta(\vec{r},h)$ acting
on the surface. In the particular case of point-like defects, such as non-magnetic
impurities, the random force is given by
\begin{equation} \eta(\vec{r},h)=-\sum_i f_p(\vec{r}-\vec{r}_i,h-h_i)
\end{equation}
where($r_i,h_i$) are the coordinates of the pinning centers and $f_p(x)$ is the
individual pinning force, which typically has a range comparable with the domain wall
width $\delta_w \simeq \sqrt{K/A}$.  After coarse-graining at a scale larger than the
typical distance between the pinning centers, this disorder can be replaced by a
Gaussian random noise with correlations
\begin{equation}
\langle\eta(\vec{r},h)\eta(\vec{r}^{\;\prime},h^\prime)\rangle
=\delta^2(\vec{r}-\vec{r}^{\;\prime})R(h-h')
\end{equation}
where $R(x)$ decays very rapidly for large values of the argument. It has been shown
that the particular form of $R(x)$ (i.e. due to random-bond or random-field types of
disorder) has not a relevant effect on the scaling laws associated the domain wall
dynamics \cite{NAR-93}.

Another possible source of pinning is due to local variations of the domain wall energy
$\gamma_w$ \cite{NEE-46}, due for instance to random anisotropies. Thus the domain wall
energy becomes a function of position $\gamma(\vec{r},h)$ which we can expand about its
average $\gamma_w+\eta(\vec{r},h)$. Introducing this expression in Eq.~\ref{eq:dw}, we
obtain to lowest order an additional random term $\eta(\vec{r},h)$, whose distribution
and correlations can be directly related to the random anisotropies.

\subsubsection{Domain wall dynamics and depinning transition}
\label{sec:depin} In most cases, the motion of the domain wall is strongly overdamped,
since eddy currents cancel inertial effects. The equation of motion for the wall is
thus given by \beq \Gamma\frac{\partial h(\vec{r},t)}{\partial t}= -\frac{\delta
E(\{h(\vec{r},t)\})}{ \delta h(\vec{r},t) }, \label{eq:v=f} \eeq where $
E(\{h(\vec{r},t)\})$ is the total energy functional, derived in the previous section,
and $\Gamma$ is an effective viscosity. We have neglected here thermal effects, since
experiments suggest that these are not relevant for the Barkhausen effect in bulk three
dimensional samples \cite{URB-95}. This could be different in thin films \cite{LEM-98}
where thermal activated motion can be described by adding an extra noise term to the
equation of motion.

Collecting all the energetic contributions, we obtain the equation of motion for the
domain wall \cite{ZAP-98}. In order to avoid a cumbersome notation all the unnecessary
factors can be absorbed in the definitions of the parameters. The equation then becomes
\beq \frac{\partial h(\vec{r},t)}{\partial t}= H-\bar{k}\tilde{h}+ \gamma_w
\nabla^2h(\vec{r},t) + \nonumber\eeq\beq \int d^2r^\prime
K(\vec{r}-\vec{r}^{\;\prime})(h(\vec{r}^{\;\prime})-h(\vec{r})) +\eta(\vec{r},h),
\label{eq:tot} \eeq where the dipolar kernel $K$ is reported Eq.~(\ref{eq:ker}), the
effective demagnetizing factor in given by $\bar{k}\equiv 4\mu_0 kM_s^2/V$
and $\tilde{h} \equiv \int d^2r^\prime
h(\vec{r}^{\;\prime},t)$. Owing to the fact that the demagnetizing field term is just
an approximation, the dependence of $k$ on the sample shape and size can be quite
complex. Variants of Eq.~(\ref{eq:ker}) have been extensively studied in the past
\cite{NEE-46,HIL-76a,HIL-77,ENO-94,ENO-94b,ELM-95,ENO-97,URB-95a,NAR-96,CIZ-97,ZAP-98,BAH-99,QUE-01}.

When the demagnetizing factor $k$ is negligible, as for instance in a frame geometry,
Eq.~\ref{eq:tot} displays a depinning transition as a function of the applied field $H$
and the domain wall moves only if the applied field overcomes a critical field $H_c$.
For fields $H>H_c$ the domain wall moves with an average velocity $v$ that scales as
\beq v \sim (H-H_c)^\beta \theta(H_c-H), \label{eq:v-H} \eeq where $\theta(x)$ is the
step function. The critical behavior associated to the depinning transition has been
studied using renormalization group methods \cite{NAT-92,NAR-93,LES-97,CHA-01}, which
show that at large length scales the critical exponents take mean-field values
\cite{ZAP-98,CIZ-97}. This result is due to the linear dependence on the momentum of
the interaction kernel (Eq.~\ref{eq:ker}) in Fourier space \cite{ERT-94,CHA-01}. In
general, if we consider an interface whose interaction kernel in momentum space scales
as $K(q)=A_K|q|^\mu$, the upper critical dimension is given by $d_c=2\mu$ and the
values of the exponents depend on $\mu$. Notice that here we define $d$ as the internal
dimension of the domain wall. Hence in the present analysis, we have a two dimensional
interface ($d=2$) moving in a three dimensional medium.  While in general we would
expect mean-field behavior, there are situations in which the dipolar coupling could be
neglected with respect to the domain wall tension $\gamma_w$ where one effectively
could observe the critical behavior associated with $\mu=2$.

\begin{figure}
   \begin{center}
   \includegraphics[width=12cm]{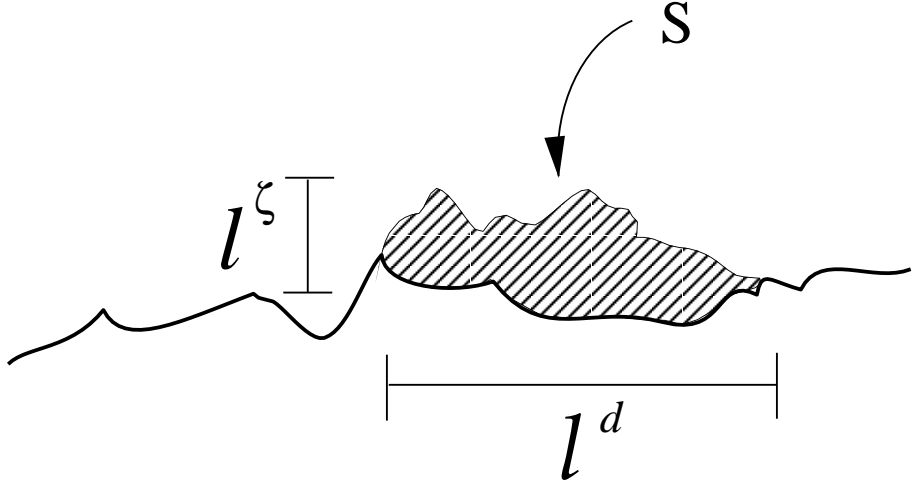}
   \end{center}
   \caption{The domain wall moves
between two pinned configuration in an avalanche of size $S \sim l^{d+\zeta}$}
\label{fig:avalanche}
   \end{figure}

In general, for interfaces close to the depinning transition, the response to small
variations of the applied field occurs by avalanches whose sizes $S$ are distributed as
\beq P(S)\sim S^{-\tau}f(S/S_0), \label{eq:ps} \eeq where the cutoff scales as $S_0
\sim (H-H_c)^{-1/\sigma}$ and is related to the correlation length $\xi$ by \beq S_0
\sim \xi^{d+\zeta}, \label{eq:s0xi} \eeq where $\zeta$ is the roughness exponent
(Fig.~\ref{fig:avalanche}). The correlation length diverges at the depinning transition
as \beq \xi\sim(H-H_c)^{-\nu}, \label{eq:nu} \eeq which implies \beq
\frac{1}{\sigma}=\nu (d+\zeta). \eeq The average avalanche size also diverges at the
transition \beq \langle S \rangle \sim (H-H_c)^{-\gamma}, \label{eq:s_av} \eeq where
$\gamma$ is related to $\tau$ and $\sigma$ by \beq \gamma=\frac{(2-\tau)}{\sigma},
\label{eq:gamma} \eeq and to $\nu$ and $z$ by \cite{NAR-93} \beq \gamma=(1+\nu\zeta),
\eeq thus implying \beq \tau=2-\frac{1+\nu\zeta}{\nu(d+\zeta)}. \eeq

The other exponent relevant for the Barkhausen effect describes the distribution of
avalanche durations \beq P(T) \sim T^{-\alpha}g(T/T_0), \eeq where the cutoff diverges
at the transition as $T_0 \sim (H-H_c)^{-1/\Delta}$. From Eq.~(\ref{eq:nu}) and the
relation $T_0 \sim \xi^z$ we obtain $\Delta = 1/z\nu$ and \beq \alpha=1+\frac{\nu
d-1}{z\nu}. \label{eq:alpha} \eeq Finally, an additional symmetry of the equation of
motion can be used to reduce the number of independent exponents through the relation
\beq \nu(\mu-\zeta)=1 \eeq

The renormalization group analysis allows for a determination of the critical exponents
in the framework of $\epsilon=d_c-d$ expansion. As discussed above, for $\mu=1$ we
expect mean-field results $\tau = 3/2$ and $\alpha =2$, while for $\mu=2$ to first
order in $\epsilon$ one obtains $\tau = 5/4$ and $\alpha =11/7$. Recently a two loop
expansion has been carried out, allowing for an estimate to order O($\epsilon^2$) of
the exponents ($\tau=1.24$ and $\alpha=1.51$) \cite{CHA-01}. Finally, numerical
simulations yield $\tau=1.27$ and $\alpha = 1.5$ for $\mu=2$.

The discussion above applies strictly to the case $k=0$ and scaling requires that $H$
is close to $H_c$. If one would instead ramp the field slowly, sampling the avalanche
distribution over all the values of the field, the result will be different.  We need
to integrate Eq.~\ref{eq:ps} over $H$ \cite{ZAP-98} \beq p_{int}(S)=\int^{H_c} dH
S^{-\tau} f(S(H-H_c)^{1/\sigma}) \sim S^{-\tau_{int}}, \eeq with
$\tau_{int}=\tau+\sigma$. Using the values of $\tau$ and $\sigma$ reported above, we
obtain $\tau_{int}=1.72$ for $\mu=2$ and $\tau_{int}=2$ for $\mu=1$. A similar
discussion can be repeated for avalanche duration, yielding
$\alpha_{int}=\alpha+\Delta$. The numerical result is $\alpha_{int}=2.3$ and for
$\mu=2$ and $\alpha_{int}=3$ for $\mu=1$.

\begin{figure}[h]
   \begin{center}
   \includegraphics[width=12cm]{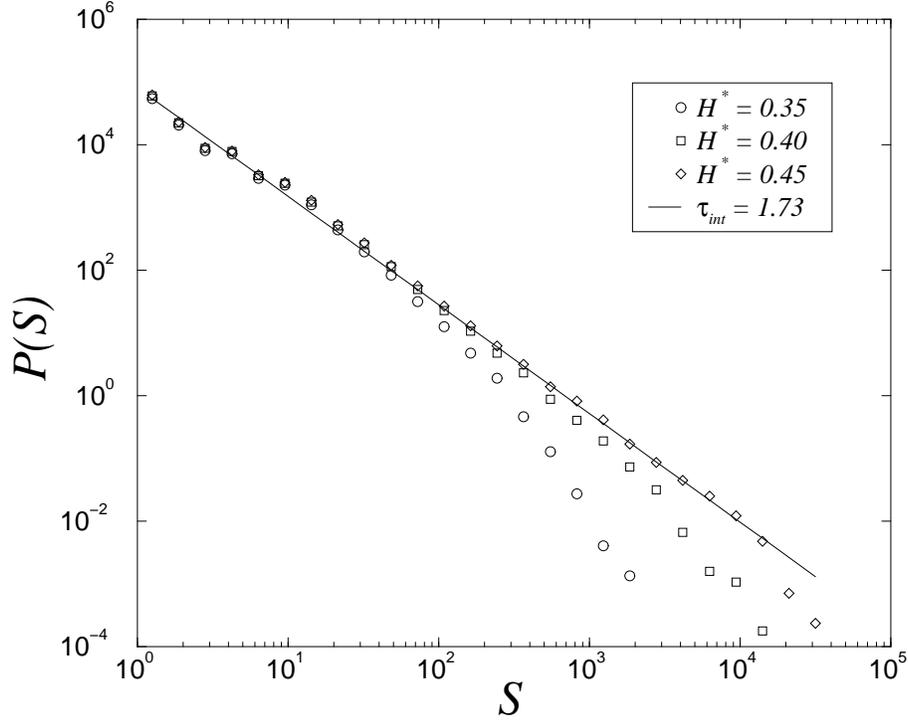}
   \end{center}
\caption{The distribution of avalanche sizes in the short-range domain wall model at
$k=0$ when the field is swept from $H=0$ to $H=H^*<H_c$. For $H^* \to H_c$, we obtain
$\tau_{int}=\tau+\sigma$.} \label{fig:psh}
   \end{figure}

In a typical Barkhausen experiment, however, the situation changes since $k>0$ and $H$
is normally increased at constant rate. The effective field acting on the domain wall
is given by $H_{eff}=ct - k\bar{h}$. Thus, according to Eq.~\ref{eq:v-H}, when
$H_{eff}<H_c$ the domain wall is pinned and the effective field increases. As soon as
$H_{eff}>H_c$ the domain wall acquires a finite velocity and $\bar{h}$ increases. As a
consequence, the effective field will be reduced until the domain wall stops. This
process keeps the domain wall close to depinning transition, but criticality is only
reached in the limit $c\to 0$ and $k\to 0$. A similar mechanism occurs in models of
self-organized criticality \cite{DIC-00}.

  \begin{figure}[h]
   \begin{center}
   \includegraphics[width=12cm]{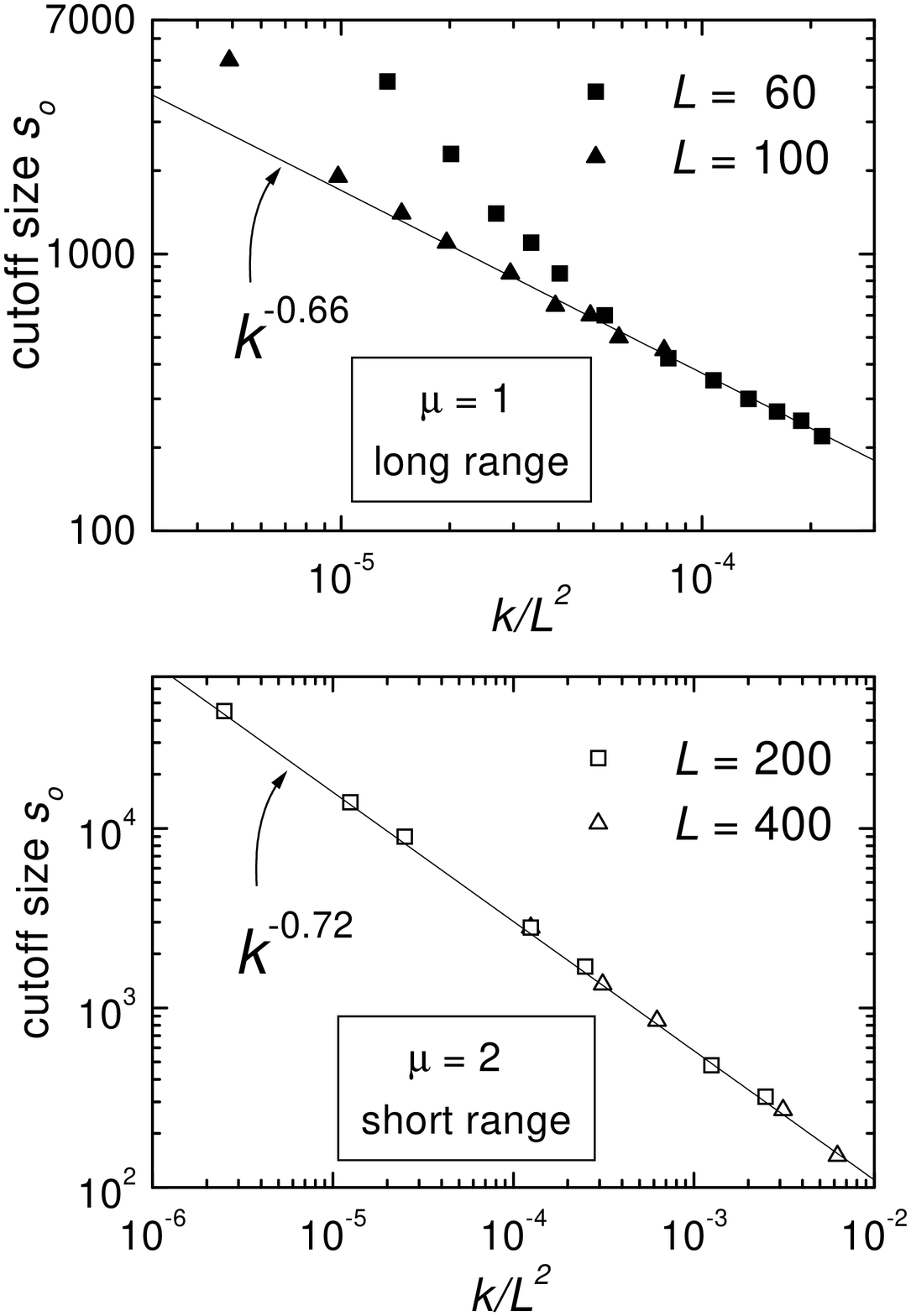}
   \end{center}
\caption{The cutoff $S_0$ of the avalanche size obtained from numerical simulations for
different demagnetizing factors. The two universality classes correspond to a kernel in
the equation of the $|q|^\mu$ type in Fourier space, with $\mu=1$ (long range dipolar
interactions - up), and $\mu=2$ (short range elastic interactions - down).}
\label{fig:cut_sim}
   \end{figure}

The demagnetizing factor $k$ represents a restoring force hindering the propagation of
Barkhausen avalanches. Thus when $k>0$, we expect to find a finite avalanche size
cutoff $S_0$. The domain wall can not propagate over distances larger than $\xi$,
defined as the length at which the interaction term $A_K|q|^\mu$ is equal to the
restoring force \beq A_K h\xi^{-\mu} \sim k\xi^d h \eeq which implies \beq \xi \sim
(k/A_K)^{-\nu_k} ~~~~~~\nu_k = 1/(\mu+d). \eeq The avalanche size and duration
distributions cutoff $S_0$ and $T_0$ can be obtained using the scaling relations
reported above \beq S_0 \sim D(k/A_k)^{-1/\sigma_k},~~~~~~1/\sigma_k =
(d+\zeta)/(\mu+d) \label{eq:s0} \eeq and similarly \beq
T_0\sim(k/A_k)^{-\Delta_k},~~~~~~~\Delta_k = z/(\mu+d), \label{eq:T0} \eeq where
$D\equiv\sqrt{\langle\eta^2\rangle}$ denotes the typical fluctuation of the disorder.
Inserting in Eqs.~(\ref{eq:s0}-\ref{eq:T0}) the renormalization group results to first
order in $\epsilon$ \cite{NAT-92,NAR-93,LES-97} $\zeta = (2\mu-d)/3$ and
$z=\mu-(4\mu-2d)/9$, we obtain $1/\sigma_k = 2/3$ and
$\Delta_k=(\mu-(4\mu-2d)/9)/(\mu+d)$. Using the results obtained to order $\epsilon^2$
\cite{CHA-01}, one obtains instead for $\mu=2$, $1/\sigma_k=0.71$ and $\Delta_k=0.345$.
The numerical values obtained from simulations are in good agreement with this
prediction (see Fig.~\ref{fig:cut_sim}).

The mean-field theory, which provides a good qualitative description of the depinning
transition and describes quantitatively the data for $d=d_c$ (apart from logarithmic
corrections), is obtained discretizing the equation of motion, coupling all the sites
with the average domain wall position $\bar{h}$ \cite{FIS-85}. The dynamics of such an
infinite range model is described by \beq \frac{d h_i}{dt}=ct-k\bar{h} +J(\bar{h}-h_i)
+\eta_i(h), \label{eq:mf1} \eeq where $J$ is an effective coupling and $i=1,...N$.
Summing over $i$ both sides of Eq.~(\ref{eq:mf1}), one obtains an equation for the
total magnetization $m$ \beq \frac{dm}{dt}=\tilde ct-k m +\sum_{i=1}^{N} \eta_i(h).
\eeq This equation has the same form of the ABBM model (see Eq.~\ref{eq:abbm})
\cite{ALE-90}.

\begin{figure}[h]
\centerline{\psfig{file=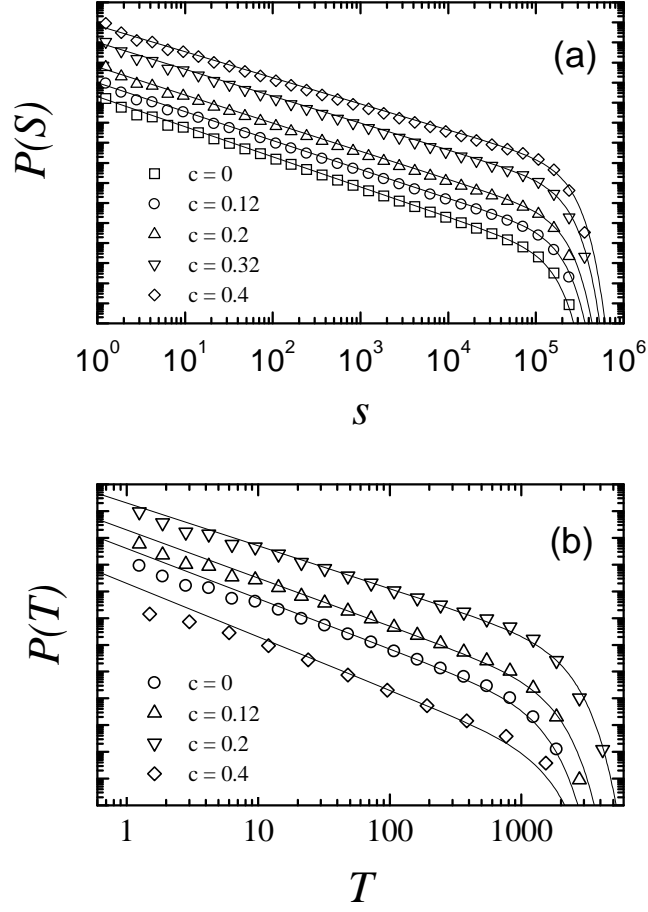,width=10cm,clip=!}} \caption{(a) The
distribution of avalanche sizes in the infinite-range model for different driving rates
for $N=32696$. The fit of the power-law part yields $\tau = 3/2-c/2$, with
$c=\tilde{c}/D$ and $1/D\simeq 1.2$. (b) The corresponding distribution of avalanche
durations. The power-law part is fit with an exponent $\alpha=2-c$.}
\label{fig:infrange}
\end{figure}

To make the similarity more stringent, we should interpret $\sum_i \eta_i$ as an
effective pinning $W(m)$, with Brownian correlations. When the domain wall moves
between two pinned configuration $W$ changes as \beq W(m^{\prime})-W(m) =
\sum_{i=1}^{n} \Delta\eta_i, \eeq where the sum is restricted to the $n$ sites that
have effectively moved (i.e. their disorder is changed). The total number of such sites
scales as $n \sim l^{d}$ and in mean-field theory is proportional to the avalanche size
$S = |m^{\prime}-m|$ (since $S \sim l^{d+\zeta}$ and $\zeta=0$). Assuming that the
$\Delta\eta_i$ are uncorrelated and have random signs, we obtain a Brownian effective
pinning field \beq \langle |W(m^{\prime})-W(m)|^2 \rangle = D|m^{\prime}-m|,
\label{eq:pf-abbm} \eeq where $D$ quantifies the fluctuation in $W$. Thus the Brownian
pinning field, introduced phenomenologically in the ABBM model, is an effective
description of the disorder resulting from the collective motion of a flexible domain
wall. As a consequence of this mapping, we expect that the infinite range model
displays the sane frequency dependence of the exponents as in the ABBM model
(Eq.~\ref{eq:exp-c}). Simulations confirm this claim as shown in
Fig.~\ref{fig:infrange}.

\section{Discussion of theoretical results}

\subsection{Avalanches distributions}
\label{sec:th-jump-distr}

The models discussed in Sec.~\ref{sec:theory} provide different result and theoretical
interpretation for the pulse distributions. While we have reported already the results
obtained for the various models, for the sake of clarity we summarize here again the
main conclusions. We have distinguished between two general classes of models: spin
models and domain walls model. Rigorously speaking the second class is contained in the
first, since a domain structure is typically present in spin systems, but the analysis
of domain wall dynamics would be needlessly complicated. It is useful instead to treat
separately the case in which domain wall propagation does not interfere with
nucleation.

The two classes of models yield different interpretation of the origin of scaling in
the Barkhausen pulse distributions. In spin models, scaling is ruled by the proximity
to a disorder induced critical point. The cutoff to the power law scaling is determined
by the distance from this critical point. The broadness of the critical region would
justify the wide occurrence of scaling in experiments, without any apparent parameter
tuning \cite{PER-95}. In domain wall models scaling is dictated by a different critical
point: the depinning transition. The interplay between slow driving ($c\to 0$) and
small demagnetization $k\to 0$ keeps the domain wall close to the depinning critical
point. The cutoff of the distributions is then controlled by the value of $k$, which in
strip geometries is typically small. A collection of the numerical estimates of the
Barkhausen noise distribution exponents for the RFIM and domain wall models is reported
in Table~\ref{tab:sim_exp}.

\begin{table}

\begin{tabular}{|c|c|c|c|c|c|c|c|c|}
\hline
  model  &    $\tau$     & $\alpha$ & $\sigma\nu z$  &   $1/\sigma$ & $1/\sigma_k$ & $\Delta_k$ & $\tau_{int}$     & $\alpha_{int}$
\\
\hline
   RFIM  $d=3$    &      1.60 $\pm$ 0.06      &   2.05 $\pm$ 0.12      &  0.57 $\pm$ 0.03  &   4.2 $\pm$ 0.3     &  - & - & 2.03 & 2.81
\\
\hline
    MF  (ABBM)   &      3/2      &   2        & 1/2  &  2 & 2 & 1 & 2 & 3
\\
\hline
    LRDW    &      1.5       &   2    & 0.5  & 2 & 0.65$\pm$0.05  &   0.3$\pm$0.1& 2 & 3\\
\hline
   SRDW     &     1.27$\pm$0.02       &   1.50 $\pm$0.05 & 0.57 $\pm$ 0.02 & 2.2 $\pm$ 0.1 & 0.72$\pm$0.03  &
0.39$\pm$0.05& 1.72 & 2.3 \\ \hline

\end{tabular}

\caption{The values of critical exponents estimated from simulations of the RFIM
\protect\cite{PER-99} and domain wall models with short-range (SRDW) and long-range
(LRDW) interactions \protect\cite{DUR-00,DUR-00a}. Mean-field (MF) results are the same
for both classes of models and coincide with ABBM exponents. The exponent $\sigma$
describes the scaling of the avalanche size cutoff with the reduced control parameter,
$R-R_c$ in the RFIM and $H-H_c$ in domain wall models. For $k>0$ and $R<R_c$ the RFIM
reduces to the SRDW model and those exponents apply.} \label{tab:sim_exp}
\end{table}

Depending on the problem at hand the appropriate model for the Barkhausen noise may
vary. In soft magnetic ribbons the domain structure is typically controlled by dipolar
interactions and the sample shape, imposing a parallel set of domain walls spanning the
sample from end to end. In this condition, the depinning scenario should apply as it is
testified by the experimental findings. The avalanche exponents are in good agreement
with the prediction of the depinning model and the cutoff is controlled by the
demagnetizing factor.

In particular, experiments on polycrystalline SiFe are well described by the ABBM model
or, equivalently, by a long-range depinning model. The exponents measured in amorphous
alloys are instead well fitted by the short range interface model. This is due to the
fact that in amorphous alloys the average anisotropy vanishes and thus dipolar
interactions have no effect on the domain wall dynamics (i.e. $Q$ in
Eq.~\ref{eq:kernel_an} is very large).  On the contrary, in polycrystalline materials,
every crystal is the source of a relevant anisotropy and long range effects occur. This
is confirmed by the fact  that long-range exponents are found when small crystallites
are induced in an amorphous matrix by proper thermal annealing \cite{DUR-00}.

The scaling exponents measured in experiments on NiFe \cite{LIE-72} wires and Vitrovac
\cite{SPA-96} could be also explained in this framework if one assumes that the
Barkhausen signal was not stationary. This could occur if the field is not swept up to
saturation so that the effective field varies and it is not kept around $H_c$ as in
most cases. Then one would obtain integrated exponents $\tau_{int}$, $\alpha_{int}$
instead of $\tau$ and $\alpha$. The striking similarity between the experimentally
measured exponents and the theoretical ones (i.e. $\tau_{int}=1.72$ and
$\alpha_{int}=2.3$) supports this conclusion.

Spin models could instead apply to cases in which the domain structure is more
intricate and the effect of dipolar forces negligible. A typical example would be an
hard magnetic materials, with strong local anisotropies.  Systematic investigation of
the Barkhausen effect in hard magnetic materials has just been started and a
quantitative comparison between experiments and theory is still not possible.

\begin{figure}
\centerline{\psfig{file=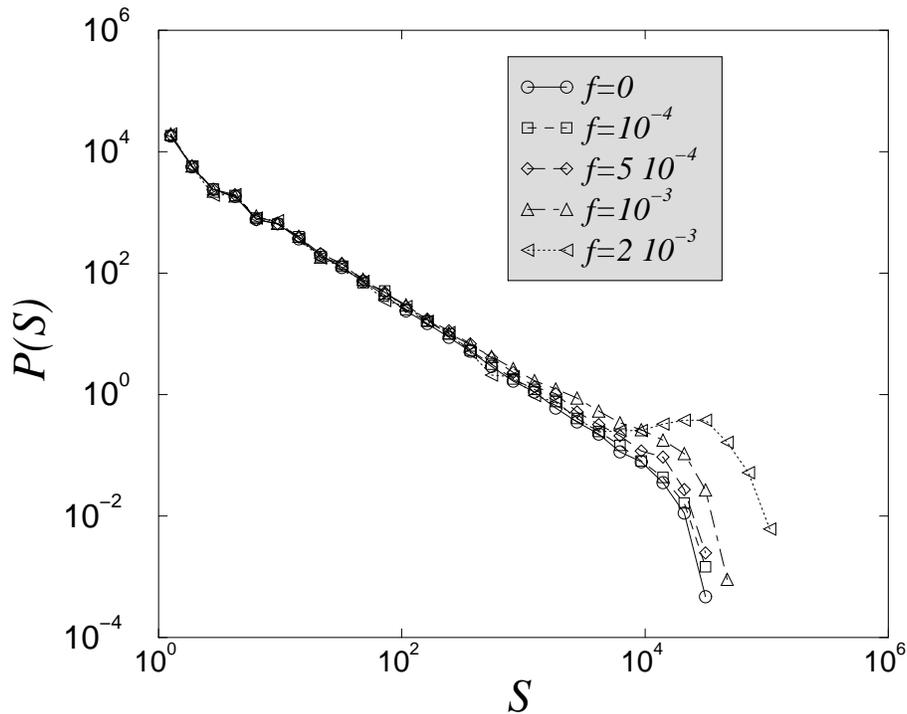,width=12cm,clip=!}} \caption{The distribution of
Barkhausen jump sizes from a short-range domain wall model for different driving field
frequencies. While the scaling regime remains essentially unchanged a peak occurs
around the cutoff. (See also Ref.~\protect\cite{QUE-01})} \label{fig:que}
\end{figure}

An important problem to address, regardless of the theoretical framework, is the effect
of the field rate on the Barkhausen noise statistics. As we discussed in section
\ref{sec:abbm} the ABBM predicts a linear dependence from the exponents $\tau$ and
$\alpha$ from the field rate, encoded in the parameter $c$. This result applies as well
for the mean-field domain wall model (see Fig.~\ref{fig:infrange}) and agrees well with
experiments on FeSi alloys. Simulations of the short-range domain wall model show that
the scaling exponents do not change continuously as in mean-field, but a peak appears
for large values of the jump size and grows with the rate \cite{QUE-01} (see
Fig.~\ref{fig:que}). While one could fit the result for higher driving rate with a
different exponents, such a fit does not appear very reliable and a constant exponent
coexisting with a growing peak seems a more appropriate approximation. A similar result
is found in amorphous soft magnetic alloys, as shown in Fig.~\ref{fig:pds-rate-dep}.

The effect of the driving rate on the avalanche distributions was analyzed in more
general terms in Ref.~\cite{WHI-03}.  A set of results can be derived under some basic
assumptions:\\ (i) The system is close to a critical point, so that in the adiabatic
limit ($c\to 0$) the size and duration distributions are a power law. (ii) The
avalanche signal is stationary under the time window considered. (iii) In the adiabatic
limit, the average number of avalanche nucleation events per unit field increase is a
smooth function of the driving field. (iv) For low $c$ the avalanche dynamics is
independent of $c$.  (v) The field increases between avalanche nucleations is
independent of the avalanche sizes. (vi) The avalanche sizes are uncorrelated in time.
(vii) Avalanches are nucleated randomly in space.

All these assumption are reasonable both for the domain wall models and for spin
models. Notice, however, that assumption (ii) restricts the derivation to a stationary
signal and thus does not apply to distributions measured integrating along the entire
hysteresis loop.

Under the assumptions above, one can show that the effect of the driving rate depends
on the value taken by the exponent $\alpha$ in the duration distribution for $c\to 0$.
For $\alpha>2$ the distributions are unaffected by the driving rate, while for
$\alpha<2$ a peak appears at large sizes and durations. The interesting case is
$\alpha=2$, which corresponds to a linear dependence of the exponents from the driving
rate $c$, recovering the ABBM result in Eq.~\ref{eq:exp-c}. This general derivation
provides an explanation of the results of experiments and numerical simulations,
predicting the conditions under which drive dependent exponents should be observed.

\subsection{The power spectrum \label{sec:th-ps}}

Explaining the power spectrum of the Barkhausen noise has been one of the main
objectives of the past theoretical activity. In earlier approaches the power spectrum
was considered as the result of a superposition of elementary independent events (see
for instance \cite{MAZ-62b,ARQ-68}) without a clear relation to the microscopic
magnetization processes. A step forward in this direction was made in the framework of
the ABBM model \cite{ALE-90,ALE-90a}.

>From Eq.~\ref{eq:abbm} one can derive a Fokker-Plank equation describing the evolution
of the probability distribution $P(v,t|v_0)$ to have a velocity $v$ at time $t$ when
the velocity at $t=0$ is $v_0$ \cite{ALE-90}
\begin{equation}
\frac{\partial P}{\partial t}=\frac{\partial }{\partial v}\left(
(kv-c)P+D\frac{\partial vP}{\partial v}\right). \label{eq:fokpl_abbm}
\end{equation}
Notice that, as in section \ref{sec:abbm}, we have rescaled time setting a damping
constant equal to one. The velocity correlation function is defined as
\begin{equation}
G(t)\equiv \langle (v(t)-c/k)(v(0)-c/k)\rangle = \int dv dv_0 (v-c/k) (v_0-c/k)
P(v,t|v_0)P(v_0), \label{eq:gt_abbm}
\end{equation}
where $P(v_0)$ is given by Eq.~\ref{eq:pv-abbm}. Deriving Eq.~\ref{eq:gt_abbm} and
using Eq.~\ref{eq:fokpl_abbm}, we obtain an evolution equation for $G(t)$
\begin{equation}
\frac{dG}{dt}= -k G,
\end{equation}
which can be easily solved yielding
\begin{equation}
G(t)=c/k \exp -kt.
\end{equation}
The power spectrum has thus a Lorenzian shape
\begin{equation}
F(\omega)= \frac{2c}{\omega^2+k^2}.
\end{equation}
This simple result contains several interesting predictions: (i) at large frequencies
the spectrum decays as $\omega^{-2}$. (ii) The scaling is cut off below a frequency
$\omega_0 =1/k$. (iii) The amplitude of the spectrum increases with the field driving
rate $c$. These features describe qualitatively the shape of the experimentally
measured power spectrum, but not quantitatively. For instance, the tail of the spectrum
often decays with an exponent different than $2$ and the cutoff frequency depends on
$k$ but not simply as $1/k$.

\begin{figure}
\centerline{\psfig{file=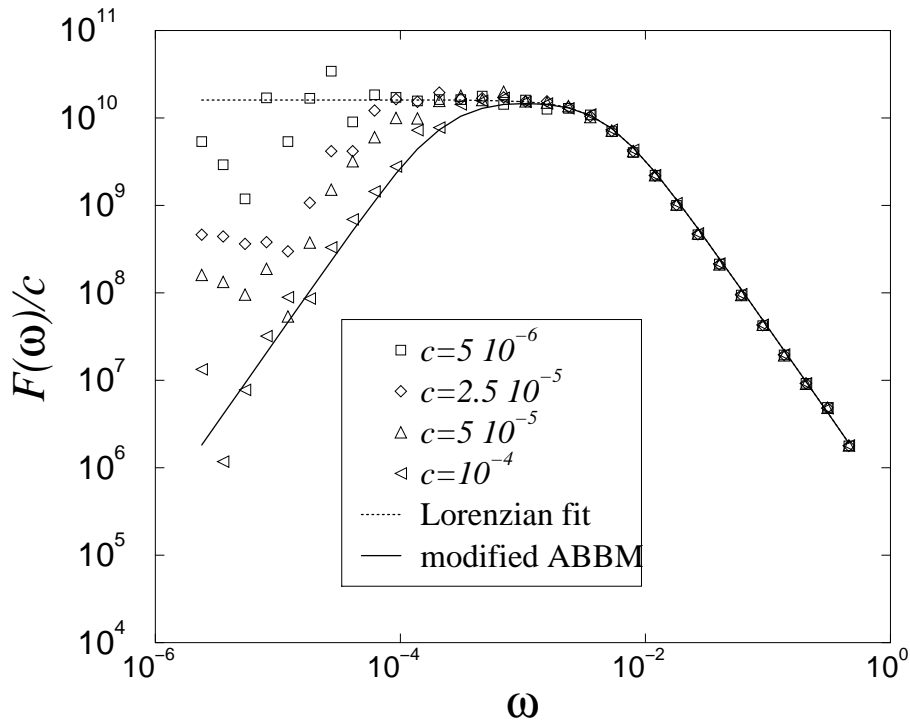,width=12cm,clip=!}} \caption{The power spectrum
measured in the infinite-range domain wall model for different field rates $c$. The
tail is well fitted by a Lorenzian function with amplitude proportional to $c$. The
entire spectrum is in perfect agreement with the ABBM result.} \label{fig:psdw-long}
\end{figure}

To reproduce the low frequency increase of the power spectrum, the ABBM model can be
modified introducing a correlation length $\xi^*$ in the pinning field \cite{ALE-90}.
This gives rise to a power spectrum of the type
\begin{equation}
F(\omega)\propto\frac{\omega^2}{(\omega^2+k^2)(\omega^2+\tau_c^{-2})},
\end{equation}
where $\tau_c \propto \xi^*$. The ABBM predictions for the power spectrum can be
compared with numerical simulations of the infinite-range domain wall model
(Eq.~\ref{eq:mf1}) for different values of $c$. The result in Fig.~\ref{fig:psdw-long}
are, as expected in good agreement with the theory.

Several attempts have been made in the past to link through scaling relations the power
spectrum exponent to the scaling exponents describing the avalanche distributions
\cite{LIE-72,DAH-96,NAR-96,SPA-96,ZAP-98}.  The comparison between these results and
experiments is, in most cases, not satisfactory. In any case, a simple exponent would
just describe the tail of the power spectrum, while a complete theory should explain
the entire shape observed experimentally. The problem was recently revisited in
Ref.~\cite{KUN-00} where all the assumptions underlying previous scaling approaches
\cite{LIE-72,DAH-96,NAR-96,SPA-96,ZAP-98} were carefully scrutinized and tested.
Through a combination of analytical methods and numerical simulations, Kuntz and Sethna
\cite{KUN-00} found a scaling relation describing the high frequency tail of the
spectrum which seems to be in good agreement with experiments \cite{DUR-02}.

The derivation of Ref.~\cite{KUN-00} follows different steps involving some assumptions
which can be tested numerically and experimentally. Close to criticality one expect
that the average size for an avalanche of duration $T$ scales as $S(T) \sim
T^{1/\sigma\nu z}$ and consistently the average shape scales as
\begin{equation} \label{eq:vshape}
v(T,t)=T^{1/\sigma \nu z-1}f_{shape}(t/T),
\end{equation}
where $v$ is the signal voltage, $t$ is the time and $f_{shape}(t/T)$ is a (possibly
universal) scaling function.

\begin{figure}
\centerline{\psfig{file=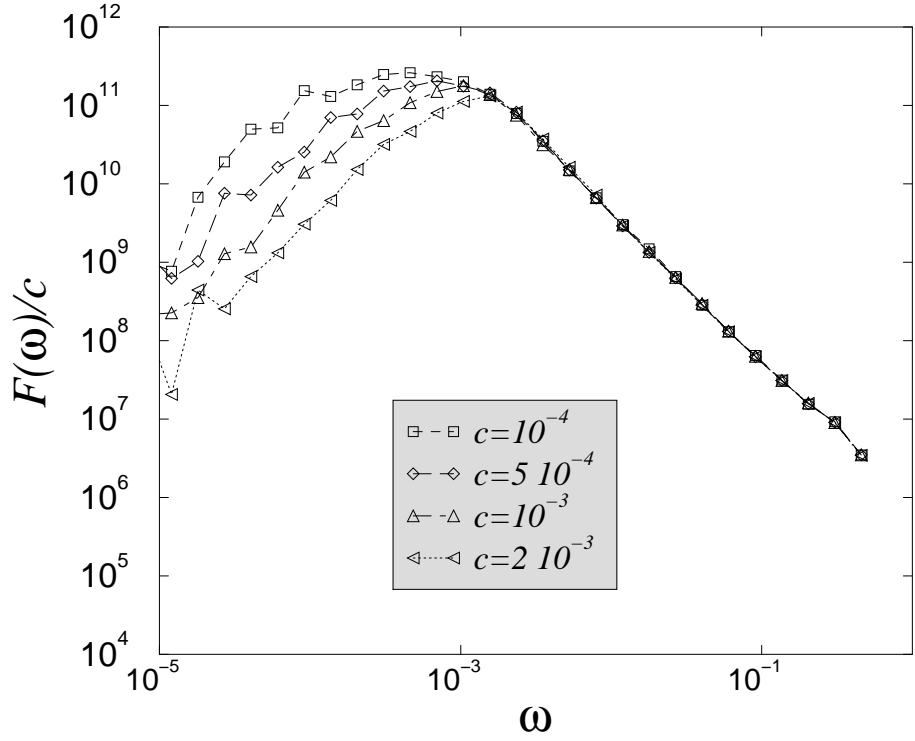,width=12cm,clip=!}} \caption{The power spectrum
measured in the short-range domain wall model for different field rates $c$. The tail
is well fitted by a power law $\omega^{-1.8}$, in good agreement with
Eq.~\protect\ref{eq:ps_kunz}, which predicts an exponent $1/\sigma \nu z =1.77$. }
\label{fig:psdw-short}
\end{figure}

The second assumption involves the probability $P(v|S)$ that a voltage $v$ occurs in an
avalanche of size $S$. This probability scales as:
\begin{equation}\label{eq:PVS}
  P(v|S)=v^{-1} f_{voltage}(vS^{\sigma \nu z-1})
\end{equation}
where $f_{voltage}$ is another universal scaling function. Using this equation one can
show that the energy $E=\langle v^2\rangle$ of an avalanche of size $s$ scales as $E(S)
\sim S^{2-\sigma\nu z}$. The analysis proceeds with the calculation of the voltage
correlation function
\begin{equation}
G(t)\equiv \int dt' \langle v(t+t')v(t) \rangle,
\end{equation}
which can be decomposed into various contributions for each avalanche $G(t|S)$. This
quantity obeys a scaling law
\begin{equation}
G(t|S)= S^{2-\sigma \nu z} f_G(tS^{-\sigma \nu z}).
\end{equation}
The contribution to the power spectrum is then obtained by a cosine transformation
\begin{equation}
F(\omega|S)=\int_0^\infty dt \cos(\omega t)G(t|S)=S^2 f_{energy} (\omega^{-1/\sigma \nu
z}S). \label{eq:Somega_s}
\end{equation}
To obtain the power spectrum of the signal one should average Eq.~\ref{eq:Somega_s}
over different avalanche sizes. A naive estimate of the integral resulting from the
average of Eq.~\ref{eq:Somega_s} using $P(S) \sim S^{-\tau}$ leads to the incorrect
prediction \cite{LIE-72,DAH-96,SPA-96} $F(\omega) \sim \omega^{-(3-\tau)/\sigma \nu
z}$. The reason behind this incorrect result comes from the fact that $f_{energy}(x)
\sim 1/x$ for large $x$. Thus, if $\tau<2$, the leading contribution to the averaged
power spectrum comes from the upper cutoff in the size distribution, ultimately leading
to the result \cite{KUN-00}
\begin{equation}
F(\omega) \sim \omega^{-1/\sigma \nu z}. \label{eq:ps_kunz}
\end{equation}
Using mean field exponents $\sigma \nu z=1/2$, we recover the ABBM result $F(\omega)
\sim \omega^{-2}$. The result in Eq.~\ref{eq:ps_kunz} agrees reasonably well with
experiments and is confirmed by numerical simulations of the domain wall model (see
Figs.~\ref{fig:psdw-long} and \ref{fig:psdw-short}).

\subsection{The avalanche shape}
\label{sec:th-shape}

\begin{figure}
\centerline{\psfig{file=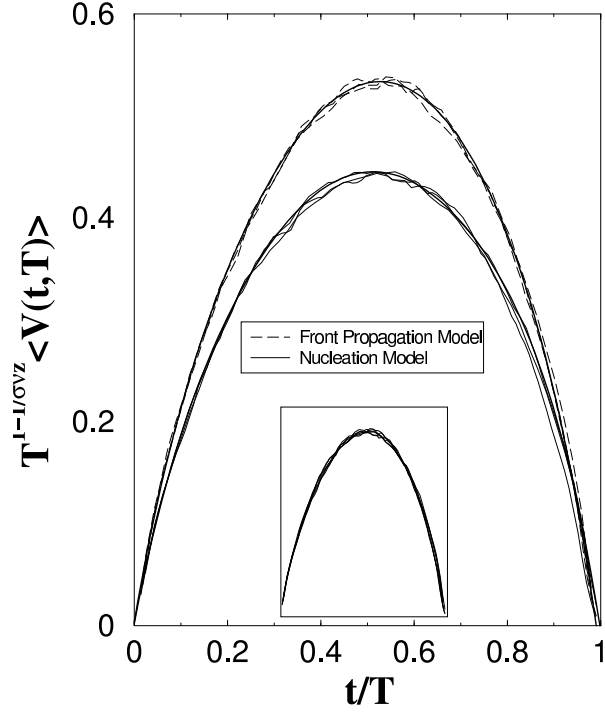,width=8cm,clip=!}} \caption{The avalanche shapes for
the RFIM with and without demagnetizing field (the second case is referred as front
propagation). The shapes for different times have been collapsed using the value
$1/z\sigma\nu=1.72$ and $1/z\sigma\nu=1.75$ in the two cases. An additional
normalization factor allows to rescale the two set of curves together. [From
\protect\cite{MEH-02}, Fig. 1, pg. 046139-3]} \label{fig:shape_MEH}
\end{figure}

The problem of the Barkhausen avalanche shape was first introduced in
Ref.~\cite{KUN-00} and proposed as a strong test to characterize the universality
classes of the Barkhausen noise. Subsequently, Ref.~\cite{MEH-02} analyzed more
systematically the results of numerical simulations for the RFIM with and without
demagnetizing field. The conclusion reached is that the average avalanche shape is in
both case symmetric and close to a parabolic shape. The pulse shape changes slightly
with and without demagnetizing field although a reasonable collapse of the two set of
curves can be obtained multiplying one set by a factor (see Fig.~\ref{fig:shape_MEH}).
Small differences emerge from a more detailed fitting procedure. A favorable comparison
with experiments and a theoretical explanation of the numerical results is, however,
still lacking.

The general problem of the shape of a sstochastic excursion was recently addressed in
more generality by Baldassarri et al. \cite{BAL-03}. The authors show that for a broad
class of stochastic processes the average shape of the excursion is a semicircle. This
occurs irrespective of the value of the scaling exponent (the equivalent of $\sigma\nu
z$). Thus in this sense the shape would appear to be more universal than the exponents.
On the other hand, the shape and symmetry of the average shape can be changed at will
introducing correlations in the stochastic process \cite{BAL-03} and it is even
possible to construct two stochastic processes with the sane scaling exponents, but
different shapes. This calls into question the idea that the shape is a universal
function \cite{COL-03}.

The approach of Ref.~\cite{BAL-03} can be extended to compute the shape exactly of the
Barkhausen pulses in the ABBM model \cite{COL-04a}.  Neglecting the contribution of the
demagnetizing factor $k$, Eq.~\ref{eq:abbm-log} describes the motion of a 1$d$ random
walk in a logarithmic potential $E(v)=-c \,log(v)$, where the magnetization $s$ plays
the role of time. The corresponding Fokker--Planck equation is
\begin{equation}
\frac{\partial P(v,s)}{\partial s}= \frac{\partial}{\partial v} \left(
-\frac{c}{v}+\frac{\partial}{\partial v} P(v,s) \right),
\end{equation}
where $P(v,s)$ is the probability to find the walk in $v$ at $s$. We are interested in
a solution of this equation with the initial condition $P(v,0)=\delta(v-v_0)$ and an
absorbing boundary at the origin $v=0$. Following Ref.~\cite{BRA-00} we can express
this solution in terms of modified Bessel functions. For the case $0<c<1$, which is the
condition of the ABBM model to have power laws in the avalanche distribution (see
Eq.~\ref{eq:pdv-def}, and Ref.~\cite{ALE-90}), the probability $P(v,s | v_0, 0;c)$ for
a walk starting at $v_0$ to be at $v$ after a ``time'' $s$, in the limit $v_0
\rightarrow 0$ is simply proportional to a power of $v$ times a Gaussian with variance
$s$, and the average excursion of the walk is then obtained as
\begin{equation}
\langle v \rangle \propto \sqrt{s(S-s)}
\end{equation}
This results means that the universal scaling function $g_{shape}$ is proportional to a
semicircle.

It is also possible to calculate the function $f_{shape}$ in the time domain
\cite{COL-04a}. By definition, the avalanche size $s$ at time $t$ is given by the
integral of $v(t,T)$ from time zero to time $t$:
\begin{equation}
s=\int_0^t dt' v(t',T)\propto T^{1/\sigma \nu z}\int_0^{t/T} f(x)dx \,,
\end{equation}
which provides an expression of $s=s(t,T)$ as a function of $t$ and $T$. If we now
impose
\begin{equation}
v(t,T)=v(s=s(t,T),S(T))
\end{equation}
we get an integral equation for $f_{shape}$  involving $g_{shape}$
\begin{equation}
f_{shape}(x) \propto g_{shape}\left(\int_0^{x} g(x')dx'\right) \,.
\end{equation}
Using the form of $g_{shape}$ computed previously, we can solve this equation with the
boundary conditions $f(0)=f(1)=1$. The solution is
\begin{equation}
f_{shape}(x) \propto \sin(\pi x) \,,
\end{equation}
i.e. the universal function $f_{shape}$ is  proportional to an arch of sinusoid.
Finally for the normalized avalanche one obtains
\begin{equation}
v(s,S)=S^{1-\sigma \nu z} \pi\sqrt{(s/S)(1-s/S)} \,,\label{eq:Vtnorm}
\end{equation}
\begin{equation}
\label{eq:Vsnorm} v(t,T)=T^{1/\sigma \nu z-1} \pi/2\sin(\pi t/T) \,.
\end{equation}

\subsection{Two dimensions\label{sec:th-2dim}}

The discussion presented so far focused on three dimensional systems, which is relevant
for most Barkhausen noise experiments. The experimental results obtained in thin
magnetic field require instead an analysis of two dimensional models. This task is not
straightforward and several new features arise in two dimensions. The demagnetizing
field, that plays a major role in determining the statistical properties of the
Barkhausen noise, vanishes for very thin samples.  For this reason, often we do not
have have a simple parallel domain structure as in three dimensional ribbons and an
analysis of single domain walls would not be appropriate. On the other hand, an
approach in terms of spin models suffers for different difficulties, since there is no
consensus on the presence of a disorder induced transition in two dimensions.

The RFIM has been extensively simulated in $d=2$, using extremely large systems sizes
(up to $(7000)^2$) \cite{PER-95,PER-99}. Despite this, it was not possible to obtain a
reliable estimate of the threshold $R_c$ for disorder induced transition. The
simulations are consistent either with a vanishing threshold or with very large
correlation critical exponents (i.e. $\nu=5.3\pm 1.4$, $1/\sigma=10\pm2$ with
$R_c=0.54$). An avalanche scaling exponent has also been estimated as
$\tau_{int}=2.04\pm 0.04$. Additional results and analysis have appeared in the
literature, but are not particularly reliable, being based on much smaller system
sizes. Typically the presence of a disorder induced transition is assumed and exponents
are evaluated \cite{TAD-00,ZHE-02,ZHE-02a}. A similar discussion would apply other
models analyzed in the literature, like the RBIM \cite{VIV-94,VIV-95}, the RFIM with
site dilution \cite{TAD-96} and dipolar interactions \cite{MAG-99}. In summary, the
large scale hysteretic behavior of disordered spin models in $d=2$ is quite
controversial and quantitative predictions seem at the moment not available. A
possibility is that $d=2$ is the lower critical dimensions, as it is expected for these
models in equilibrium \cite{IMR-75}. If this is the case, power law scaling may still
be observed for low disorder, with the conjecture that $\tau=3/2$ \cite{PER-95}.
Nevertheless, we feel that at present the results of both experiments and models are
still unreliable to a make a comparison.

Even in thin films there are several cases in which a domain wall model may be
appropriate. In particular, in the magnetoptical measurements of
Ref.~\cite{KIM-03,KIM-03a} one observes directly Barkhausen jumps due to domain wall
motion. The equation of motion for a single domain wall in two dimensions is similar to
the one in three dimensions (e. g. Eq.~\ref{eq:tot}) with some notable differences. For
thin films with in plane magnetization the demagnetizing factor $k$ is virtually zero
and in Fourier space the dipolar kernel (Eq.~\ref{eq:ker}) does not scale as $q$ but as
$q^2 \log(aq)$ where $a$ is a small scale cutoff \cite{NAT-83,CHU-95}. Apart from the
logarithmic correction, the dipolar term is similar to the domain wall energy and thus
in principle should not significantly affect the universality class of the depinning
transition.  Numerical simulations, however, yield slightly different results in the
two cases (for instance $\zeta=1$ with long-range forces and $\zeta=1.2$ without
\cite{TAN-98,ZAP-01a}). While the short-range interface model has been studied
extensively in two dimensions, less is known for the long-range problem. A result that
could be relevant for the Barkhausen effect would be the avalanche distribution
exponent which has been numerically estimated for short-range models: $\tau=1.13\pm
0.02$ \cite{PAC-96}, $\tau=1.02\pm 0.29$ \cite{CHE-95}. The most reliable and precise
estimate is probably $\tau=1.115$ \cite{ALA-02}.

Before comparing simulation data with the experiments one has to be certain of the
quantity effectively measured. When the demagnetizing factor is zero, a ramp up of the
field does not lead to a Barkhausen stationary signal with $H_{eff}=H_c$, but the
avalanches grow up to $H_c$. One would then measure an effective exponent
$\tau_{int}=\tau+\sigma$. Given the relation $\sigma=\nu/(d+\zeta)$ and $\zeta=1.25$
and $\nu=1.33$ \cite{LES-97} we obtain $\tau_{int}\simeq 1.6$. If instead of ramping up
the field, one would instead keep it fixed just below $H_c$ as in
Ref.~\cite{KIM-03,KIM-03a}, then we would expect that the avalanche distribution scales
with $\tau$. Notice, however, that the avalanche ``size'' $S$ defined in
Ref.~\cite{KIM-03,KIM-03a} is not what we have defined, but is instead proportional to
the change in the displacement $\Delta h$. Avalanches are only counted when the
transverse length $l$ (see Fig.~\ref{fig:avalanche}) is larger than the system size
$L$. Hence $S=L \Delta h$ with a constant $L$. Using scaling relation one can show that
the exponent measured $\tau_h = ((\zeta+1)\tau-1)/\zeta \simeq 1.2$, which is still
lower than the estimated value but could be consistent with the data.

Finally, we mention that spin models have been also used to study the dynamics of a
single domain wall including overhangs. Simulations have shown a crossover from flat to
fractal depending on the disorder strength \cite{JI-91}. The relation between these
phenomenon and the Barkhausen noise still remains to be explored.

\section{Conclusions and perspectives}

The Barkhausen effect has been studied for almost a century. While the essence of the
phenomenon, i.e. an irregular magnetization reversal process occurring at the
microscopic scale, was already indicated in the original paper by Barkhausen, a
quantitative and detailed understanding was gained only recently. The key to this
achievement is twofold: the injection in the field of methods and theories drawn from
statistical mechanics of non-equilibrium critical phenomena and the identification of a
standard experimental setup to collect reproducible and statistically significant data.
These two aspects are tightly linked, since the necessity of rationalizing the
experimental setup and to perform a detailed statistical analysis of the data was
dictated by the theory, which was stimulated on its turn by the indication of scaling
behavior coming from experiments.

The successful understanding outlined above is restricted mostly to the Barkhausen
noise measured in bulk soft magnetic alloys, ribbons in particular. These are
particularly instructive from the theoretical point of view because of the relatively
simple parallel domain structure, induced by the interplay between dipolar forces and
the sample geometry. The magnetization process is mainly due to domain wall motion in a
disordered landscape which is responsible for the avalanche dynamics leading to the
observed noise. The scaling behavior describing the statistical properties of the noise
(avalanche distributions, power spectrum) can be related to an underlying domain-wall
depinning transition. Critical phenomena are typically associated to a certain degree
of universality: the scaling exponents do not depend on the microscopic details of the
system but only on general symmetries, dimensionality, and long-range properties of the
interactions. One would thus expect to be able to classify materials in broad
universality classes, sharing the same critical exponents, irrespective to the
microscopic details, such as sample composition, grain size or disorder strength.

To some extent experimental results confirm this expectation: polycrystalline SiFe
alloys with different grain sizes and Si content, together with partially crystallized
Co based alloys were all found to share the same critical exponents. Amorphous alloys
of different composition are instead grouped in a different class characterized by a
different set of exponents. These two groups of exponents coincide with those
associated to the depinning transition of a domain wall, with or without long-range
dipolar interactions. The latter are enhanced by local anisotropies, induced by the
polycrystalline grains and absent in the amorphous structure. The domain wall depinning
scenario successfully predicts additional features of the Barkhausen effect, such as
the dependence on the field rate and the demagnetizing factor, and can thus be used as
a basis to understand the role of the microstructure on the macroscopic magnetization
properties.

The new frontier of the Barkhausen effect is represented mainly by thin films, as
testified the recent literature. Inductive methods, that have played a leading role in
noise measurements in bulk materials, become less effective as the thickness of the
film is reduced to smaller and smaller sizes. Magneto optical techniques do not suffer
of this problem and the improved resolution gained in time and space is allowing for
precise noise measurements, joined to a direct visual inspection of domain wall
dynamics. The effort in this direction, however, is still in process and a systematic
analysis of the noise statistics in different materials has not yet been completed.
Other techniques, such as magnetic force microscopy \cite{SCH-04} or Hall sensors
\cite{DAM-87}, are emerging as powerful tools to study the Barkhausen noise at the
nanoscale. For instance, a recent experiment revealed Barkhausen jumps due to a domain
wall depinning from the atomic Peierls potential \cite{NOV-03}. As a note of warning,
we note that these techniques reveal the magnetic properties of the surface and are
thus not effective to study the crossover from bulk three-dimensional to thin two
dimensional behavior. To this end, it would be desirable to improve inductive
techniques, as recently proposed.

Apart from the experimental problems outlined above, thin films pose new theoretical
challenges. It is unlikely that a simple translation of three-dimensional models to two
dimensions will provide a significant understanding of the Barkhausen effect in thin
films. The domain structure is in most cases more complicated than in the bulk and a
single domain wall model will be inappropriate. The general framework of
non-equilibrium critical phenomena and the understanding gained in domain wall and spin
models could nevertheless lead to a successful theoretical understanding.

Finally, we would like to mention here hard ferromagnetic materials for which few
Barkhausen noise measurements have been reported in the literature
\cite{CUN-96,THO-97,BAS-04}. In these nucleation-type materials, the noise is not due
to the motion of domain walls, but exclusively to grain reversal. Correlation between
neighboring grains play a significant role \cite{CUN-96}, as well as the temperature.
These systems, having large local anisotropies and nucleation fields, could be well
represented by disordered spin models such as the RFIM.  Recent studies of these models
have contributed to renew the interest in the Barkhausen effect, but a clear
experimental realization is still lacking.

\section{Acknowledgments}
We wish to thank all our collaborators who have greatly contributed to our current
understanding of the Barkhausen effect: M. J. Alava, V. Basso, G. Bertotti, P. Cizeau,
F. Colaiori, C. Castellano,  R. da Silveira, A. Magni, L. Santi, R. L. Sommer, H. E. Stanley, M. B. Wiessmann. SZ  thanks Ingeborg Walter and Michael Zaiser for their help in the translation.

\appendix
\section{Translation of the original paper by Barkhausen}
\centerline{Two phenomena discovered with the help of the new amplifier} \centerline{by
H. Barkhausen}
\bigskip
\centerline{1. Noise during magnetization of Iron}

With the improvement of the new vacuum tube amplifier it is possible to generate easily
a 10000 times larger current, which means a 100 million gain in power. With this
apparatus, one can reveal electric or magnetic alternate fields that otherwise, due to
their weakness, would be unaccessible to our knowledge. In some sense it is as if we
discovered a microscope, and one with almost 10000 times magnification! Shortly after
W. Shottky has indicated that, through a large amplification, one can, in some sense,
hear the electrons flying in the amplifier tube, since the current they generate reveal
the spontaneous fluctuations predicted by the kinetic gas theory.

I discovered a similar phenomenon two years ago in occasion of an experiment that I did
in collaboration with Dr. T. Tuczek. Iron produces a noise when magnetized. As the
magnetomotive force is smoothly varied, the molecular magnets flip in jumps to their
new position. Because of this, they generate irregular induction pulses in a coil wound
around the sample, that can then be heard as a noise in a telephone.

\begin{figure}
\begin{center}
   \includegraphics[width=8cm]{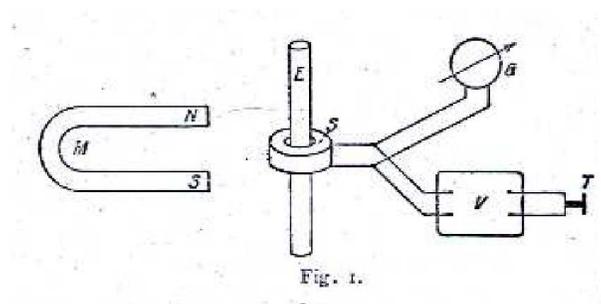}
   \end{center}
\caption{Figure 1 of the original paper \label{fig:a1}}
\end{figure}

I have now examined this phenomenon more closely. Fig.~\ref{fig:a1} shows the
experimental apparatus. The iron sample E to be studied was put inside a small coil S
with a diameter of 25mm and 300 turns. One end of the coil was connected to a telephone
T through a 10000 times amplifier V and the other end to a mirror galvanometer G. The
latter was so strongly damped that a deflection was staying almost unaltered for
several seconds. Then each modification of the amplitude in practice is proportional to
the corresponding modification of the induction flux of the coil. Displacing the
U-shaped 10cm long magnet M, one could observe at the same time the modification of the
induction and the noise strength. I determined the contribution of the iron sample to
the former, measuring and subtracting the smallest observable deflection of the
galvanometer for a given displacement of the magnet, but in absence of the iron sample.
The displacement was made by hand. For this reason, the precision was not so large.

Next, it was shown that the noise became weaker for larger samples. 20mm large samples
yielded effectively no more noise. As will be shown later, the reason for this is
related to the too small variation of the induction (per unit area). The small magnet
was not sufficient to magnetize strongly enough the large samples. So a pile of slabs
gave rise to a weaker noise than a single slab coming from the same pile. The most
suitable samples were found to be 1 to 1/2 mm large wires or 5 to 10mm thin ribbons.
Employing stronger magnets one could also use stronger iron samples.

Later it was established on a large number of different samples that the noise was
stronger as the iron was softer. Hardened steel gave almost no noise, while soft
annealed iron yielded a particularly strong noise. The cause of this, however, was not
the well known difficulty in magnetizing steel, that showed itself through a somewhat
smaller deflection of the galvanometer given  the same position of the magnet. The
distinction in the noise existed also when soft iron was magnetized two to three times
more weakly. Very soft iron still produced a noise when the magnet was moved at a 1/2m
distance. When it was placed in the close vicinity, the noise was so strong that it
could be heard clearly in the telephone even without the amplifier. The distinction
between different kinds of iron is so large that one could really build on this a
method to investigate iron.

\begin{figure}
\begin{center}
   \includegraphics[width=8cm]{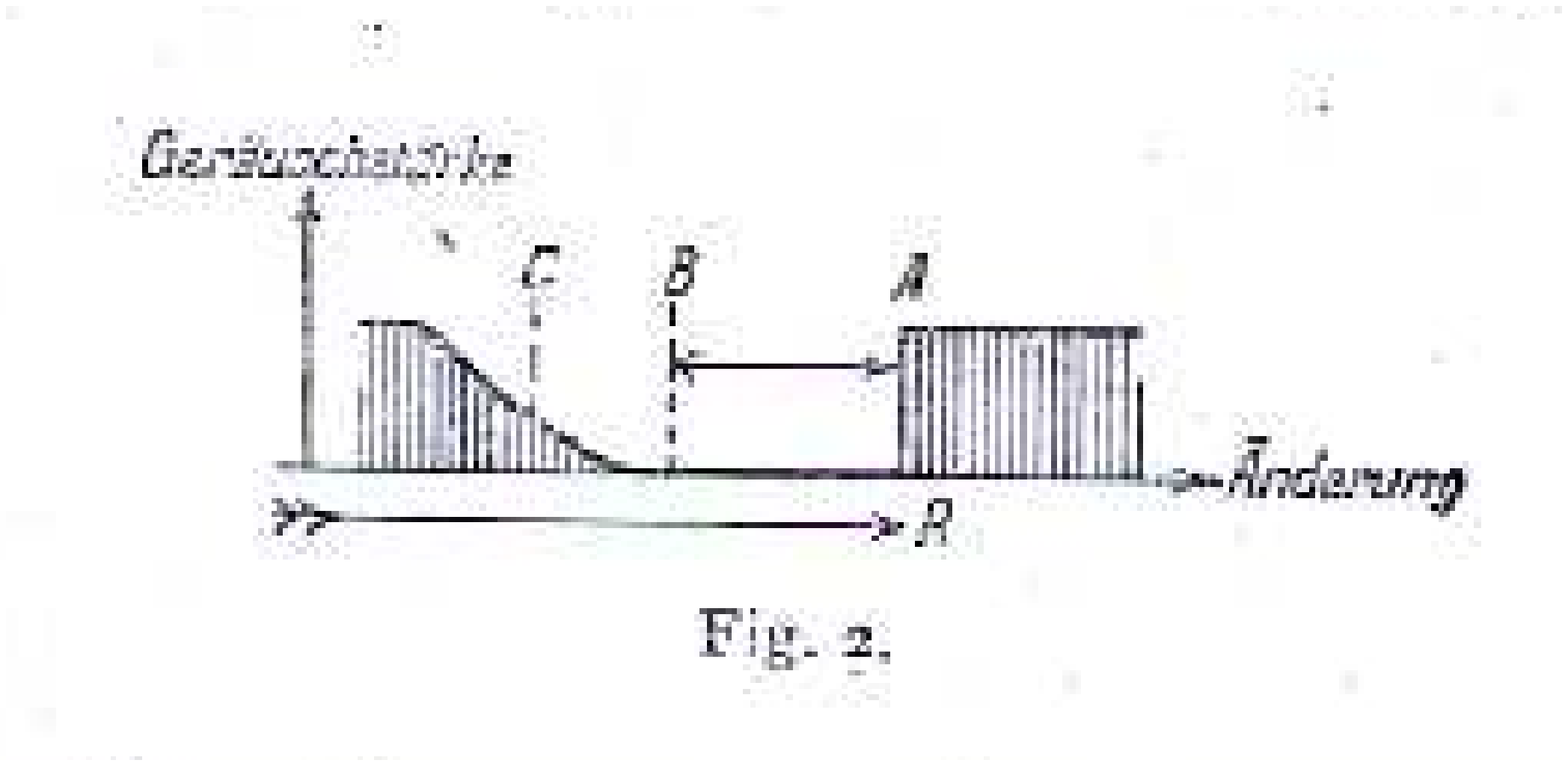}
   \end{center}
\caption{Figure 2 of the original paper} \label{fig:a2}
\end{figure}

\begin{figure}
\begin{center}
   \includegraphics[width=8cm]{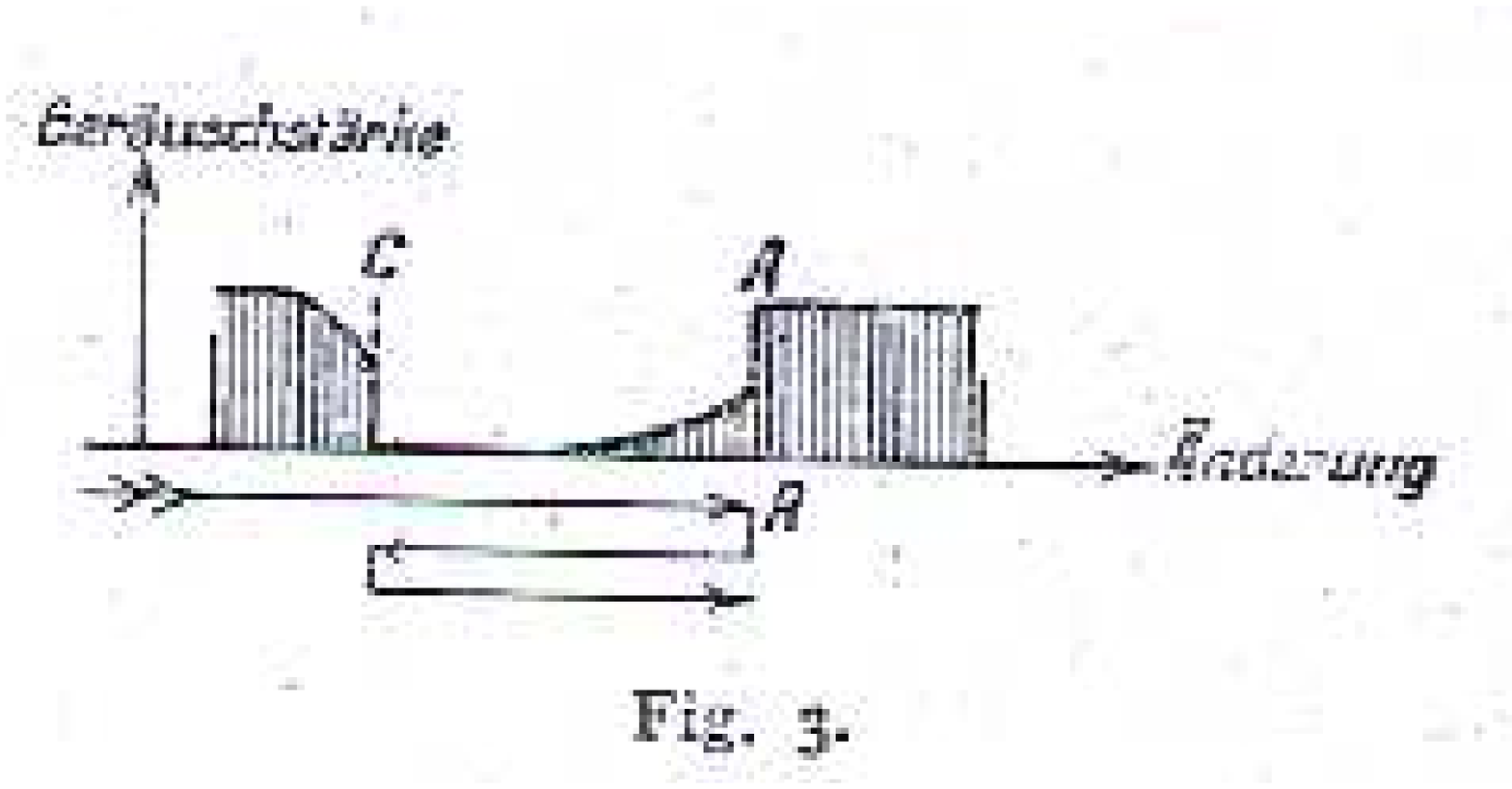}
   \end{center}
\caption{ \label{fig:a3}Figure 3 of the original paper}
\end{figure}

The magnetic field variation was mostly done in such a way that the magnet was either
displaced away by the side, turning it at 180$^\circ$ and then moving it back, or
slowly turned away for 180$^\circ$ from a given position, in particularly close
vicinity. The noise due to this motion was very strong in particular positions. The
variations of the deflection of the galvanometer did not always correspond to the noise
strength. When a variation in the direction R stopped at a point A (Fig.~\ref{fig:a2}),
all was suddenly quite and stayed perfectly quite as one made a variation back from A
to B, even when the magnet was moved at will back and forth between A and B. As soon as
the magnet overcome even slightly the position A, suddenly the full noise started
again. In addition, by a further displacement in the direction R, the boundaries A and
B were in some sense shifted away to the right. Then the boundary A remained again
surprisingly sharp in correspondence to any motion between A and B. The boundary B was
not sharp. The noise grew up slowly when one went back through B until C. At the same
time, the boundary A became as well less sharp, in the sense sketched in
Fig.~\ref{fig:a3}. The galvanometer also showed quite strong variations in the region
between A and B. No jump could be detected during its motion, when A was crossed and
the noise suddenly started. Clearly the observation method was not very precise.

For thick samples it would be perhaps the same for a stronger field, but the region A-B
was not reachable with the changes produced by the present magnet. Thus, as remarked
above, with this no noise could be generated.

One could maybe explain this peculiar behavior of iron in the following way: the
individual molecular magnets group themselves into assemblies of various sizes with
different degrees of stability. Noise results only from the disruption or formation of
greater assemblies. For a motion corresponding to Fig.~\ref{fig:a2}, only small
assemblies are modified in the region A-B and noise is does not start. For a motion
across B in the direction -R, little by little also medium and large assemblies get
involved. In the direction +R, all small and middle size assemblies have been already
dissolved in the first change coming from far away up to A. Only the largest and more
stable assemblies in this direction are still remaining. Those will be first dissolved
by a further change beyond A.

In good agreement with this idea, moving across B one hears a uniform, low sough, while
a loud crackling noise starts across A. For very slow variations, for instance when the
magnet is moved far away from the sample, one can clearly distinguish isolated snapping
hits. It appears that in steel the molecular magnets do not have the ability to join in
large assemblies. In this material one only hears the low sough. Also the behavior
described in Fig.~\ref{fig:a3} is in agreement with this idea. The weak assemblies,
magnetized in the backward motion (from B to C in Fig.~\ref{fig:a2}), are overcome
again in the direction R, before A is reached.

A phenomenon occurring in a stripe, thin as paper, of soft iron sheet is still non
clearly understood. A turn of the magnet by 180$^\circ$ gave rise to the stronger noise
when the magnet was placed at a distance of 10cm.  By turning it at shorter distance,
the noise was surprisingly very weak, even when the magnet was turned very slowly.

A similar analysis would be desirable with the magnetodetector of Marconi and with the
telegraph, also in view of practical applications.

Dresden, Institut f\"{u}r Schwachstromtechnik, Mai 1919.



\end{document}